\documentclass{emulateapj}
\shorttitle{A Steep Faint-End Slope at $z\sim 2-3$}
\shortauthors{Reddy \& Steidel}
\slugcomment{Received 2008 July 15; Accepted 2008 October 10}

\begin{document}

\newcommand{\umg}{U_{\rm n}-G}
\newcommand{\rmk}{{\cal R}-\ks}
\newcommand{\ebmv}{E(B-V)}
\newcommand{\sfr}{{\rm M}_{\odot} ~ {\rm yr}^{-1}}
\newcommand{\ks}{K_{\rm s}}
\newcommand{\ugr}{U_{\rm n}G{\cal R}}
\newcommand{\bzk}{BzK}
\newcommand{\rs}{{\cal R}}
\newcommand{\gmr}{G-\rs}
\newcommand{\lya}{Lyman~$\alpha$}
\newcommand{\lyb}{Lyman~$\beta$}
\newcommand{\za}{$z_{\rm abs}$}
\newcommand{\ze}{$z_{\rm em}$}
\newcommand{\cmtwo}{cm$^{-2}$}
\newcommand{\nhi}{$N$(H$^0$)}
\newcommand{\degpoint}{\mbox{$^\circ\mskip-7.0mu.\,$}}
\newcommand{\kms}{\,km~s$^{-1}$}      
\newcommand{\minpoint}{\mbox{$'\mskip-4.7mu.\mskip0.8mu$}}
\newcommand{\peryr}{\mbox{$\>\rm yr^{-1}$}}
\newcommand{\secpoint}{\mbox{$''\mskip-7.6mu.\,$}}
\newcommand{\sqdeg}{\mbox{${\rm deg}^2$}}
\newcommand{\squig}{\sim\!\!}
\newcommand{\subsun}{\mbox{$_{\twelvesy\odot}$}}
\newcommand{\et}{{\rm et al.}~}

\def\ltsima{$\; \buildrel < \over \sim \;$}
\def\simlt{\lower.5ex\hbox{\ltsima}}
\def\gtsima{$\; \buildrel > \over \sim \;$}
\def\simgt{\lower.5ex\hbox{\gtsima}}
\def\arcs{$''~$}
\def\arcm{$'~$}
\def\erf{\mathop{\rm erf}}
\def\erfc{\mathop{\rm erfc}}

\title{A STEEP FAINT-END SLOPE OF THE UV LUMINOSITY FUNCTION AT $z\sim 2-3$:
IMPLICATIONS FOR THE GLOBAL STELLAR MASS DENSITY AND STAR FORMATION IN LOW
MASS HALOS\altaffilmark{1}} \author{\sc Naveen
A. Reddy\altaffilmark{2,3} \& Charles C. Steidel\altaffilmark{4}}

\altaffiltext{1}{Based, in part, on data obtained at the W.M. Keck
Observatory, which is operated as a scientific partnership among the
California Institute of Technology, the University of California, and
NASA, and was made possible by the generous financial support of the
W.M. Keck Foundation.}

\altaffiltext{2}{National Optical Astronomy Observatory, 950 N. Cherry 
Ave, Tucson, AZ 85719}
\altaffiltext{3}{Hubble Fellow.}
\altaffiltext{4}{California Institute of Technology, MS 105--24, Pasadena,
CA 91125}

\begin{abstract}

We use the deep ground-based optical photometry of the Lyman Break
Galaxy (LBG) Survey to derive robust measurements of the faint-end
slope ($\alpha$) of the UV luminosity function (LF) at redshifts
$1.9\le z\le 3.4$.  Our sample includes $>2000$ spectroscopic
redshifts and $\approx 31000$ LBGs in $31$ spatially-independent
fields over a total area of $3261$~arcmin$^{2}$.  These data allow us
to select galaxies to $0.07L^{\ast}$ and $0.10L^{\ast}$ at $z\sim 2$
and $z\sim 3$, respectively.  A maximum-likelihood analysis indicates
steep values of $\alpha(z=2)=-1.73\pm0.07$ and
$\alpha(z=3)=-1.73\pm0.13$.  This result is robust to
luminosity-dependent systematics in the Ly$\alpha$ equivalent width
and reddening distributions, is similar to the steep values advocated
at $z\ga 4$, and implies that $\approx 93\%$ of the unobscured UV
luminosity density at $z\sim 2-3$ arises from sub-$L^{\ast}$ galaxies.
With a realistic luminosity-dependent reddening distribution, faint to
moderately luminous galaxies account for $\ga 70\%$ and $\ga 25\%$ of
the bolometric luminosity density and present-day stellar mass
density, respectively, when integrated over $1.9\le z<3.4$.  We find a
factor of $8-9$ increase in the star formation rate density between
$z\sim 6$ and $z\sim 2$, due to both a brightening of $L^{\ast}$ and
an increasing dust correction proceeding to lower redshifts.
Combining the UV LF with stellar mass estimates suggests a relatively
steep low mass slope of the stellar mass function at high redshift.
The previously observed discrepancy between the integral of the star
formation history and stellar mass density measurements at $z\sim 2$
may be reconciled by invoking a luminosity-dependent reddening
correction to the star formation history combined with an accounting
for the stellar mass contributed by UV-faint galaxies.  The steep and
relatively constant faint-end slope of the UV LF at $z\ga 2$ contrasts
with the shallower slope inferred locally, suggesting that the
evolution in the faint-end slope may be dictated simply by the
availability of low mass halos capable of supporting star formation at
$z\la 2$.

\end{abstract}

\keywords{galaxies: evolution --- galaxies: formation --- galaxies:
high redshift --- galaxies: luminosity function}

\section{INTRODUCTION}
\label{sec:intro}

The last decade has seen significant advances in our understanding of
the history of star formation and stellar mass assembly.  Today, one
can find several hundred determinations of the star formation rate
density (SFRD) estimated from observations at many wavelengths across
a large range of lookback time.  Taken together, these observations
suggest a rapid increase in the SFRD from the epoch of reionization to
$z\sim 1-2$, after which time the SFRD steadily decreased over the
last $\sim 10$~billion years.  This picture is generally understood in
the context of hierarchical buildup at early times and gas exhaustion
or heating at late times.  While this characterization of the star
formation history is broadly accepted, there are several key details
that are missing from this interpretation, including the potentially
important contribution of UV-faint (sub-$L^{\ast}$) galaxies to the
census of star formation and baryon budget.  If rest-UV/optical light
is a biased tracer of galaxy formation --- particularly at high
redshift where most of the baryons in galaxies are likely to reside in
cold gas \citep{prochaska08} --- then faint galaxies may constitute an
important population for studying the process of star formation and
feedback.  Further, the number density of both bright and faint
galaxies departs significantly from expectations based on the
$\Lambda$CDM model, suggesting a regulation of star formation in both
luminosity regimes.  In this paper, we present an extension of our
previous work on the UV luminosity function at $z\sim 2-3$ in order to
provide robust constraints on the prevalence of UV-faint galaxies at
epochs when galaxies were forming most of their stars.

The luminosity function (LF) is a fundamental probe of galaxy
formation and evolution, and can be used to address the relative
importance of bright and faint galaxies to the energy budget at a
given epoch.  Furthermore, comparison with the dark matter halo
distribution informs us of the efficiency of star formation and
effects of feedback at different mass scales \citep{rees77, silk98,
dekel06}.  Therefore, constraining accurately the shape of the LF is a
necessary step in acquiring a more complete census of galaxies and
elucidating the relationship between the baryonic processes that
govern galaxy evolution and the dark matter halos that host them.

The UV LF is relevant in several respects.  Rest-frame UV emission is
a direct tracer of massive star formation, modulo the effects of dust.
Rest-frame UV observations of high redshift galaxies are generally not
limited by spatial resolution and the deepest observations are up to a
factor of $\approx 2000$ times more sensitive than those in the
infrared and longer wavelengths.  The combination of resolution,
sensitivity, and the accessibility of UV wavelengths over almost the
entire age of the Universe makes the UV LF a unique tool in assessing
the star formation history.  The relative efficiency of UV-dropout
selection has enabled a number of investigations of the UV LF at high
redshifts based on photometrically-selected samples \citep{steidel99,
adel00, yan04b, bunker04, dickinson04, bouwens07, sawicki06a, reddy08,
ouchi04, beckwith06, yoshida06, iwata07}.

Apart from the uncertainties that can be constrained from photometry
alone, such as photometric errors and field-to-field variations,
spectroscopy is a critical means of quantifying several important
systematics that can affect the LF.  These include contamination from
low redshift objects and high redshift AGN/QSOs, attenuation of UV
emission due to the opacity of the intergalactic medium (IGM), and
perturbation of galaxy colors due to Ly$\alpha$, reddening, and
stellar population ages of galaxies.  The relevance of these
systematic effects is underscored by the fact that while there are
numerous studies of the UV LF, the results have been inconsistent,
both at low ($z\la 3$; e.g., \citealt{reddy08,lefevre05}) and at high
($z\ga 4$; e.g., \citealt{beckwith06, bouwens07}) redshifts.

Unfortunately, spectroscopic surveys have been limited to UV-bright
($\rs\la 25.5$) galaxies at relatively low redshifts ($z\la 4$) and
spectroscopic constraints on the number density of faint sources at
$z\ga 2$ are still lacking.  Deeper spectroscopy is expensive and will
remain so until the next generation of large ($\ga 10$~m) ground-based
telescopes come online.  Given the present practical limitations and
the complexity of systematics involved in computing LFs, it seems
prudent to revisit and extend our initial estimate of the UV LF
(\citealt{reddy08}, hereafter R08) by evaluating the impact of these
systematics on the inferred number density of faint galaxies at $z\sim
2-3$.  We go beyond the initial analysis of R08 by quantifying several
important effects relevant in the computation of star formation rate
and stellar mass densities, including the effects of
luminosity-dependent dust corrections and the integrated stellar mass
of low mass galaxies.

To this end, we combine what we know about the spectroscopic
properties of LBGs at $z\sim 2-3$ with deep ground-based optical data
in 31 spatially independent fields to determine the faint-end slope
with greater precision.  A brief description of the LBG survey,
photometry, and followup spectroscopy is given in
\S~\ref{sec:selection}.  Our method for computing the UV LF is
presented in \S~\ref{sec:maxlik}, and results are discussed in
\S~\ref{sec:results}.  Our results are compared with those in the
literature and analyzed in the context of the early Hubble Deep Field
(HDF) results, and we assess the impact of sample variance in
\S~\ref{sec:lsc}.  The evolution of the UV LF is quantified in
\S~\ref{sec:lfevol}.  The contribution to the faint-end population
from dusty, star-forming galaxies as well as quiescently-evolving
galaxies with large stellar masses is discussed in
\S~\ref{sec:nature}.  \S~\ref{sec:sfrd} and \ref{sec:sfrdevol} present
constraints on the star formation rate density (SFRD) and its
evolution.  We also reassess the stellar mass density at $z\sim 2$ and
compare it to inferences from integrating the star formation history.
Finally, the evolution of the faint-end slope of the UV LF is
discussed briefly in \S~\ref{sec:alpha}.  All magnitudes are expressed
in AB units, unless stated otherwise.  Unless indicated, a
\citet{kroupa01} IMF is assumed.  A flat $\Lambda$CDM cosmology is
assumed with $H_{0}=70$~km~s$^{-1}$~Mpc$^{-1}$,
$\Omega_{\Lambda}=0.7$, and $\Omega_{\rm m}=0.3$.

\section{DATA: SAMPLE SELECTION AND SPECTROSCOPY}
\label{sec:selection}

\subsection{Fields and Photometry}
\label{sec:fields}

The Lyman Break Galaxy (LBG) survey is being conducted in fields
chosen primarily for having relatively bright background QSOs with
which to study the interface between the IGM and galaxies at $z\sim
2-3$ \citep{adel03, adelberger04, adelberger05b, steidel03,
steidel04}.  Additionally, the survey was extended to include fields
that are the focus of multi-wavelength campaigns, including the
Groth-Westphal \citep{steidel03,groth94} and GOODS-North fields
\citep{dickinson03,giavalisco04}.  Presently, the survey includes 31
fields, 29 of which have been targeted spectroscopically (see R08 for
further details).  As emphasized throughout this paper, the large
number of spatially independent fields provides a precise handle on
the magnitude of sample variance, an effect that has limited the
conclusions that could be drawn from previous determinations of the
LF.  For this analysis, we have included two additional fields beyond
the 29 that were presented in R08, ``Q1603'' and ``Q2240''.
Instruments used and dates of observation are presented in
\citet{steidel03, steidel04}.  For ease of reference, the fields are
listed in Table~\ref{tab:fields}; together they include an area of
3261 arcmin$^{2}$, or $0.91$~square degrees.

\begin{deluxetable*}{lcrrrrr}
\tabletypesize{\footnotesize}
\tablewidth{0pc}
\tablecaption{Survey Fields}
\tablehead{
\colhead{} &
\colhead{$\alpha$\tablenotemark{a}} &
\colhead{$\delta$\tablenotemark{b}} &
\colhead{Field Size} &
\colhead{} &
\colhead{} & 
\colhead{} \\
\colhead{Field Name} &
\colhead{(J2000.0)} &
\colhead{(J2000.0)} &
\colhead{(arcmin$^{2}$)} &
\colhead{$\rs_{\rm lim}$} & 
\colhead{$N_{\rm BX}$\tablenotemark{c}} &
\colhead{$N_{\rm LBG}$\tablenotemark{d}}}
\startdata
Q0000	&	00 03 25 & -26 03 37 &	\*18.9	& 25.5 &	78 &	29 \\	
CDFa	&	00 53 23 & 12 33 46 &	\*78.4	& 26.0 &	490 &	192 \\	
CDFb	&	00 53 42 & 12 25 11 &	\*82.4	& 25.5 &	347 &	123 \\	
Q0100	&	01 03 11 & 13 16 18 &	\*42.9	& 26.5 &	579 &	230 \\	
Q0142	&	01 45 17 & -09 45 09 &	\*40.1	& 26.0 &	379 &	158 \\	
Q0201	&	02 03 47 & 11 34 22 &	\*75.7	& 26.0 &	289 &	114 \\	
Q0256	&	02 59 05 & 00 11 07 &	\*72.2	& 25.5 & 	325 &	105 \\	
Q0302	&	03 04 23 & -00 14 32 &	244.9	& 25.5 &       2113 &	1025 \\	
Q0449	& 	04 52 14 & -16 40 12 &	\*32.1	& 26.5 &	287 &	138 \\	
B20902	&	09 05 31 & 34 08 02 &	\*41.8	& 25.5 &	229 &	72 \\	
Q0933	&	09 33 36 & 28 45 35 &	\*82.9	& 26.0 &	723 &	313 \\	
Q1009	&	10 11 54 & 29 41 34 &	\*38.3	& 26.5 &	512 &	305 \\	
Q1217	&	12 19 31 & 49 40 50 &	\*35.3	& 26.0 &	311 &	83 \\	
GOODS-N	&	12 36 51 & 62 13 14 &	155.3	& 26.0 &	496 &	154 \\	
Q1307	&	13 07 45 & 29 12 51 &	258.7	& 26.0 &	2011 &	718 \\	
Westphal &	14 17 43 & 52 28 49 &	226.9	& 25.5 &	783 &	289 \\	
Q1422	&	14 24 37 & 22 53 50 &	113.0	& 26.0 &       1041 &	518 \\	
3C324	&	15 49 50 & 21 28 48 &	\*44.1	& 25.5 &	187 &	56 \\	
Q1549	&	15 51 52 & 19 11 03 &	\*37.3	& 26.0 &	329 &	153 \\	
Q1603   &       16 04 56 & 38 12 09 &   \*38.8    & 26.5 &        396 &   160 \\
Q1623	&	16 25 45 & 26 47 23 &	290.0	& 26.0 &	1878 &	735 \\	
Q1700	&	17 01 01 & 64 11 58 &	235.3	& 26.0 &	2263 &	609 \\	
Q2206 	&	22 08 53 & -19 44 10 &	\*40.5	& 26.0 &	257 &	70 \\	
SSA22a	&	22 17 34 & 00 15 04 &	\*77.7	& 25.5 &	274 &	183 \\
SSA22b	&	22 17 34 & 00 06 22 &	\*77.6	& 26.0 &	435 &	217 \\
Q2233	&	22 36 09 & 13 56 22 &	\*85.6	& 26.0 &	420 &	173 \\
DSF2237b &	22 39 34 & 11 51 39 &	\*81.7	& 26.5 &	1004 &	474 \\
Q2240   &       22 40 02 & 03 17 50 &   \*35.9    & 26.0 &        421 &   176 \\ 
DSF2237a &	22 40 08 & 11 52 41 &	\*83.4	& 26.5 &	553 &	183 \\
Q2343	&	23 46 05 & 12 49 12 &	212.8	& 25.5 &	1209 &	436 \\
Q2346	&	23 48 23 & 00 27 15 &	280.3	& 26.0 &	1547 &	472 \\
\\
{\bf TOTAL}	&	{\bf ...} & {\bf ...} &	{\bf 3260.8} & {\bf ...} & {\bf 22166} &	{\bf 8663} \\
\enddata
\tablenotetext{a}{Right ascension in hours, minutes, and seconds.}
\tablenotetext{b}{Declination in degrees, arcminutes, and arcseconds.}
\tablenotetext{c}{Number of BX candidates to the limiting magnitude given in column (5).}
\tablenotetext{d}{Number of LBG candidates to the limiting magnitude given in column (5).}
\label{tab:fields}
\end{deluxetable*}

Excepting Q1603, a modified version of FOCAS \citep{valdes82} was used
to extract photometry from the optical ($\ugr$) images of the survey
fields.  Source Extractor \citep{bertin96} was used for photometry in
Q1603.  We took care to minimize systematics between the FOCAS and
Source Extractor results for Q1603 from an examination of the color
distribution of recovered LBG candidates.  Object detection was done
at $\rs$ band, and $\gmr$ and $\umg$ colors were computed by applying
the isophotal aperture from the $\rs$ band image to the $U_n$ and $G$
images.  Further details on the photometry are discussed in
\citet{steidel03} and \citet{steidel04}.  The images have a
$5$~$\sigma$ depth of $27.5-29.5$~AB as measured through a $\sim
3\arcsec$ diameter aperture.

\subsection{Color Selection}
\label{sec:colorselection}

We used the BX and LBG criteria which are based on the rest-frame UV
colors expected of galaxies at redshifts $1.9\la z\la 2.7$ and $2.7\la
z\la 3.4$, respectively \citep{steidel04, steidel03}.  For
spectroscopic followup, the sample of candidates was limited to
$\rs=25.5$.  However, because we are interested in using the entire
photometric sample to constrain the LF at $1.9\la z\la 3.4$, we did
not impose this restriction.  Rather, the detection significance and
color distribution of candidates were examined to establish a faint
limit.  The limits applied to each field and numbers of candidates are
listed in Table~1.  Fields with the deepest imaging in the survey
allow us to select candidates to $\rs\sim 26.5$ with a detection
completeness of $\approx 60\%$, based on the simulations described
below.  Our maximum-likelihood method for computing the LF allows us
to extend the absolute magnitude limit $\approx 0.5$~mag fainter given
the broadness of the redshift distribution, $N(z)$, of the sample as
constrained from extensive spectroscopic followup.  Even at these
depths, galaxies are detected with typically $\ga 5$~$\sigma$
significance in the $\rs$-band.  The photometric sample used here
includes BX candidates in the original LBG survey fields where no
spectroscopic followup of BX candidates was undertaken.

\subsection{Spectroscopy}
\label{sec:specfollow}

The spectroscopy of UV-selected candidates including LBGs and BXs is
discussed in \citet{steidel03, steidel04}.  To date, roughly $24\%$
and $35\%$ of BX and LBG candidates, respectively, with $\rs<25.5$
have been targeted spectroscopically.  The resulting sample includes
$2023$ star-forming galaxies with $1.9\le z_{\rm spec}<3.4$, the
largest spectroscopic sample of star-forming galaxies at these
redshifts.  The spectroscopic statistics, including the number of
spectroscopic redshifts, for each field of the LBG survey are given in
R08.

The spectroscopic sample is used to estimate the overall contamination
rate in the photometric sample.  This contamination can arise from
stars, low redshift galaxies and AGN, as well as QSOs and AGN at
$1.9\le z<3.4$.  The contamination statistics and the extent to which
they can be applied to determine the number of contaminants in the
photometric sample are discussed in R08.  Note that the contamination
rate is a strong function of magnitude (being largest at bright
magnitudes) and quantifying it is a crucial step in computing the
bright-end of the LF.

\section{INCOMPLETENESS CORRECTIONS}
\label{sec:maxlik}

\subsection{Method}

The approach that we adopted to correct the LBG sample for
incompleteness and compute the UV LF is described in detail by R08.
For convenience, here we summarize a few of the key features of our
method.  The primary goal is to construct a set of transformations
that relate the observed properties of galaxies (e.g., observed
luminosity, rest-frame UV slope, and redshift) to their true
properties (e.g., intrinsic luminosity, reddening, and redshift).
Using X-ray and mid-IR data for a sample of LBGs at $z\sim 2-3$,
\citet{reddy06a} demonstrate that the rest-frame UV slope can be used
to measure the amount of dust reddening for typical LBGs, and we will
assume this for the subsequent discussion.

We first used a Monte Carlo simulation to add galaxies of varying
sizes and colors to our $\ugr$ images.  The distribution of colors
reflects that expected for star-forming galaxies at $z\sim 2-3$ with
constant star formation for $>100$~Myr and varying amounts of dust
reddening.  Specifically, we added objects with redshifts $1<z<4$ and
reddening of $0.0<\ebmv<0.6$.  The simulated redshift is used to apply
an IGM opacity to the colors using the \citet{madau95} prescription.
To make the simulation as realistic as possible, we forced the
simulated galaxies to abide by a \citet{schechter76} luminosity
distribution and added just $100-200$ of them at a time to the images.
The latter restriction maintains the deblending statistics which in
turn affect the photometric errors.  Photometry was performed in the
same manner as was used on the real data and the detection rate and
recovered magnitudes and colors of the simulated galaxies were
recorded.  We repeated this procedure until $\approx 2\times 10^5$
galaxies were added to the images in each field of the survey.

It is common to use such simulations to determine what fraction of
galaxies with a given magnitude will be detected with colors that
satisfy the LBG selection criteria.  However, strictly speaking, the
simulation will only tell us the probability that galaxies with a
given {\em simulated} magnitude will be detected with a given {\em
  recovered} magnitude, and there may not be a monotonic
correspondence between simulated and recovered magnitude.  More
generally, it is not necessarily true that the average simulated
properties of galaxies are equivalent to their average observed
properties.  This is particularly true if photometric errors have
significant biases and are comparable to the bin sizes used to compute
the LF and the selection window spans a region of color space not much
larger than the typical photometric errors.  Other systematic effects,
such as the Ly$\alpha$ equivalent width distribution ($W_{\rm
  Ly\alpha}$) of the population, may scatter galaxies in certain
directions of color space.  Also, some galaxies with simulated colors
that do not initially satisfy the color criteria may have recovered
colors that do: by definition, these galaxies' simulated properties
will not, on average, reflect their observed properties.  Because of
these systematic effects, the number of galaxies that lie in a
particular bin of observed properties will be some weighted
combination of the number of galaxies in any number of bins of
simulated properties.

Because of these systematic effects, we approach the problem of
incompleteness by using the maximum likelihood method described in
R08.  Using this formalism, our goal is to maximize
the likelihood of a given set of luminosity, reddening, and redshift
distributions, denoted by $\cal{L}$, according to the following
expression:
\begin{eqnarray}
-\ln{\cal{L}} \propto \sum_{ijk} \bar{n}_{ijk} - \sum_{ijk} n_{ijk}\ln 
\bar{n}_{ijk},
\label{eq:minlik}
\end{eqnarray}
where $\bar{n}_{ijk}$ is the expected number of galaxies in the
i$^{\rm th}$ bin of luminosity, j$^{\rm th}$ bin of reddening, and
k$^{\rm th}$ bin of redshift that the values of the luminosity,
reddening, and redshift distributions imply and $n_{ijk}$ is the
observed number of galaxies in that bin.  The Monte Carlo simulations
give the probability that a galaxy in the $i'j'k'$ bin of simulated
properties will lie in the $ijk$ bin of recovered properties.  The set
of probabilities, defined as the transitional probability function,
\begin{eqnarray}
\xi \equiv \{p_{i',j',k'\rightarrow ijk}\},
\label{eq:xi}
\end{eqnarray}
is used to compute $\bar{n}_{ijk}$.  The basic procedure is to then
vary the input distributions of luminosity, reddening, and redshift
until the differences between the expected and observed numbers of
galaxies in each $ijk$ bin are minimized.

\subsection{Ly$\alpha$ Equivalent Width ($W_{\rm Ly\alpha}$) Distribution}

\begin{figure}[t]
\plotone{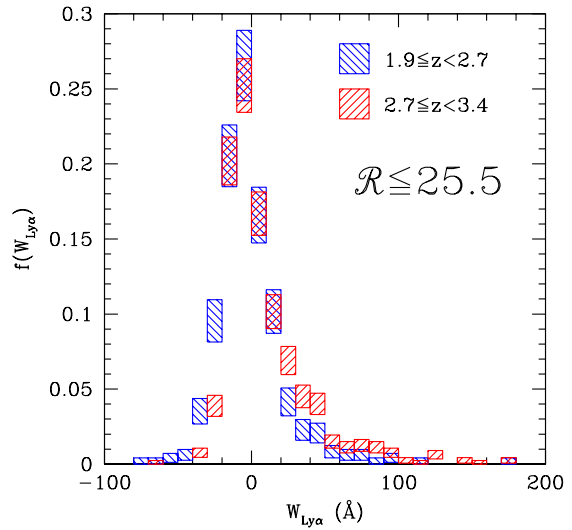}
\caption{Intrinsic rest-frame Ly$\alpha$ equivalent width ($W_{\rm
Ly\alpha}$) distribution for $\rs\le 25.5$ star-forming galaxies
at $1.9\le z<2.7$ and $2.7\le z<3.4$, from R08.
Ly$\alpha$ in emission is represented as $W_{\rm Ly\alpha}>0$.  Error
bars are determined from simulations and reflect the variance in the
distributions allowed by the errors in the UV LF and reddening
distribution (see R08 for discussion).}
\label{fig:lyaew}
\end{figure}

An important systematic effect to consider is the scattering of colors
due to the presence of Ly$\alpha$ emission and absorption: the
Ly$\alpha$ line falls within the $U_n$ and $G$ bands at redshifts that
are targeted with the BX and LBG criteria, the same bands that are
used to select the galaxies.  Rather than adding the $W_{\rm
Ly\alpha}$ distribution as another free parameter in the
maximum-likelihood analysis --- thus complicating our ability to
determine the LF --- we performed simulations where we made various
assumptions regarding the intrinsic $W_{\rm Ly\alpha}$ distribution at
$z\sim 2-3$ and observed the effects on the best-fit LF (R08).  We did
this by adding a random $W_{\rm Ly\alpha}$ drawn from a distribution
of $W_{\rm Ly\alpha}$ to each galaxy, and then recomputing the colors
of each galaxy.  Effectively, the addition of Ly$\alpha$ will perturb
the colors and thus modulate the transitional probability function,
$\xi$.  R08 showed that the BX and LBG color criteria did little to
alter the intrinsic $W_{\rm Ly\alpha}$ distribution at $z\sim 2$ and
$z\sim 3$.  Therefore, we assume the $W_{\rm Ly\alpha}$ distribution
observed for BXs and LBGs (Figure~\ref{fig:lyaew}).  Here, we repeat
the simulations of R08, but also allow for changes in the shape of the
$W_{\rm Ly\alpha}$ distribution proceeding from UV-bright to UV-faint
galaxies (see Appendix).

\section{RESULTS: THE UV LF AT $1.9\le z<3.4$}
\label{sec:results}

\subsection{Computation of the LF and Errors}

The value of the luminosity distribution that maximizes the likelihood
of observing our data (Eq.~\ref{eq:minlik}) is computed using the
method discussed in the previous section.  Initially, we assumed that
the intrinsic distribution of (1) rest-frame UV slopes, (2) redshifts,
and (3) $W_{\rm Ly\alpha}$ remain constant as a function of apparent
magnitude.  In R08, we justified these assumptions when computing the
LF to our spectroscopic limit of $\rs = 25.5$.  However, it is not
unreasonable to suspect that, for example, the reddening and $W_{\rm
Ly\alpha}$ distributions of galaxies fainter than our spectroscopic
limit may be different than those for galaxies where we are able to
directly measure the distributions with spectroscopy (``spectroscopic
distributions'').  Such differences will change $\xi$ and thus affect
our incompleteness corrections.  First, we first made the simplified
assumptions that all of these distributions remain unchanged as a
function of apparent magnitude.  LFs derived in this case are referred
to as the ``fiducial'' LFs.  In the appendix, we discuss how
the LF would change if we adopt more realistic assumptions for the
properties of UV-faint galaxies.  In our analysis, the effect of
increasing photometric error for fainter galaxies is already
incorporated in the calculation of $\xi$.

LFs were computed separately for star-forming galaxies in the redshift
ranges $1.9\le z<2.7$ and $2.7\le z<3.4$ using the photometric BX and
LBG samples, respectively.  For the lower redshift range, the LF is
computed in terms of a (composite) absolute magnitude that is the
average of the $G$ and $\rs$-band fluxes.  For the higher redshift
range, the LF is computed using the $\rs$-band magnitude.  This method
provides the closest match between rest-frame wavelengths, roughly
$1700$~\AA.

The total error in the LF is computed using the following method.  The
observed number counts of galaxies in each field were adjusted
randomly in accordance with a Poisson distribution and the
maximum-likelihood LF was computed for each field.  This procedure was
repeated many times for all the fields.  The dispersion in the LF
values for each bin in absolute magnitude is taken as the total error
which, as a consequence of the way in which it is computed, includes
both Poisson and field-to-field variations.

\subsection{Summary of Systematic Effects and Final Results}
\label{sec:summary}

The details of the systematic tests performed to judge the effects of
luminosity-dependent $W_{\rm Ly\alpha}$ and reddening distributions on
the LF are presented in the Appendix.  To summarize, we analyzed the
influence of galaxies with (1) strong Ly$\alpha$ emission and (2) zero
or declining reddening with apparent UV magnitude.  Employing current
estimates of the mean stellar population, reddening, and number
density of galaxies with strong Ly$\alpha$ emission as a function of
UV luminosity at high redshifts, we find that the inferred number
density of galaxies on the faint-end of the UV LF increases by $\la
3\%$ at $1.9\le z<2.7$ and decreases by $\la 4\%$ at $2.7\le z<3.4$.
Because these changes are not negligible compared to the Poisson and
field-to-field errors on the faint-end number densities, they should
be included in any proper assessment of the LF.  Nonetheless, these
changes in number density can be accommodated by Schechter
parameterizations that vary within the uncertainties of the individual
parameters, $\alpha$, $M^{\ast}$, and $\phi^{\ast}$.

\begin{figure*}[t]
\plotone{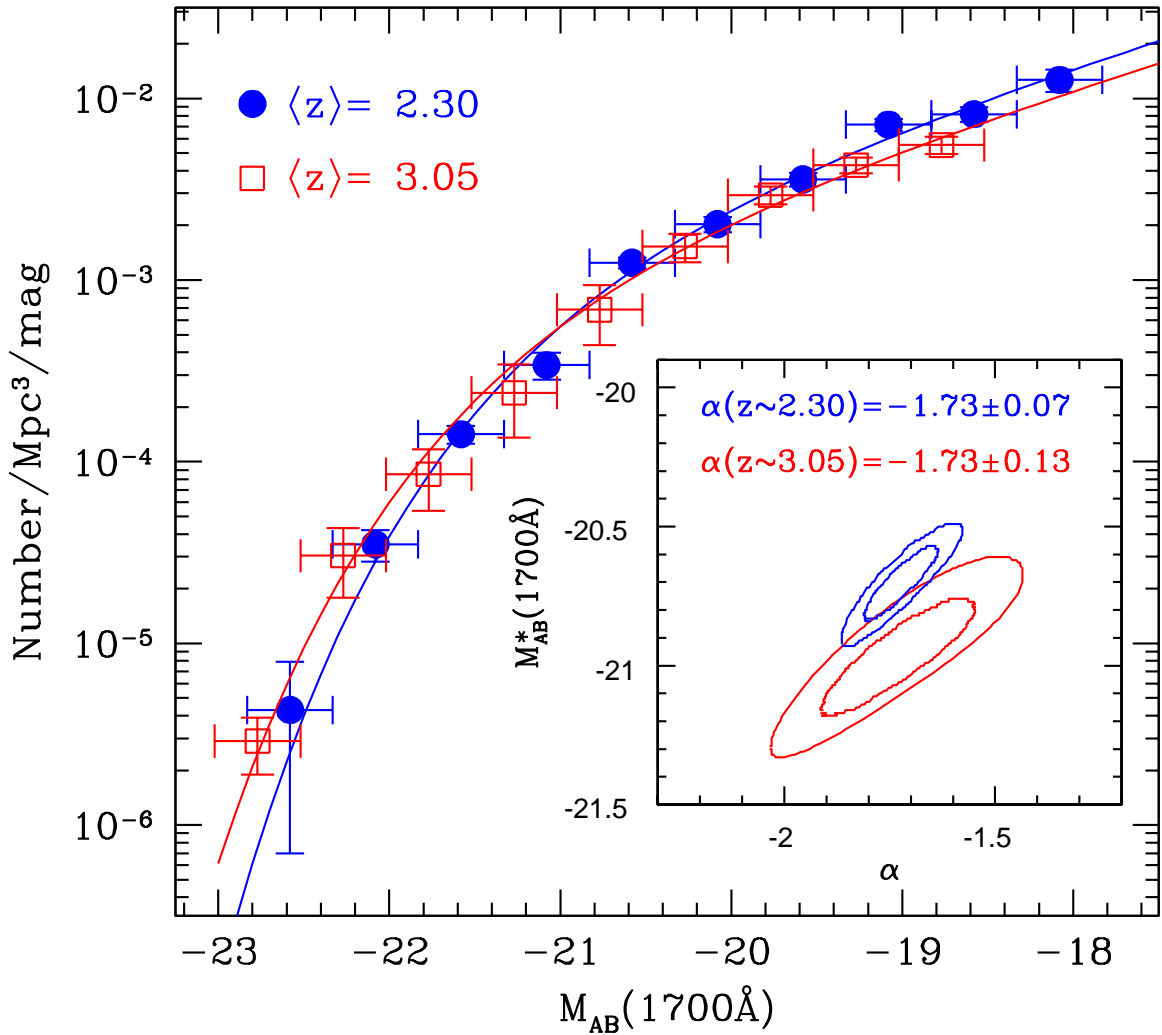}
\caption{Rest-frame UV luminosity functions of star-forming galaxies
at $1.9\le z<2.7$ ({\em circles}) and $2.7\le z<3.4$ ({\em squares}),
along with the best-fit \citet{schechter76} functions.  The $68\%$ and
$95\%$ likelihood contours between $M^{\ast}$ and $\alpha$ for our
final determinations of the LFs are shown in the inset panel.}
\label{fig:uvlf}
\end{figure*}

We have also examined how changes in the mean reddening of galaxies as
function of UV luminosity affects our measurement of the UV LF.  We
considered two scenarios, one in which the extinction drops to zero
for galaxies fainter than $\rs=25.5$ and one in which the extinction
decreases monotonically with UV luminosity, and approaches zero in the
faintest luminosity bin considered here.  The latter scenario is more
realistic than the former, and is parameterized as a linear relation
between $\ebmv$ and magnitude (see the appendix; we refer to
this latter scenario as the ``luminosity-dependent reddening model'').
In this case, we find appreciable increases of $\approx 10\%$ in the
inferred number density between $1.9 \le z<2.7$.  In the higher
redshift range $2.7\le z<3.4$, there is little change in the inferred
number densities.  Our final determinations of the LF and the
corresponding Schechter fits are shown by the data points and solid
lines, respectively, in Figure~\ref{fig:uvlf}, with the values and
Schechter parameterization listed in Tables~\ref{tab:uvlf} and
\ref{tab:schechterfinal}.

\begin{deluxetable}{lcc}
\tabletypesize{\footnotesize}
\tablewidth{0pc}
\tablecaption{Rest-Frame UV Luminosity Functions of $1.9\la z\la 3.4$ Galaxies}
\tablehead{
\colhead{} &
\colhead{} &
\colhead{$\phi$} \\
\colhead{Redshift Range} &
\colhead{M$_{\rm AB}(1700\AA)$} &
\colhead{($\times 10^{-3}$~$h^{3}_{0.7}$~Mpc$^{-3}$~mag$^{-1}$)}}
\startdata
$1.9\le z<2.7$ & $-22.83$ --- $-22.33$ & $0.004\pm0.003$ \\
& $-22.33$ --- $-21.83$ & $0.035\pm0.007$ \\
& $-21.83$ --- $-21.33$ & $0.142\pm0.016$ \\
& $-21.33$ --- $-20.83$ & $0.341\pm0.058$ \\
& $-20.83$ --- $-20.33$ & $1.246\pm0.083$ \\
& $-20.33$ --- $-19.83$ & $2.030\pm0.196$ \\
& $-19.83$ --- $-19.33$ & $3.583\pm0.319$ \\
& $-19.33$ --- $-18.83$ & $7.171\pm0.552$ \\
& $-18.83$ --- $-18.33$ & $8.188\pm0.777$ \\
& $-18.33$ --- $-17.83$ & $12.62\pm1.778$ \\
\\
$2.7\le z<3.4$ & $-23.02$ --- $-22.52$ & $0.003\pm0.001$ \\
& $-22.52$ --- $-22.02$ & $0.030\pm0.013$ \\
& $-22.02$ --- $-21.52$ & $0.085\pm0.032$ \\
& $-21.52$ --- $-21.02$ & $0.240\pm0.104$ \\
& $-21.02$ --- $-20.52$ & $0.686\pm0.249$ \\
& $-20.52$ --- $-20.02$ & $1.530\pm0.273$ \\
& $-20.02$ --- $-19.52$ & $2.934\pm0.333$ \\
& $-19.52$ --- $-19.02$ & $4.296\pm0.432$ \\
& $-19.02$ --- $-18.52$ & $5.536\pm0.601$ \\\enddata
\label{tab:uvlf}
\end{deluxetable}

\begin{deluxetable}{lccc}
\tabletypesize{\footnotesize}
\tablewidth{0pc}
\tablecaption{Best-fit Schechter Parameters for UV LFs of $1.9\la z\la 3.4$ Galaxies}
\tablehead{
\colhead{Redshift Range} &
\colhead{$\alpha$} &
\colhead{M$^{\ast}_{\rm AB}(1700{\rm \AA})$} &
\colhead{$\phi^{\ast}$ ($\times 10^{-3}$~Mpc$^{-3}$)}}
\startdata
$1.9\le z<2.7$ & $-1.73\pm0.07$ & $-20.70\pm0.11$ & $2.75\pm0.54$ \\
$2.7\le z<3.4$ & $-1.73\pm0.13$ & $-20.97\pm0.14$ & $1.71\pm0.53$ \\
\enddata
\label{tab:schechterfinal}
\end{deluxetable}

Our determinations of the bright-end of the UV LFs to $M(1700\AA) =
-18.83$ and $-19.52$ at $z\sim 2$ and $z\sim 3$, respectively,
incorporate data over $3261$~arcmin$^{2}$ in $31$ independent fields.
Data from 22 fields and $2239$~arcmin$^{2}$ are used to constrain the
LFs at $-18.83\le M(1700\AA)<-18.33$ and $-19.52\le M(1700\AA) <
-19.02$ at $z\sim 2$ and $z\sim 3$, respectively.  Finally, data from
6 spatially independent fields and $317$~arcmin$^{2}$ are used to
constrain the LF in the faintest magnitude bin to $M(1700\AA) =
-17.83$ and $-18.52$ at $z\sim 2$ and $z\sim 3$, respectively.  To
ensure that spatial variance in these 6 deep fields are not driving
the observed faint-end slope, we recalculated $\alpha$ by fitting
Schechter functions to the LFs excluding the faintest bin.  Allowing
$\phi^{\ast}$ and $M^{\ast}$ to vary, we calculate $\alpha = -1.75\pm
0.09$ and $-1.94\pm0.18$ at $z\sim 2$ and $z\sim 3$, respectively,
still significantly steeper than the shallower $\alpha>-1.6$ found in
previous studies.  The similarity in $\alpha$ obtained with or without
data from the faintest bin is not surprising given that the
uncertainty in the LF includes the sample variance from the $6$ fields
used to constrain the number density in this bin.

\begin{figure}[t]
\plotone{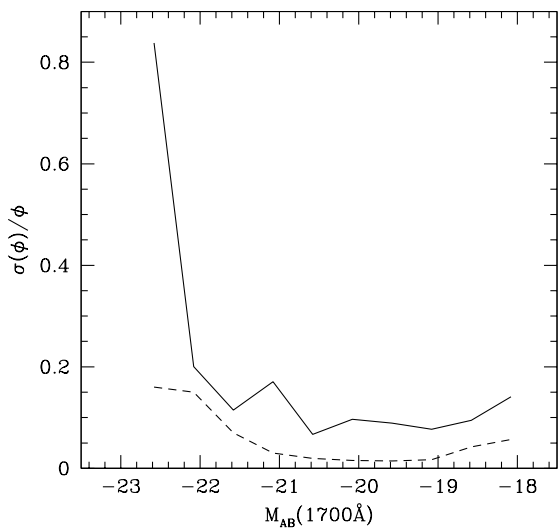}
\caption{Poisson ({\em dashed line}) and total ({\em solid line})
error in the UV LF at $z\sim 2$.  The Poisson error increases to
fainter magnitudes given the smaller survey area used to constrain the
faint-end.  At all magnitudes, however, the error in the LF is
dominated by field-to-field variations.  Similar results are obtained
for the $z\sim 3$ UV LF.}
\label{fig:frac}
\end{figure}

The degeneracy between the faint-end slope ($\alpha$) and
characteristic magnitude ($M^{\ast}$) --- illustrated by the
likelihood contours in Figure~\ref{fig:uvlf} (inset) --- is reduced
significantly compared to that computed in R08.  Our analysis extends
to luminosities that are $4$ times fainter than the limit dictated by
efficient spectroscopy, and $\approx 14$ and $10$ times fainter,
respectively, than the characteristic luminosity $L^{\ast}$ at $z\sim
2$ and $3$.  Our sample is large enough so that the error in the LF at
all magnitudes is dominated by field-to-field variations
(Figure~\ref{fig:frac}).  Within the total errors, the UV LFs at
$z\sim 2$ and $z\sim 3$ are virtually indistinguishable, indicating
little change between the two in the number density of both UV-bright
and UV-faint galaxies.

\section{Discussion: Large-Scale Context}
\label{sec:lsc}

\begin{figure*}[t]
\plottwo{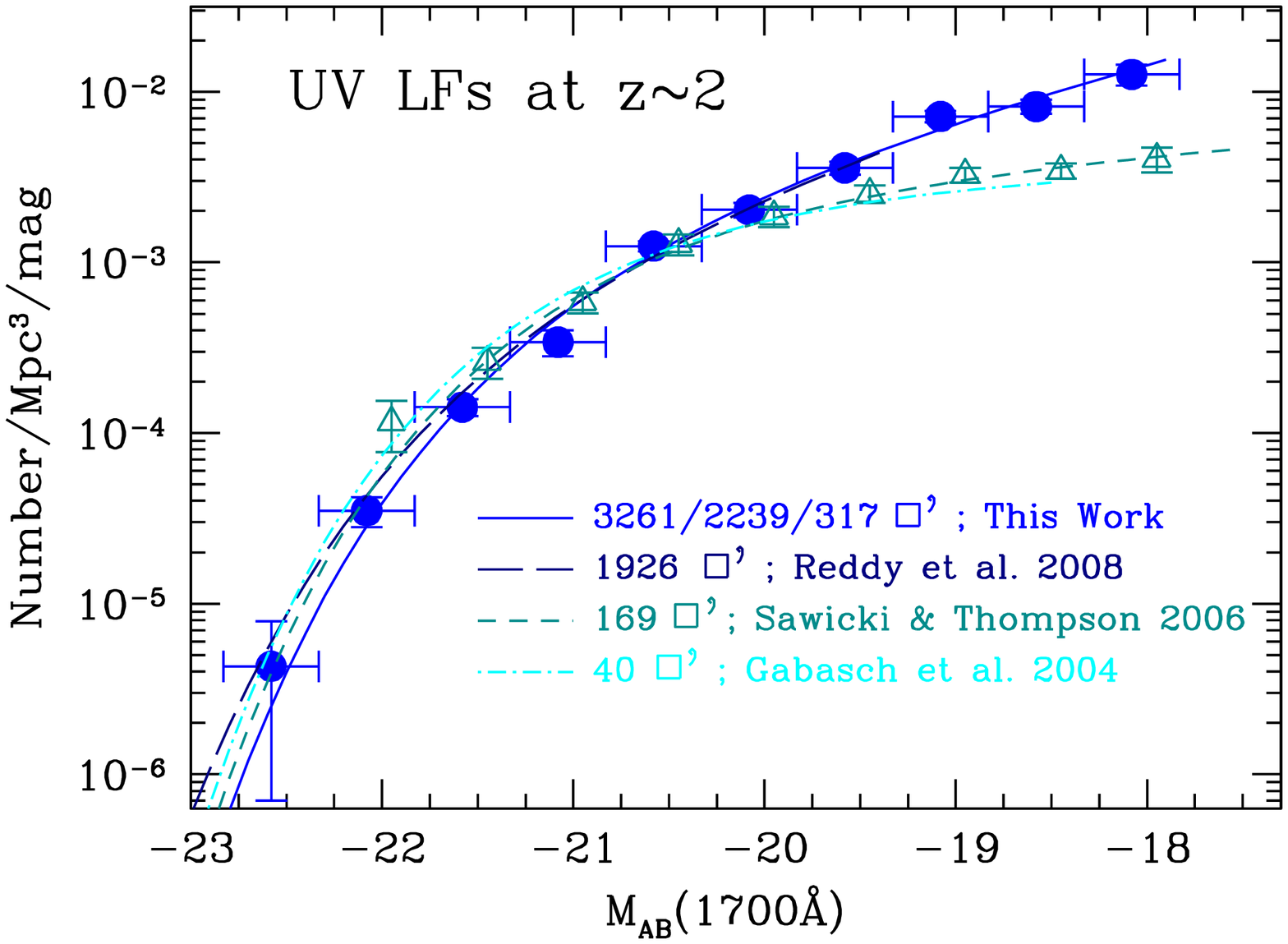}{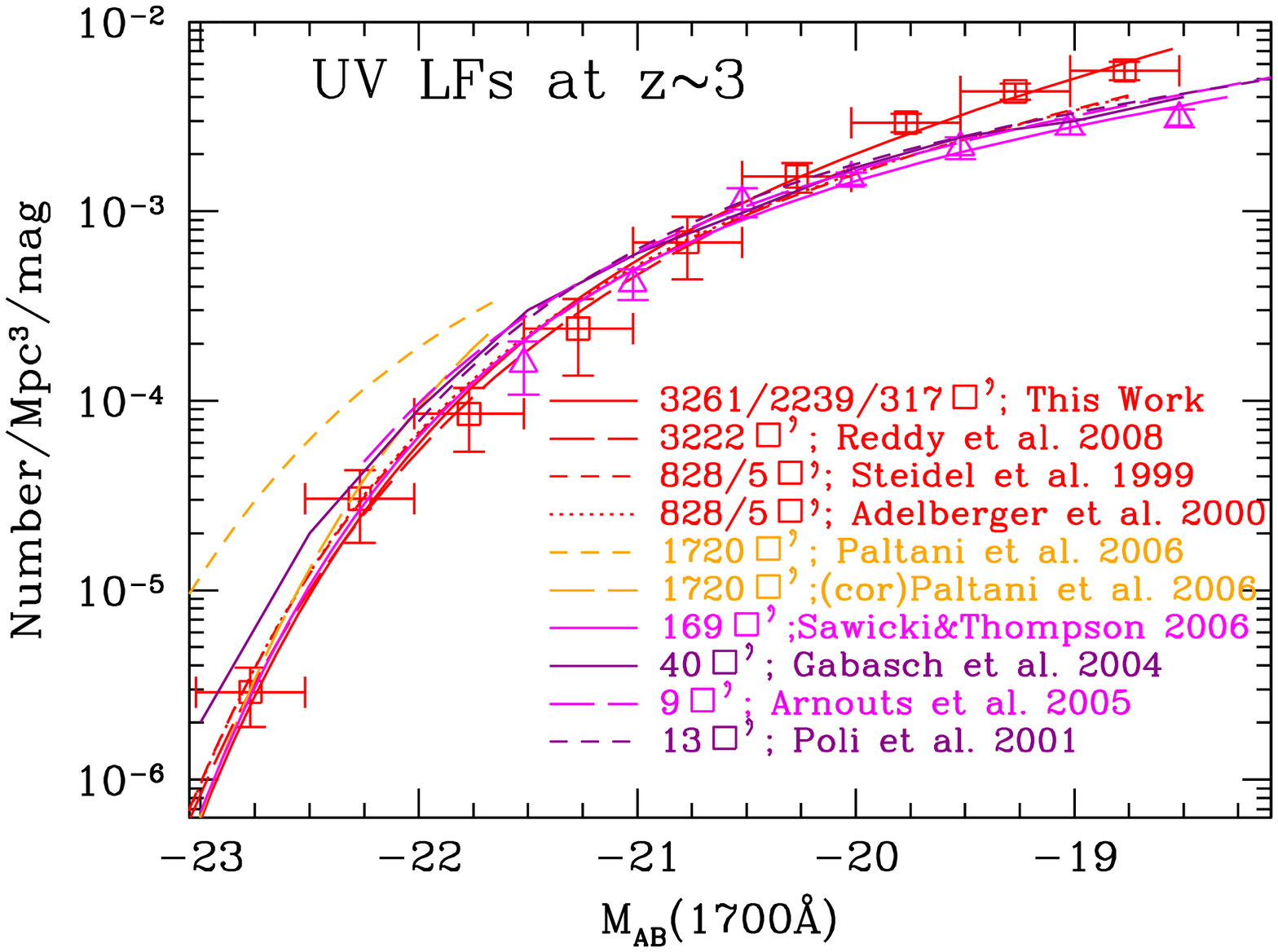}
\caption{Comparison of UV LFs at $z\sim 2$ ({\em left}) and $z\sim 3$
  ({\em right}).  For clarity, data points are excluded on all but our
  current determinations and those of \citet{sawicki06a}, but the
  errors are typically smaller than the observed differences discussed
  in the text.  Also shown are the survey areas over which the LF is
  derived, with some surveys using a combination of wider shallower
  data to anchor the bright-end of the LF and deeper data in smaller
  areas to constrain the faint-end slope.  Included are are data from
  \citet{reddy08, sawicki06a, paltani07, arnouts05, gabasch04, poli01,
  adel00, steidel99}.}
\label{fig:uvlfcomp}
\end{figure*}

In the following sections, we discuss our results in the context of
previous determinations of the UV LF (\S~\ref{sec:context}) and its
evolution with redshift (\S~\ref{sec:lfevol}).  To gain further
insight into the nature of sub-$L^{\ast}$ galaxies, we assess the
contribution to the faint-end population from dusty, star-forming
galaxies and those with large stellar masses (\S~\ref{sec:nature}).
In \S~\ref{sec:sfrd}, we discuss the implications of a
luminosity-dependent reddening distribution and the average
corrections required to recover the bolometric star formation rate
density.  These findings are then discussed in the context of the star
formation history and the buildup of stellar mass in
\S~\ref{sec:sfrdevol}.  Finally, we discuss briefly the evolution of
the faint-end slope in \S~\ref{sec:alpha}.

\subsection{Comparisons at $z\sim 2-3$}
\label{sec:context}

In this section, we place our results in the context of previous
determinations of the UV LFs, starting with those around $z\sim 2$.
Figure~\ref{fig:uvlfcomp} summarizes the results of several previous
studies including those of \citet{gabasch04, sawicki06a} and R08,
along with our current determination, at $z\sim 2$.  The redshift
intervals over which the LF is computed are similar between these
studies, but we note the almost two orders of magnitude difference in
the areas probed, from $\approx 40$~arcmin$^{2}$ at the low end to
$3261$~arcmin$^{2}$ in the current determination.  There are
significant differences between the LFs at faint magnitudes
($M(1700\AA)\ga -20$).  In general, it is possible that the
determinations of the smaller surveys (e.g., from the FORS-Deep Field;
\citealt{gabasch04}) could be mimicked by an overall under-density in
the small area probed combined with an overestimation of the
bright-end due to contamination from low redshift interlopers.
\citet{gabasch04} do not specify the contamination fractions for their
higher redshift samples at $z\ga 2$, so a fair comparison with our
findings at the bright-end is not possible.  We also note that
\citet{gabasch04} relied on photometric redshifts which could not be
well-calibrated due to the lack of spectroscopically-confirmed
galaxies at $z\sim 2$ examined in their study (see their Figure~2).
R08 showed that biases in photometric redshifts can easily boost the
bright-end of the luminosity function with respect to the faint-end,
and the overall shape of their LF may be a result of this effect.

Perhaps a fairer comparison can be made with \citet{sawicki06a} since
they use exactly the same filter set to select BX candidates at $z\sim
2.3$ in the Keck Deep Fields (KDFs).  For their fiducial model, they
assumed no perturbation of colors due to Ly$\alpha$ and a constant
$\ebmv=0.15$ with no dispersion, and they compute the LF using the
standard $V_{\rm eff}$ method.  Their LF suggests a much shallower
slope of $\alpha\sim -1.2$ compared to our result
(Figure~\ref{fig:uvlfcomp}).  However, \citet{sawicki06a} point out
that their LF derived at $z\sim 2$ is sensitive to the assumed
$\ebmv$, and that bluer values of $\ebmv$ will tend to yield larger
inferred number densities (see their Figure~7).  This observation is
consistent with our finding and, in particular, if the $\ebmv$
distribution becomes significantly bluer proceeding to fainter
galaxies, this effect would manifest itself as a steepening of the
faint-end slope (appendix).  However, this systematic
effect alone cannot account for all of the difference between our
result and that of \citet{sawicki06a}, since even in the fiducial case
of a luminosity-invariant $\ebmv$ distribution (but not
constant-valued) we find a steep $\alpha=-1.67\pm0.06$
(appendix).  In any case, regardless of how the mean
$\ebmv$ varies with magnitude, the distribution itself is not a delta
function, of course, and has intrinsic dispersion; those galaxies at
the blue end of the distribution (i.e., less reddening) will tend to
escape the selection criteria more frequently than galaxies with
redder $\ebmv$.  Hence, a bluer mean $\ebmv$, the intrinsic scatter in
$\ebmv$ for UV-faint galaxies, and a general perturbation of colors
due to Ly$\alpha$ will all result in larger corrected number densities
at the faint-end.

Focusing on the higher redshift range, we find again reasonable
agreement among the various determinations of the bright-end of the UV
LF at $z\sim 3$.\footnote{\citet{sawicki06a} use the results from
\citet{steidel99} to constrain the bright-end of the UV LF at $z\sim
3$.}  The only significantly discrepant points are from the VVDS that
imply significant numbers of UV-bright galaxies \citep{paltani07}.
However, applying the correct contamination fractions (based on
spectroscopy) to their points brings them in accordance with the other
determinations (R08).  As at lower redshifts, we find a
substantially steeper faint-end slope at $z\sim 3$ than suggested by
previous results.  Most determinations have found $\alpha>-1.6$,
shallower than the canonical $\alpha=-1.6$ from \citet{steidel99},
although most of these studies (including \citealt{steidel99})
constrained $\alpha$ using deep data from only one or two small deep
fields (e.g., Hubble Deep Field, FORS Deep Field) where large-scale
structure may be an issue.  \citet{sawicki06a} find $\alpha =
-1.43^{+0.17}_{-0.09}$ based on the Keck Deep Field data over an area
of $169$~arcmin$^{2}$.

\subsection{Differences in LF Computation}
\label{sec:difflfcomp}

What could be the reason for the disparity in the faint-end number
densities between our study and previous determinations?  Without a
more detailed comparative analysis incorporating the data used in
these other studies, it is difficult to pinpoint a single cause for
the discrepancy.  There are, however, a number of differences between
our analysis and others that may lead to the observed variance in
$\alpha$.  Our analysis (1) uses over $2000$ spectroscopic redshifts
to evaluate and correct for contamination as a function of luminosity;
(2) models the systematic effects of a luminosity dependence in the
intrinsic Ly$\alpha$ equivalent width and reddening distribution of
galaxies, likely to be the two dominant sources of systematic error in
the LF; (3) employs a maximum-likelihood method that is more robust
than the $V_{\rm eff}$ method against biases in photometry and other
non-uniform sources of scatter; and (4) takes advantage of data in
$31$ spatially uncorrelated fields over a total area of close to a
square degree.  Even at the faint-end, our determinations are based on
$6$ independent fields with a total area of $317$~arcmin$^{2}$, a
roughly $88\%$ larger area than used in the previous faint-end
determination at $z\sim 2-3$ (but see next section).  For all of these
reasons, we believe our LFs to be the most robust determinations to
date.

The differences in faint-end slope derived between studies with
similar depth is not particularly significant within the marginalized
errors on $\alpha$.  For example, the $\alpha=-1.43^{+0.17}_{-0.09}$
of \citet{sawicki06a} is still consistent within the $1$~$\sigma$
(marginalized) error of our determination of $\alpha=-1.73\pm 0.13$ at
$z\sim 3$.  Yet, the difference in the actual number density of faint
galaxies is significant at the $2-3$~$\sigma$ level.  This emphasizes
why comparisons between $\alpha$ derived in different studies should
perhaps not be taken too seriously without placing them in the context
of the errors on the actual number density of UV-faint galaxies.

\subsection{Cosmic Variance}

In spite of the care used in the present sample, even
$317$~arcmin$^{2}$ is a relatively small area over which to constrain
$\alpha$.  As noted above, the uncertainties in the LF are dominated
by field-to-field variance at all magnitudes.  We can assess how the
empirically-constrained errors on the UV LF compare to expectations
based on the correlation function.  Following the procedure outlined
by \citet{trenti08}, we can estimate the combined uncertainty due to
cosmic variance and Poisson statistics by integrating the two point
correlation function for dark matter halos with some average galaxy
bias.  The basic premise is that the spatial correlation function of
halos gives information on the variance in the spatial distribution of
galaxies along different lines of sight given various assumptions for
the cosmology and halo filling factor.  For this calculation, we
assumed a number density of objects as implied by the
maximum-likelihood LF at $z\sim 2$ and a sample ``completeness''
fraction of $0.47$.  This number is the ratio of star-forming galaxies
that satisfy the color selection criteria to the total number of
star-forming galaxies as determined from the LF (see R08 for a
discussion of this fraction).  The cosmology is set as follows:
$\Omega_{\lambda} = 0.74$, $\Omega_{\rm m}=0.26$, $H_{\rm o} =
70$~km~s$^{-1}$~Mpc$^{-1}$, a spectral index $n_{\rm s} = 1$, and
$\sigma_{\rm 8} = 0.9$ \citep{spergel07}.  We must also make some
assumption for the halo filling factor.  Star-forming galaxies are
scattered out of the LBG selection window due to primarily random
processes such as photometric scatter \citep{reddy05a}, and must
therefore cluster in the same way as galaxies that do satisfy the LBG
criteria \citep{conroy08}.  Further, the comoving number density of
LBGs is similar to the number density of halos that have similar
clustering strength, suggesting a halo filling factor of $\approx 1$
\citep{conroy08}.  Assuming this remains valid for UV-faint galaxies,
we find a fractional error in number counts of $\approx 9\%$ over a
survey area of $317$~arcmin$^{2}$.  In general, we would expect this
calculation to yield a lower limit to the uncertainty since other
effects (e.g., uncertainties in zeropoints and systematics in the
color distributions from field to field) contribute to the error in
the LF, and indeed our empirically-derived error in the faintest bin
of the $z\sim 2$ UV LF is $\approx 50\%$ larger than the value
obtained from the two point correlation function.  For comparison,
this calculation implies that the field-to-field variance is $\approx
17\%$ lower than what we would have obtained over the area probed by
the KDFs of $169$~arcmin$^{2}$ \citep{sawicki06a}.  This difference is
not large enough to explain the apparent discrepancy at the faint-end,
and some of the systematics discussed above are also likely to play a
role.  Deep UV imaging over areas of close to a square degree (similar
to that used to estimate the bright-end of the LF) will be required to
constrain the fractional error in number counts to $\la 5\%$.

\subsection{Hubble Deep Field (HDF)}
\label{sec:hdf}

A comparison between the present work and those of the early HDF-based
studies of the UV LF is useful, particularly in light of the
often-used argument that the HDF presents a biased view of the
Universe, and one that is invoked to explain the divergent results on
the UV LF at $z\sim 3$ \citep{steidel99, dickinson03, giavalisco04b,
gabasch04, sawicki06a}.  The bright-end of the LFs computed here and
by \citet{steidel99} are in excellent agreement.  Within the
$1$~$\sigma$ marginalized errors, the faint-end slope derived at
$z\sim 3$ agrees with the slope found by \citet{steidel99} and
\citet{adel00}, and there is essentially no significant difference in
$\phi^{\ast}$ and $M^{\ast}$.  However, given the widespread use of
the \citet{steidel99} results, it is important to note that their
determination of $\alpha$ --- constrained from a $U$-dropout sample in
the HDF --- does not take into account incompleteness from photometric
scatter.  As discussed in \S~\ref{sec:maxlik}, the effect of such
scatter is to make the incompleteness corrections larger at the
faint-end, thus steepening the faint-end slope.  In summary, contrary
to the suggestion that the HDF contained an over-density of faint
galaxies relative to bright ones when compared with other fields, our
results imply that the HDF is reasonably representative of the $z\sim
2-3$ universe.

\begin{figure*}[t]
\plottwo{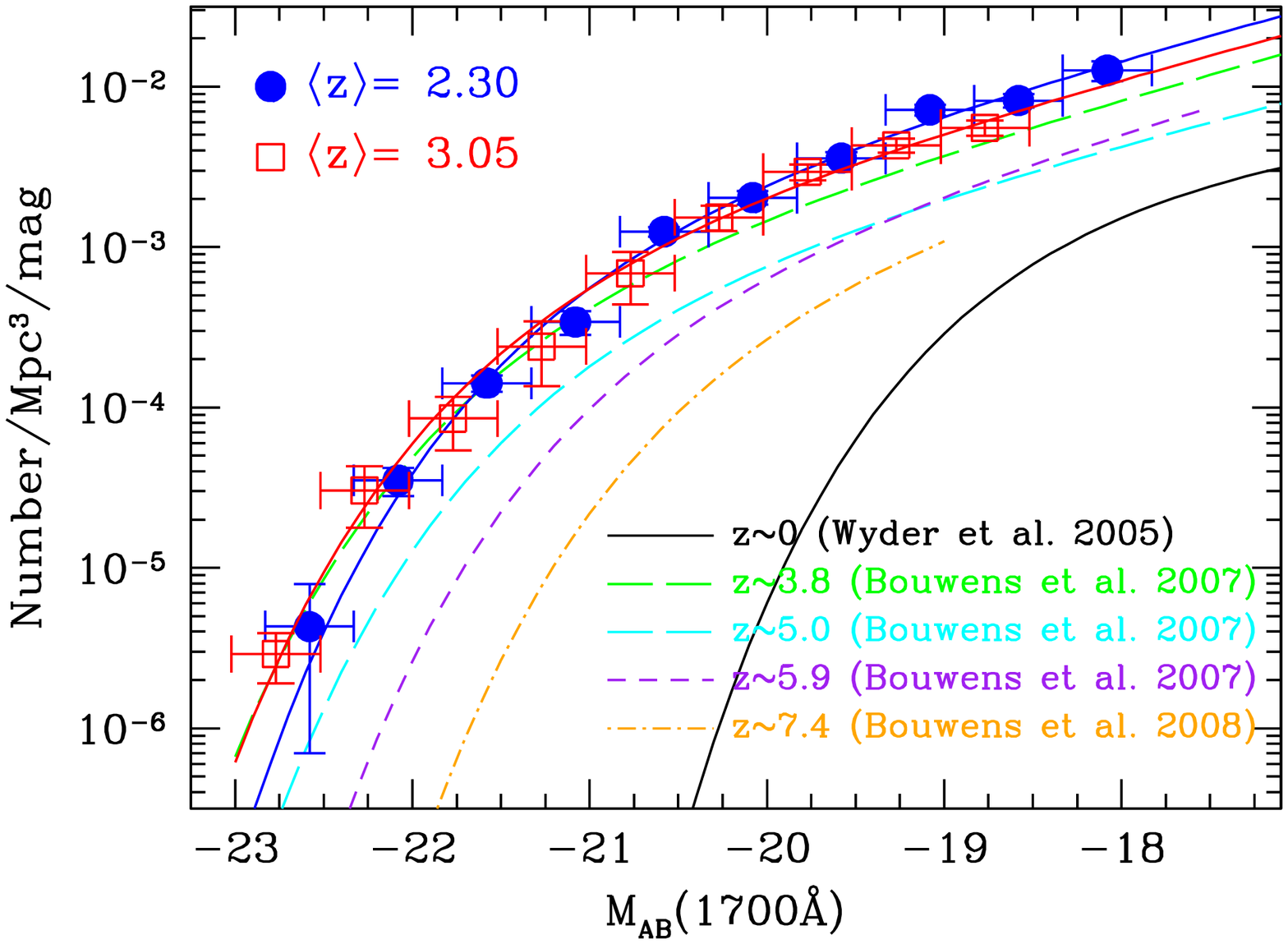}{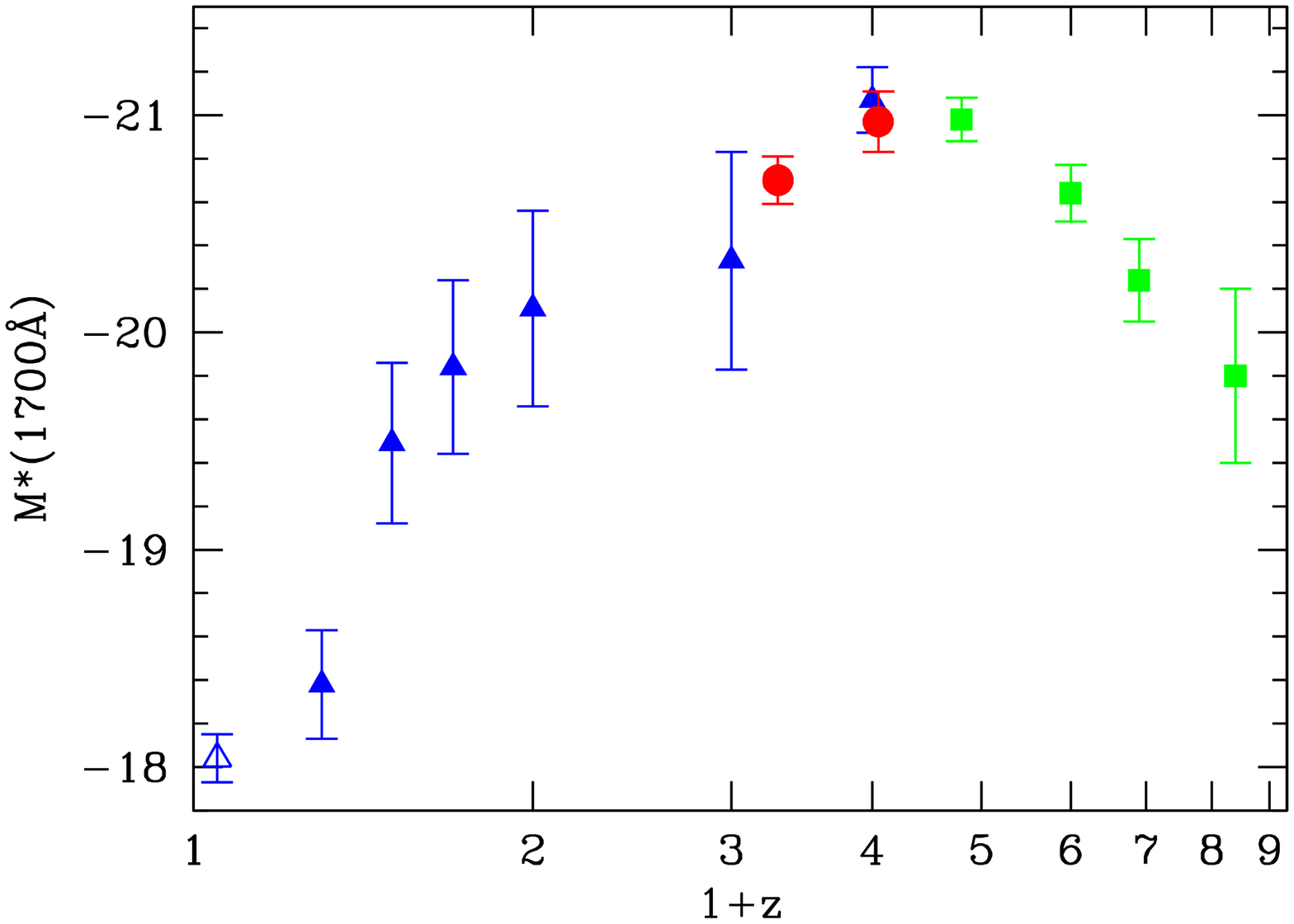}
\caption{({\em Left}) Evolution of the UV LFs from $z\sim 7$ to $z\sim
2$.  For clarity and consistency, we show only LFs at $z\ga 4$ from
\citet{bouwens07, bouwens08} since they are calculated using a
maximum-likelihood technique similar to the one used here.  For
comparison, the local UV LF from \citet{wyder05} is also shown.  ({\em
Right}) Evolution of the characteristic UV luminosity or magnitude,
$M^{\ast}$, with redshift.  Points are from \citet{wyder05} at $z\sim
0$ ({\em open triangle}), \citet{arnouts05} at $0\la z\la 3.0$ ({\em
filled triangles}), \citet{bouwens08} at $z\ga 4$ ({\em squares}), and
our determinations at $z\sim 2-3$ ({\em circles}).}
\label{fig:uvlfevol}
\end{figure*}

\begin{figure*}[t]
\plottwo{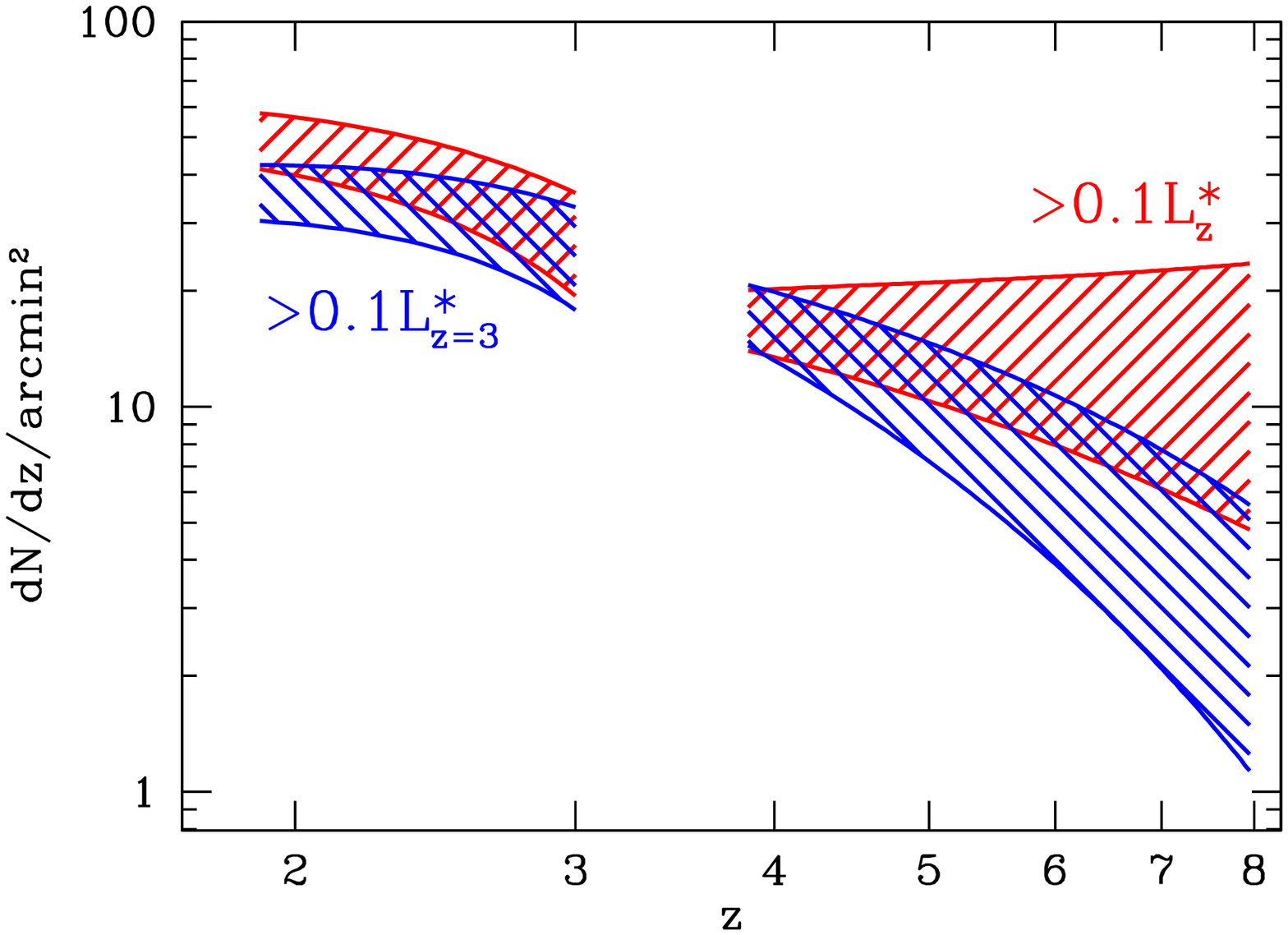}{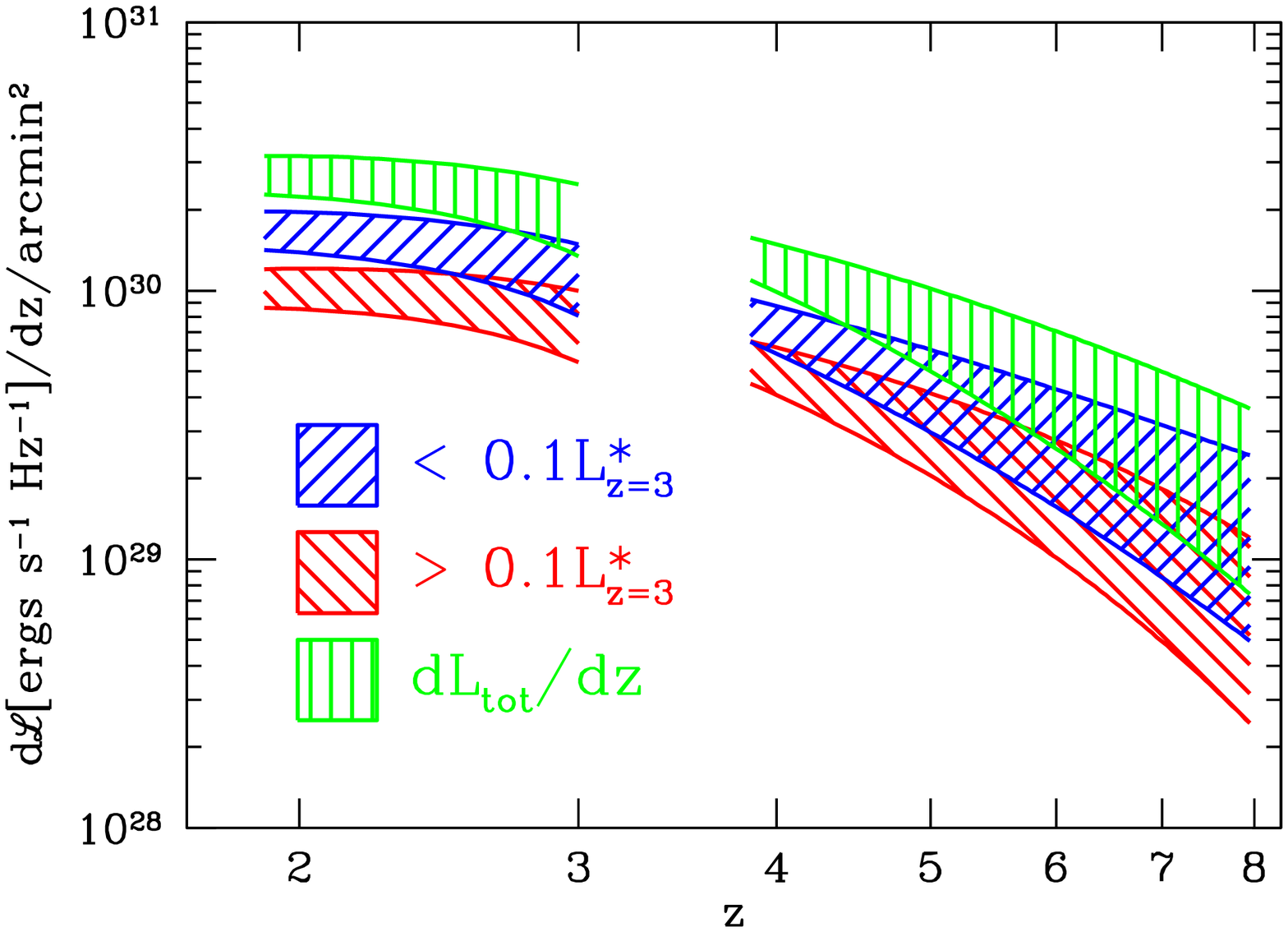}
\caption{({\em Left}) $dN/dz$ as a function of redshift, assuming our
determinations of the UV LF at $z\sim 2-3$ and those of
\citep{bouwens07} at $z\ga 4$, integrated to $0.1L^{\ast}_{z=3}$ ({\em
blue}) and $0.1L^{\ast}_{z}$ ({\em red}).  The shaded regions indicate
approximately the uncertainty based on the errors in the Schechter parameters.
({\em Right}) Total $dL/dz$ as a function of redshift ({\em green})
and $dL/dz$ brighter and fainter than $0.1L^{\ast}_{z=3}$ ({\em red} and {\em
blue}, respectively).}
\label{fig:dndldz}
\end{figure*}

\section{Discussion: Evolution of the UV LF}
\label{sec:lfevol}

Figure~\ref{fig:uvlfevol} summarizes our UV LFs at $z\sim 2-3$ along
with higher redshift determinations.  For clarity and consistency, we
included the findings from \citet{bouwens07} only since those authors
use a maximum-likelihood method for determining the LF that is similar
to the method we use.  These authors provide a detailed comparison of
UV LFs at $z\ga 4$ from different studies.

\subsection{Evolution in $M^{\ast}$}
\label{sec:mstarevol}

Despite the large number of investigations at $z\ga 4$, there is still
a fair amount of uncertainty regarding the parameterization of the
evolution in the UV LF.  Some have claimed that the evolution occurs
primarily in $L^{\ast}$ \citep{bouwens07}, while others find an
evolution in $\phi^{\ast}$ \citep{beckwith06} or the faint-end slope
$\alpha$ \citep{iwata07}.  Some have also suggested an evolution in
both $L^{\ast}$ and $\phi^{\ast}$ (e.g., \citealt{dickinson04,
giavalisco04b}) such that the total luminosity density could remain
constant from $z\sim 3 - 6$.  Of course, the reach of some of these
conclusions is limited by the depth of data used to derive the LF.
Because the LFs at $z\ga 2$ shown in Figure~\ref{fig:uvlfevol} are
derived using data of comparable depth and analyzed in a similar
manner --- although we note that our LFs at $z\sim 2-3$ are anchored
by spectroscopy and photometry in an area roughly an order of
magnitude larger than used at $z\ga 4$ --- hereafter we will assume
that the evolution of the LF at $z\ga 4$ can be accommodated by a
change in $L^{\ast}$ as advocated by \citet{bouwens07}.  Given the
observed fading of galaxies at $z\la 2$ (e.g., \citealt{dickinson03,
madau96, lilly95, lilly96, steidel99}), it is useful to examine our
results in the context of this evolution in $L^{\ast}$
(Figure~\ref{fig:uvlfevol}).  In particular, we find that $L^{\ast}$
is brightest at $z\sim 2-3$, with this average unobscured UV
luminosity decreasing at $z\ga 4$ (earlier cosmic time) and decreasing
by a factor of $\approx 16$ between $z\sim 2$ and the present-day.
Quantitatively, \citet{bouwens08} found a linear parameterization
between $M^{\ast}$ and $z$ at $z\ga 4$ that appears to follow that
generally expected for the growth of the halo mass function ---
assuming an evolution in the mass-to-light ratio for halos of $\sim
(1+z)^{-1}$ --- given standard assumptions for the matter power
spectrum, indicating that hierarchical assembly of halos may be
dominating the evolution in $M^{\ast}$, or equivalently $L^{\ast}$.
In the context of this study, the linear parameterization can be ruled
out at the $8$~$\sigma$ level at $z=2.3$ (in the sense that it would
predict a significantly larger $L^{\ast}$ at $z=2.3$ than is
observed), indicating that by these redshifts, some other effect(s)
modulate $L^{\ast}$ away from the value expected from pure
hierarchical assembly.

These observations are illustrated more clearly by examining $dN/dz$
as a function of redshift (Figure~\ref{fig:dndldz}), which is
extrapolated based upon linearly fitting the relationship between
$L^{\ast}$ and $z$ and $\phi^{\ast}$ and $z$, and assuming a fixed
$\alpha=-1.73$ as indicated by the Schechter fits
(Table~\ref{tab:schechterfinal}).  Integrating the number counts to a {\em
fixed} luminosity shows that bright galaxies with
$L>0.1L^{\ast}_{z=3}$ increase in number density by an order of
magnitude with cosmic time from $z\sim 7$ to $z\sim 2$.
Alternatively, the number counts are flatter at $z\ga 4$ when
integrating to $0.1L^{\ast}(z)$ (i.e., $L^{\ast}$ appropriate at the
redshift $z$ where $dN/dz$ is calculated) suggesting that
$\phi^{\ast}$ is relatively constant at these redshifts (e.g.,
\citealt{bouwens07}).  There may be a slight increase in $\phi^{\ast}$
between $z\sim 2-3$, though the magnitude of the errors on
$\phi^{\ast}$ are large enough that we cannot rule out non-evolution
in the number density.

Also shown is $dL/dz$, both above and below a fixed luminosity, in
this case $0.1L^{\ast}_{z=3}$, along with the total luminosity
density.  The evolution implied by our LFs suggests that the
approximately order of magnitude increase in luminosity density
between $z\sim 7$ and $z\sim 4$ is followed by a flattening between
$z\sim 2-3$.  This result itself is hardly surprising (see
\citealt{giavalisco04b}, R08), but its significance is constrained
robustly given that our LFs are determined over two orders of
magnitude in luminosity.  We will return to a discussion of these
findings in the context of the cosmic star formation history
(\S~\ref{sec:sfrd}).

\subsection{Evolution in $\alpha$}
\label{sec:alphaevol}

\begin{figure}[t]
\plotone{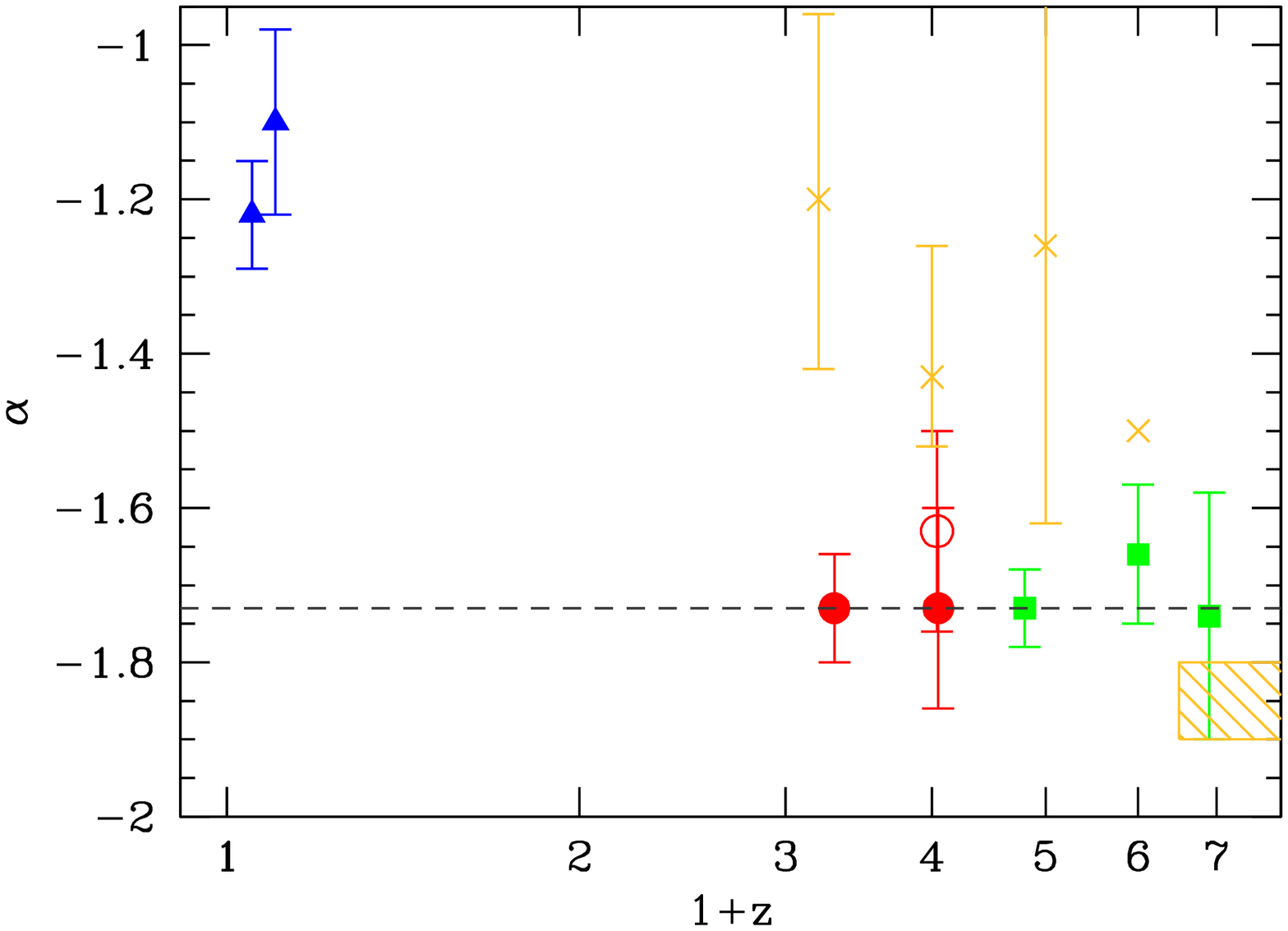}
\caption{Faint-end slope $\alpha$ as a function of redshift.  At $z\ga
2$, we include our results ({\em filled circles}), that of
\citet{steidel99} ({\em open circle}), and those of \citet{bouwens07}
({\em squares}).  At lower redshifts, we included only points in the
\citet{ryan07} compilation that were derived from the rest-UV LF and
that relied on data extending at least two magnitudes fainter than
$M^{\ast}$, including results from \citet{budavari05} and
\citet{wyder05} ({\em triangles}).  Also shown are points from
\citet{sawicki06a}, \citet{iwata03} (errors in $\alpha$ are not
provided by these authors; {\em crosses}), and \citet{yan04b} (range
of likely $\alpha$ indicated by hashed box).  The dashed line marks
the mean value of $\alpha$ found at $z \ga 2$ from our study and that
of \citet{bouwens07} ($\langle\alpha\rangle\sim -1.73$).}
\label{fig:alpha}
\end{figure}

Perhaps the most striking result of our analysis --- and one that is
possible to address with confidence given the depth of data considered
here --- is a very steep faint-end slope of $\alpha \sim -1.73$ at
$z\sim 2-3$ that is robust to the luminosity-dependent systematics
discussed in the Appendix.  The $\alpha$ we derive at $z=2.30$ is
virtually identical to that derived at $z=3.05$, and is remarkably
similar to the steep faint-end slopes favored at $z\ga 4$
(Figure~\ref{fig:alpha}).  Given the rapid evolution in $L^{\ast}$ and
the luminosity density at $z\ga 2$, the invariance of $\alpha$ over
the same $\sim 3$~Gyr timespan and the shallow $\alpha$ found locally
\citep{wyder05, budavari05} pose interesting constraints on models of
galaxy formation.  We revisit this issue in \S~\ref{sec:alpha}.

\section{Discussion: Nature of Galaxies on the Faint-End of the UV LF}
\label{sec:nature}

Before proceeding to discuss the implications of our results, it is
useful to assess the contribution of galaxies selected with different
methods to the UV LF.  R08 demonstrate that the BX and LBG criteria to
$\rs=25.5$ select the majority of galaxies on the bright-end of the UV
LF, namely those with $L_{\rm UV}\ga 0.1L^{\ast}$.  We show in the
appendix that these criteria recover the majority of star-forming
galaxies fainter than $0.1L^{\ast}$.  The tests discussed in the
Appendix assume that the vast majority of galaxies on the faint-end of
the UV LF are relatively unreddened, young galaxies.  The aim of this
section is to quantify the fraction of galaxies on the faint-end that
are (1) UV-faint simply because they are heavily-reddened or (2) older
galaxies that have passed their major phase of star formation.  The
latter investigation is relevant if we are to make inferences on the
connection between the dark matter halo mass distribution and the
luminosity function.

\subsection{Bolometrically-Luminous Galaxies}

Deep mid-to-far IR surveys have uncovered a sizable population of
dusty and infrared luminous galaxies at $z\sim 2-3$ (e.g.,
\citealt{yan07, caputi07, papovich07, reddy05a, chapman05,
vandokkum04, smail97, barger98}).  The first such galaxies were
discovered via their submillimeter emission \citep{smail97, barger98,
hughes98}, and are now commonly referred to as submillimeter galaxies
(SMGs).  \citet{chapman05} estimated that $\approx 65\%$ of such
spectroscopically-confirmed bright SMGs (e.g., with $S_{\rm 850\mu
m}\ga 5$~mJy) at $z\sim 2-3$ have rest-frame UV colors similar to
those of BXs and LBGs, yet are on average a factor of $\approx 10$
times more luminous.  There is some uncertainty in the luminosities
related both to the conversion between mid and IR luminosities to
total bolometric luminosities and the fraction of the luminosity that
arises from an AGN \citep{alexander05}.  Taking the far-IR estimates
of the SFRs of SMGs at face value then suggests that SMGs are examples
of galaxies whose UV slopes typically under-predict their total
attenuation and hence total bolometric luminosities \citep{reddy06a}.

Measuring the frequency of such dusty galaxies among UV-faint sources
requires that we estimate the former's space density.
\citet{coppin06} determine a surface density of SMGs with $S_{\rm
850\mu m}>5$~mJy of $0.139$~arcmin$^{-2}$.  The spectroscopic study of
\citet{chapman05} found that $50\%$ of bright SMGs lie at redshifts
$1.9\le z<2.7$, implying a space density at these redshifts of
$2.63\times 10^{-5}$~Mpc$^{-3}$.  These authors also found $30-50\%$
of them have $25.5< \rs < 28.0$, corresponding to $L_{\rm UV} \la
0.34L^{\ast}$ at the mean redshift of the BX sample ($z=2.30$).
According to our UV LF, the total number density of galaxies over this
same apparent magnitude range is $3.28\times 10^{-2}$~Mpc$^{-3}$,
implying that UV-faint SMGs with $\rs>25.5$ constitute $0.02-0.04\%$
of sources on the faint-end.  Even in the most conservative case where
we assume that all SMGs with $S_{850\mu m}>5$~mJy lie at $1.9\le
z<2.7$ and all have $\rs>25.5$, we find a fractional contribution of
only $0.16\%$.  The results of \citet{chapman05} indicate that this
SMG fraction would be even lower among $z>2.7$ galaxies, although we
note that their adoption of a radio-preselection may have biased the
distribution of their sources to lower redshifts.  We make note of the
fact that the exact contribution will depend on the limit of what we
consider to be ``bright'' SMGs, and extending the limit to fainter
submillimeter fluxes will undoubtedly include galaxies that are less
attenuated, on average, and more likely to be recovered via their UV
colors (e.g., \citealt{reddy05a, adel00}).  In any case, the current
best estimates for SMGs that are observed routinely in the first
generation of submm surveys imply that by number they make a very
small contribution to the number density of sub-$L^{\ast}$ galaxies.

\citet{reddy06a} demonstrate that the vast majority of luminous
infrared galaxies (LIRGs) at $z\ga 2$ will have rest-frame UV colors
that satisfy the BX/LBG criteria.  While such criteria also pick up a
non-negligible number of ultraluminous infrared galaxies (ULIRGs), the
best method of accounting for these galaxies is via their infrared
emission.  The launch of {\em Spitzer} enabled observations that are
sensitive to the warmer dust in high redshift starburst galaxies.
Such galaxies are luminous in the infrared and appear to account for
an increasing fraction of galaxies at $z\ga 1$ (e.g., \citealt{dey08,
caputi07, lefloch05}).  Based on such studies, there have been a few
estimates of the number density of ultraluminous infrared galaxies at
$z\sim 2-3$.  For instance, \citet{caputi07} find a density of
$(1.5\pm0.2)\times 10^{-4}$~Mpc$^{-3}$ for $24\mu$m-selected galaxies
with $L_{\rm bol}\ga 10^{12}$~L$_{\odot}$ (excluding AGN) in the GOODS
fields.  Similarly, $24\mu$m-bright galaxies ($f_{\rm 24\mu m}\ge
0.3$~mJy) with red optical to mid-IR colors ($R-[24]\ge 14$) have a
space density in the redshift range $0.5\le z\le 3.5$ of
$(2.82\pm0.05)\times 10^{-5}$~Mpc$^{-3}$ \citep{dey08}, almost all of
which lie below $L^{\ast}$, the characteristic unobscured UV
luminosity.  Assuming conservatively that all of the ULIRGs of
\citet{caputi07} are fainter than $\rs=25.5$ and as faint as $\rs \sim
28.0$, we find a ULIRG fraction on the UV faint-end of $0.46\%$.  In
terms of the \citet{dey08} objects, assuming their space density does
not evolve between $0.5\le z\le 3.5$, this fraction is $0.086\%$.
Hence, while such dusty, star-forming galaxies contribute
significantly to the total IR luminosity density, they must be vastly
outnumbered by galaxies with fainter bolometric luminosities.  This
result is not surprising given the close-to-exponential drop-off in
number counts of such infrared luminous galaxies according to the
Schechter function, combined with the steep faint-end slope of the UV
LF.

Taken another way, if we make the supposition that a large fraction of
galaxies on the faint-end of the UV LF are indeed very dusty
star-forming ULIRGs, then by virtue of the sheer numbers of UV-faint
galaxies, we would predict a number density of ULIRGs significantly in
excess of the measured value.  These calculations indicate that
rapidly star-forming, dusty galaxies constitute a very small fraction
of the total number density of star-forming galaxies on the faint-end
of UV LF.  Moreover, they support our premise that the $\ebmv$
distribution is unlikely to be redder for UV-faint galaxies than for
UV-bright ones (appendix).

\subsection{Galaxies with Large Stellar Masses}

Another population of galaxies at $z\ga 2$ characterized by their
faint UV luminosities are those that have undergone their major
episode(s) of star formation and are evolving quiescently
\citep{franx03}, commonly referred to as ``Distant Red Galaxies,'' or
DRGs.  Such galaxies have low specific star formation rates
\citep{papovich06, reddy06a} and are inferred to have low gas
fractions \citep{reddy06a} relative to UV-selected galaxies.  The bulk
of BX/LBGs have stellar masses in the range $10^{9} -
10^{11}$~M$_{\odot}$ \citep{erb06b, reddy06b, shapley05}, while $K <
20$ DRGs have typical stellar masses of $\ga 10^{11}$~M$_{\odot}$
(e.g., \citealt{vandokkum04, vandokkum06}), although there is some
small overlap in the stellar mass distribution between BX/LBGs and
DRGs \citep{shapley05, vandokkum06}, particularly at fainter $K$-band
magnitude \citep{reddy05a}.

\citet{vandokkum06} find that galaxies with stellar masses
$>10^{11}$~M$_{\odot}$ are also typically faint in the optical, with
$>2/3$ fainter than $\rs = 25.5$.  Conservatively assuming that all
such galaxies are fainter than $\rs=25.5$ and as faint as $\rs \sim
28.0$, and have an estimated space density of $(2.2\pm0.6)\times
10^{-4}$~Mpc$^{-3}$ \citep{vandokkum06}, then we compute a fractional
contribution to the faint-end of the UV LF in the same magnitude range
of $0.67\%$.  \citet{vandokkum06} noted that only $1/3$ of these
massive galaxies had the colors of BX/LBGs.  However, examination of
their $\ugr$ color distribution shows that a large fraction of the
``missing'' $2/3$ have colors that hug the BX/LBG selection
boundaries.  Our incompleteness corrections will take into account
objects that scatter into the BX/LBG samples because of stochastic
effects like photometric errors, but conservatively assuming that
$2/3$ of massive galaxies are missed even after these corrections
would imply a massive galaxy fraction among UV-faint sources of
$\approx 2\%$.

These estimates imply that like dusty star-forming galaxies, those
with large stellar masses ($>10^{11}$~M$_{\odot}$) comprise a very
small fraction ($\la 2\%$) of all UV-faint galaxies.  Hence, virtually
all sub-$L^{\ast}$ galaxies have smaller stellar masses and are less
dusty than the types of galaxies considered above.  From a broader
perspective, several studies have shown that galaxies with large
stellar masses tend to cluster more strongly than less massive
galaxies \citep{quadri07, adelberger05a}.  This is consistent with the
expectation that galaxies with large stellar masses formed stars
earlier in more massive potential wells which are expected to be the
most clustered.  Furthermore, \citet{adelberger05c} demonstrated that
UV-bright galaxies cluster more strongly than UV-faint ones, at least
at $z\ga 2$.  Given the sheer number of UV-faint galaxies, these
observations suggest that galaxies on the faint-end of the UV LF are
likely to be less clustered than their brighter counterparts, and
hence associated with lower mass halos.

\section{Discussion: Constraints on the Star Formation Rate Density}
\label{sec:sfrd}

\begin{deluxetable*}{lcc}
\tabletypesize{\footnotesize}
\tablewidth{0pc}
\tablecaption{Total UV Luminosity Densities at $1.9\le z<3.4$}
\tablehead{
\colhead{} &
\colhead{Unobscured\tablenotemark{a}} &
\colhead{Dust-Corrected\tablenotemark{b}} \\
\colhead{Redshift Range} &
\colhead{(ergs~s$^{-1}$~Hz$^{-1}$~Mpc$^{-3}$)} &
\colhead{(ergs~s$^{-1}$~Hz$^{-1}$~Mpc$^{-3}$)}}
\startdata
$1.9\le z<2.7$ & $(3.89\pm 0.24)\times 10^{26}$ & $(1.36\pm0.30)\times 10^{27}$ \\ 
$2.7\le z<3.4$ & $(3.28\pm 0.24)\times 10^{26}$ & $(8.74\pm2.55)\times 10^{26}$ \\
\enddata
\tablenotetext{a}{Uncorrected for extinction, integrated to $0.04$~L$^{\ast}_{z=3}$.}
\tablenotetext{b}{Corrected for luminosity-dependent extinction, including both obscured and
unobscured UV luminosity, integrated to $0.04$~L$^{\ast}_{z=3}$.}
\label{tab:lumdenstab}
\end{deluxetable*}

\begin{figure}[h]
\plotone{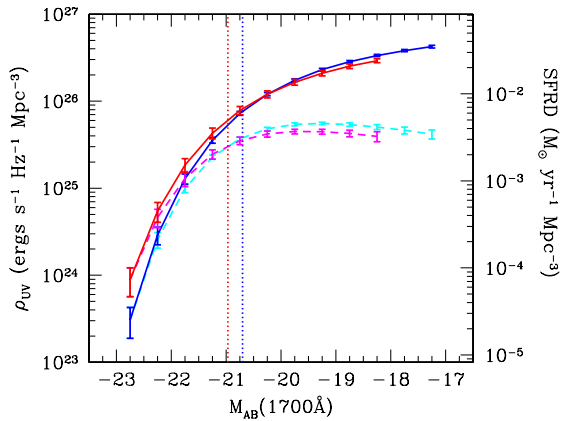}
\caption{Unobscured UV luminosity density, $\rho_{\rm UV}$, per $0.5$
magnitude interval ({\em dashed lines}) and integrated ({\em solid
lines}) at $1.9\le z<2.7$ ({\em cyan, blue}) and $2.7\le z<3.4$ ({\em
magenta, red}), respectively.  Dotted lines indicate $M^{\ast}$ at
$z\sim 2$ and $z\sim 3$.  The equivalent star formation rate density
assuming the \citet{kennicutt98} relation and a Kroupa IMF is shown on
the right-hand axis.}
\label{fig:lumdens}
\end{figure}

As is customary, the \citet{kennicutt98} relation is used to convert
UV luminosity to star formation rate (SFR), adopting a Kroupa IMF from
0.1 to 100~M$_{\odot}$ (Figure~\ref{fig:lumdens},
Table~\ref{tab:lumdenstab}, Table~\ref{tab:sfrdtab}).  This results in
factor of $\sim 1.7$ decrease in SFR for a given luminosity owing to
the larger fractional contribution of high-mass stars to the Kroupa
relative to the \citet{salpeter55} IMF.  For consistency with previous
investigations, the luminosity density is calculated to a limiting
luminosity of $0.04L^{\ast}_{z=3}$ unless stated otherwise.  The
differential and cumulative unobscured UV luminosity densities to
$0.04L^{\ast}_{z=3}$, $\rho_{\rm UV}(>0.04L^{\ast}_{z=3})$, are
$\approx 6\%$ and $42\%$ larger at $z\sim 2$ and $3$, respectively,
than reported by R08.  This difference is attributable to
the steeper faint-end slope and slightly brighter $L^{\ast}$ derived
in this study.  Below, we consider the effects of a
luminosity-dependent dust correction, the bolometric luminosity
functions at $z\sim 2-3$, and implications for the star formation
history.

\begin{deluxetable*}{lcccccccc}
\tabletypesize{\footnotesize}
\tablewidth{0pc}
\tablecaption{SFRD Estimates and Dust Correction Factors}
\tablehead{
\colhead{} &
\colhead{} &
\colhead{$z\sim 2$} &
\colhead{} &
\colhead{\,\,\,\,\,} & 
\colhead{} &
\colhead{$z\sim 3$} &
\colhead{} \\
\colhead{} &
\colhead{$L_{\rm lim}=0.04L^{\ast}$}\tablenotemark{a} &
\colhead{} &
\colhead{$L_{\rm lim}=0$} &
\colhead{} &
\colhead{$L_{\rm lim}=0.04L^{\ast}$}\tablenotemark{a} &
\colhead{} &
\colhead{$L_{\rm lim}=0$}}
\startdata
(1) UV SFRD$_{\rm uncor}$\tablenotemark{b} & $0.032\pm0.002$ & & $0.064\pm0.003$ & & $0.027\pm0.002$ & & $0.055\pm0.003$ \\
(2) UV SFRD$_{\rm cor}$ (LDR)\tablenotemark{bc} & $0.112\pm0.025$ & & $0.122\pm0.027$ & & $0.072\pm0.021$ & & $0.080\pm0.023$ \\
(3) UV Dust Correction (LDR)\tablenotemark{c} & $3.50\pm0.78$ & & $1.91\pm0.42$ & & $2.67\pm0.78$ & & $1.45\pm0.42$ \\
(4) UV SFRD$_{\rm cor}$ (CR)\tablenotemark{bd} & $0.144\pm0.009$ & & $0.288\pm0.014$ & & $0.122\pm0.009$ & & $0.248\pm0.014$ \\
(5) UV Dust Correction (CR)\tablenotemark{d} & $4.5$ & & $4.5$ & & $4.5$ & & $4.5$ \\
(6) Total SFRD$_{\rm cor}$\tablenotemark{be} & $0.142\pm0.036$ & & $0.152\pm0.038$ & & $0.102\pm0.032$ & & $0.110\pm0.034$ \\
(7) Total Dust Correction\tablenotemark{f} & $4.44\pm1.13$ & & $2.38\pm0.59$ & & $3.78\pm1.19$ & & $2.00\pm0.62$ \\
\enddata
\tablenotetext{a}{Integrated to include all galaxies with unobscured UV luminosities brighter than $0.04L^{\ast}_{z=3}$, or $M(1700\AA)\approx -17.48$.}
\tablenotetext{b}{In $M_{\odot}$~yr$^{-1}$ assuming a Kroupa IMF.}
\tablenotetext{c}{Invokes luminosity-dependent reddening (LDR).}
\tablenotetext{d}{Invokes luminosity-invariant reddening (CR).}
\tablenotetext{e}{Sum of LDR-corrected star formation rate density from row(2) and the 
contribution of $L_{\rm bol}>10^{12}$~L$_{\odot}$ galaxies from \citet{caputi07}.}
\tablenotetext{f}{Dust correction required to recover total star formation rate density in row (6) from the 
unobscured star formation rate density in row (1).}
\label{tab:sfrdtab}
\end{deluxetable*}

\subsection{Luminosity-Dependent Dust Corrections}

As a consequence of the steep faint-end slopes at $z\sim 2-3$,
$\approx 93\%$ of the unobscured UV luminosity density (integrated to
zero luminosity) is contributed by galaxies fainter than $L^{\ast}$
(Figure~\ref{fig:lumdens}).  The abundance of UV-faint galaxies and
their cumulative luminosity makes them ideal candidates for the
sources responsible for most of the ionizing flux at $z\ga 3$.
However, the luminosity dependence of reddening implies that their
contribution to the {\em bolometric} luminosity is likely to be
diminished compared to their contribution to the unobscured luminosity
density.  The bolometric luminosity density can be expressed simply
as
\begin{eqnarray}
\rho^{\rm bol}_{\rm UV} = \int_{L}L\phi(L)10^{[0.4k'(\lambda){\sc A}(L)]}dL,
\label{eq:rhobol}
\end{eqnarray}
where $A(L)$ is the reddening, parameterized by $\ebmv$, as a function
of luminosity and $k'(\lambda)$ is the starburst attenuation relation
defined in \citet{calzetti00}.  For this calculation, we have assumed
that the bolometric luminosity can be recovered from the rest-frame UV
colors --- as motivated by {\em Spitzer} mid-IR observations of
UV-selected galaxies \citep{reddy06a} --- via the \citet{calzetti00}
relation.  This has been shown to be valid for moderately luminous
galaxies (i.e., LIRGs; \citealt{reddy05a, reddy06a}).  We will
consider shortly the contribution from high redshift galaxies that do
not follow the local starburst attenuation relations \citep{meurer99,
  calzetti00}.

Defining $N(\ebmv,L)\equiv N(\ebmv)$ will of course leave the relative
contribution of UV-faint galaxies to $\rho^{\rm bol}_{\rm UV}$
unchanged from their contribution to the unobscured UV luminosity
density.  The bolometric luminosity density is calculated under the
more realistic case of a declining average reddening with unobscured
luminosity (appendix), with the results tabulated
in Tables~\ref{tab:lumdenstab} and \ref{tab:sfrdtab}.  In this case,
we find that galaxies fainter than $L^{\ast}$ --- defined as the
characteristic {\em unobscured} UV luminosity --- contribute $0.62$
and $0.78$ at $z\sim 2$ and $3$, respectively, to $\rho^{\rm bol}_{\rm
UV}$ integrated to $0.04L^{\ast}_{z=3}$.  These fractions are likely
to be lower limits since there is a non-negligible number of very
dusty and bolometrically luminous UV-faint galaxies at these redshifts
(\S~\ref{sec:nature}).  Below, we revisit our estimate of the
bolometric luminosity density after incorporating the effect of the
most luminous galaxies at $z\sim 2-3$.

\subsection{Bolometric Luminosity Functions}

\begin{figure}[t]
\plotone{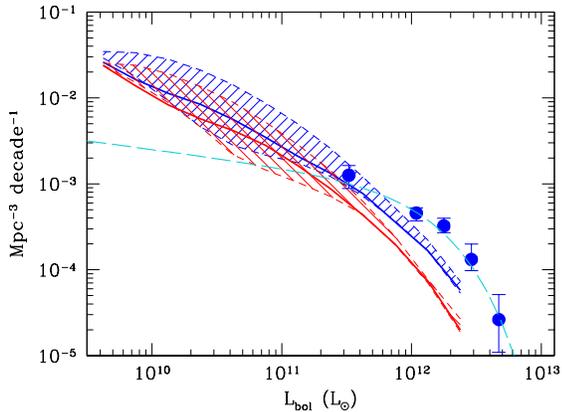}
\caption{Bolometric luminosity functions at $z\sim 2$ ({\em blue}) and
$z\sim 3$ ({\em red}), computed by combining the measurement of the UV
luminosity function with a luminosity-dependent $\ebmv$ distribution
(see text).  The upper limits of the shaded regions indicate the LF
derived assuming a constant $\ebmv$ distribution.  The lower limits
indicate the LF derived assuming that all galaxies with apparent
magnitude fainter than $\rs = 25.5$ have zero reddening.  These limits
encompass the range of likely LFs and give an indication as to the
systematic uncertainty in the bolometric LF.  The solid lines denote
the bolometric LF obtained using our model of the luminosity-dependent
$\ebmv$ distribution that gradually falls to zero reddening for the
faintest galaxies.  At $z\sim 2$, the higher luminosity points ({\em
circles}) from \citet{caputi07} are shown, along with their Schechter
extrapolation to fainter luminosities ({\em long-dashed line}).}
\label{fig:bollf}
\end{figure}

Although so far we have defined $L^{\ast}$ in terms of the knee of the
{\em UV} LF uncorrected for extinction, we can also examine the
fractional contributions as a function of luminosity to the bolometric
luminosity density.  This is accomplished by reconstructing the UV LF
corrected for extinction using a method similar to that presented in
R08.  Briefly, a large number of galaxies are simulated with
magnitudes and $\ebmv$ drawn randomly from the LF and
luminosity-dependent $\ebmv$ distribution.  The \citet{calzetti00}
relation is used to recover the bolometric luminosities, which are
then binned to produce a luminosity function (Figure~\ref{fig:bollf}).
We allow for the LF to vary within the errors and add a $0.3$~dex
scatter to the dust correction implied by $\ebmv$, reflecting the
approximate scatter in both the local relations \citep{meurer99,
calzetti00} and those found at $z\sim 2$ \citep{reddy06a}.  This
scatter results in a $5\%$ random error in the faint-end of the
bolometric LF, significantly smaller than the systematic errors that
result from assuming different relations between dustiness and UV
luminosity (Figure~\ref{fig:bollf}).

Note that we use only the $\ebmv$ distribution found for UV-selected
galaxies to reconstruct the bolometric luminosity functions.  The
range in attenuation factors obtained for such galaxies will be
smaller than the intrinsic range of reddening among all galaxies at
$z\sim 2-3$.  One obvious reason for this bias is the incompleteness
for objects that never scatter into our sample because of their red
colors.  Another reason is that even if such red, dusty galaxies do
satisfy the LBG color criteria, their bolometric luminosities may be
underestimated severely based on the UV colors alone.  Hence, the
method for recovering bolometric LFs based on the $\ebmv$ distribution
of galaxies that scatter into the BX/LBG windows will underestimate
the contribution of galaxies to the bright-end of the bolometric
luminosity function.  Because of this, the contribution of these dusty
galaxies is based on results published elsewhere.  Specifically, we
adopt the value of the bright-end of the infrared LF at $z\sim 2$
(after exclusion of bright AGN) presented by \citet{caputi07} since
the bright-end of the bolometric LF should track the bright-end of the
IR LF.  The results are summarized in Figure~\ref{fig:bollf} assuming
no evolution in the bright-end of the LF ($L_{\rm bol}\ga
10^{12}$~L$_{\odot}$) in the redshift range $1.9\le z<3.4$.  It is
important to keep in mind that $\nu L_{\nu}$ at $1700$~\AA\, scales
with SFR in a different way than the infrared luminosity ($L_{\rm IR}
\equiv L(8-1000\mu)$).  Hence, the bolometric luminosity --- the sum
of the UV and IR luminosities as defined in this paper --- will scale
in a non-linear way with SFR.  In the present context, star formation
rate densities are computed separately for (1) galaxies where $L_{\rm
bol}$ is determined from the UV-corrected values and (2) galaxies with
$L_{\rm bol}\ga 10^{12}$~L$_{\odot}$ where the bolometric luminosity
is determined purely from the infrared luminosity \citep{caputi07}.
The star formation rate densities from the two contributions are then
added to estimate the total.  With the appropriate scalings, this
calculation implies that $\approx 70-80\%$ of the bolometric
luminosity density arises from galaxies with $L_{\rm bol}\la
10^{12}$~L$_{\odot}$, consistent with findings of R08.

Taken together, these findings can be summarized as follows.
Including ULIRGs --- those galaxies whose UV slopes tend to
under-predict their bolometric luminosities (e.g.,
\citealt{papovich06, papovich07, reddy06a}) --- does not change the
fact that a large portion of the bolometric luminosity density arises
from faint galaxies, either those that are fainter than the
characteristic unobscured UV luminosity or those that are fainter than
the characteristic bolometric luminosity.  Placing these results in a
wider context will require more precise estimates of the bright-end of
the bolometric luminosity function that (1) take into account the
luminosity-dependent conversion between mid-IR luminosity (upon which
most estimates are based) and the total infrared luminosity and (2)
the potential contamination from AGN that are prevalent among galaxies
with such high IR luminosities.  Nonetheless, combining the most
recent estimate of the bright-end of the bolometric luminosity
function \citep{caputi07} with our results at the faint-end points to
a luminosity density that is dominated by bolometrically faint to
moderately luminous galaxies.  The implications for a
luminosity-dependent reddening distribution on the average dust
correction factors applied to high redshift samples and the evolution
of the star formation rate density are discussed in the following
sections.

\subsection{Average Dust Correction Factors}

A luminosity dependent dust correction and the large ratio of UV-faint
to UV-bright galaxies implies an average UV dust correction that is
sensitive to the limit of integration used to compute the luminosity
density.  It seems prudent to consider such a systematic effect given
that estimates of the star formation rate density imply stellar mass
densities in excess of what are actually measured \citep{wilkins08}.
This effect is mentioned in R08; here, we proceed to quantify the
average dust correction factors relevant for luminosity densities
computed to different limits based on our new determination of the
faint-end slope.

The calculated dust corrections and star formation rate densities are
listed in Table~\ref{tab:sfrdtab}.  We have assumed a contribution of
$L_{\rm bol}>10^{12}$~L$_{\odot}$ galaxies to the star formation rate
density at $z\sim 2$ as computed from \citet{caputi07}.  We also
assume this same contribution at $z\sim 3$, though it has not been
measured directly at these higher redshifts, in order to place
conservative estimates on the effect of a luminosity-dependent dust
correction on the average dust correction factors required to convert
UV luminosity densities to star formation rate densities.

The luminosity-dependent reddening model implies dust corrections of a
factor of $3.5$ and $2.7$ at $z\sim 2$ and $3$, respectively,
integrated to $0.04L^{\ast}$ (Table~\ref{tab:sfrdtab}), which are up
to a factor of two smaller than the typical $4.5-5.0$ dust corrections
found for $\rs\le 25.5$ galaxies \citep{steidel99, reddy04}.  Aside
from differences in the luminosity range probed, this difference in
average extinction is mitigated somewhat by the fact that a
significant fraction ($\sim 0.2-0.3$) of the bolometric luminosity
density arises from ULIRGs, where the usual dust conversions do not
apply (see discussion above).  The expectation is that the lower
dust-corrected luminosity densities inferred in the
luminosity-dependent reddening case are compensated by the inclusion
of galaxies where $\ebmv$ tends to under-predict the reddening.  The
total dust corrections required to recover the bolometric luminosity
density, including that contributed by ULIRGs, are $4.4$ and $3.8$,
somewhat larger than the values quoted above.  While the differences
between these dust corrections may seem small at $z\sim 2-3$, they do
result in up to a factor of two in systematic scatter in star
formation rate density measurements, comparable to the dispersion in
the local calibrations between luminosity and star formation rate, and
so should be taken into account.  The dependency of the average dust
correction on the integration limit will be even greater for steeper
faint-end slopes given the larger fractional contribution of
less-reddened faint galaxies to the luminosity density.  The average
dust correction factors stated above are relevant when integrating the
UV luminosity function to $0.04L^{\ast}_{z=3}$.  Integrating to zero
luminosity alters the corrections to be a factor of $2.4$ and $2.0$ at
$z\sim 2$ and $3$, respectively.  To reiterate, these extinction
corrections account for not only the dust-obscuration among moderately
luminous galaxies prone to UV-selection, but also for those
ultraluminous galaxies that may either escape UV-selection or simply
have rest-UV slopes that under-predict their bolometric luminosities.
These effects underscore the various subtleties that can affect
extinction corrections for UV-selected samples.

Of course, these dust corrections are equally important at higher
redshifts $z\ga 3$ where the only constraints on the star formation
rate density come from UV observations.  Evidence suggests that
UV-selected galaxies become bluer at redshifts $z\ga 3$, relative to
galaxies at lower redshifts \citep{yan04b, bouwens07}.\footnote{It is
generally accepted that the highest redshift star-forming populations
exhibit bluer UV colors {\em on average} than similarly selected
galaxies at lower redshifts, although the result has not been verified
independently with longer wavelength observations.  \citet{bouwens07}
present a discussion of why there may not be a strong bias against
dusty galaxies with the higher redshift dropout criteria.}  This trend
may be attributable to two effects.  First, as noted above, $L^{\ast}$
evolves strongly as a function of redshift at $z\ga 3$, such that the
average UV luminosity of galaxies decreases with increasing redshift.
If sub-$L^{\ast}$ galaxies have lower dust reddening than UV-bright
ones, the trend in UV color may be interpreted as a decrease in dust
reddening.  A consequence of the luminosity-dependent reddening model
is that, when examined over a large dynamic range of luminosity, the
unobscured UV luminosity must track the bolometric luminosity, or SFR.
This leads to the second effect which is tied to the observation that
high redshift galaxies are less attenuated than lower redshift
galaxies of the same bolometric luminosity, resulting in a trend of
decreasing extinction per unit SFR proceeding to higher redshifts
(\citealt{reddy06a}, R08).  Stated another way, extrapolating the
results of \citet{reddy06a} to redshifts $z\ga 4$ implies that higher
redshift galaxies are less opaque at UV wavelengths than lower
redshift galaxies of the same bolometric luminosity.  Hence, the
combined effect of a lower $L^{\ast}$ and lower average dust
attenuation to a given bolometric luminosity implies lower dust
corrections at higher $z$ (R08, \citealt{bouwens07}).  The
implications for this evolution in the average dust correction are
discussed in the next section.

There have been several explanations put forth to explain the
discrepancy between the integrated star formation history and stellar
mass density measurements, including missing stellar mass, an
evolution of the IMF \citep{dave08, vandokkum08}, or more generally an
evolving conversion between luminosity and star formation rate.  In
light of these effects, we consider the potential impact of an
evolving dust correction on the star formation history, as described
below.

\section{Discussion: Star Formation History and Buildup of Stellar Mass}
\label{sec:sfrdevol}

We have already touched upon a few of the implications of a steep
faint-end slope on the star formation rate density.  In particular, we
noted that a very large fraction of the unobscured UV luminosity
density ($\ga 90\%$) arises from galaxies fainter than the
characteristic unobscured UV luminosity.  Similarly, our results
suggest that even assuming a lower reddening among UV-faint galaxies
relative to UV-bright ones implies a bolometric, or dust-corrected,
luminosity density dominated by galaxies fainter than the
characteristic bolometric luminosity.  There are several important
consequences of these results that we discuss in the next few
sections.

\subsection{Contributions at $z\sim 2-3$ to the Global Stellar Mass Density}
\label{sec:smdcont}

A significant fraction of the stellar mass density that formed between
$z=1.9$ and $z=3.4$ (the redshift limits of our analysis) ---
corresponding roughly to the epoch when galaxies were forming most of
their stars (\S~\ref{sec:intro}) --- occurs in galaxies with $L_{\rm
bol}\la 10^{12}$~L$_{\odot}$.  Using a linear interpolation of the
contributions of galaxies with different luminosities to the
bolometric luminosity density between $z\sim 3$ and $z\sim 2$ and
multiplying by the time between $z=3.4$ and $z=1.9$, yields a total
SMD of $\Omega_{\ast}(1.9\le z < 3.4) = 0.0014\pm 0.0003$ in units of
the critical density.  This value is already $0.57$ times that of the
present-day value reported in \citet{cole01}.  As mentioned above,
there is still a fair amount of uncertainty regarding the bright-end
of the bolometric luminosity function.  Irrespective of the number
density of bolometrically-luminous galaxies, our calculations suggest
that $43\pm 9\%$ of the present-day stellar mass density was formed in
galaxies with $6\times 10^{8}<L_{\rm bol}<10^{12}$~L$_{\odot}$ between
redshifts $z=3.4$ and $1.9$.\footnote{The limit of $6\times
10^{8}$~L$_{\odot}$ is adopted for consistency with R08.  The
bolometric LF exhibits a slope that is somewhat shallower than the UV
LF, so changing the limit of integration to zero bolometric luminosity
will add roughly $10\%$ to the luminosity density.}  While much
attention recently has been focused on the stellar mass buildup
associated with luminous galaxies at high redshift, it is clear that
fainter galaxies, those that are routinely picked up in UV surveys but
may be missing from rest-optical and far-IR ones, also play an
important role.  It suggests that much of the stellar mass assembly at
a time when galaxies where forming most of their stars occurred in the
typical and more numerous galaxies that populate these redshifts.

\subsection{Evolution of the Star Formation Rate Density}

\begin{figure}[t]
\plotone{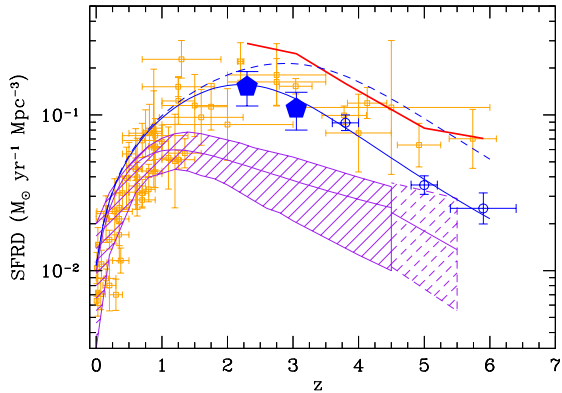}
\caption{Cosmic star formation history, including the
luminosity-dependent dust-corrected determinations at $z\sim 2-3$ from
this analysis ({\em large pentagons}) and those of \citet{bouwens07}
at $z\ga 3.8$ ({\em open circles} at high $z$), and the compilation
from \citet{hopkins04} ({\em open squares}) at low $z$.  Note that our
estimates {\em include} the directly-measured contribution to the star
formation rate density from ultraluminous infrared galaxies and assume
that this contribution is non-evolving between $z\sim 3$ and $z\sim
2$.  Also, for consistency with the \citet{hopkins04} compilation, our
points and those of \citet{bouwens07} are computed by integrating the
UV LF to zero luminosity.  The {\em solid} red line shows the star
formation history assuming a constant dust correction of $4.5$ to the
unobscured UV luminosity densities at $z\sim 2-6$.  The {\em
short-dashed} line shows the fit to the star formation history
including this constant dust correction model.  The {\em solid} blue
line indicates the best-fit star formation history assuming a
luminosity-dependent dust correction to the $z\ga 2$ measurements.
The {\em solid hatched purple} region indicates the $\pm 1$~$\sigma$
star formation history inferred from the evolution of the stellar mass
density \citep{wilkins08}, with an extrapolation at $z\ga 4.5$ based
on stellar mass density measurements at $z\sim 5-6$ from
\citet{mclure08, eyles07, stark07, verma07, yan06b} ({\em dashed
purple} region).  As discussed in the text, much of the discrepancy
between the stellar mass density measurements and the integral of the
star formation history may be due to incompleteness of low mass
objects in the stellar mass estimates.  A Kroupa IMF is assumed
throughout.}
\label{fig:sfrd}
\end{figure}

A luminosity-dependent dust correction not only has consequences for
the total star formation rate density measured at a given redshift,
but the strong evolution of $L^{\ast}$ suggests that it will induce a
systematic effect with redshift.  Figure~\ref{fig:sfrd} summarizes the
star formation rate densities inferred with the luminosity-dependent
reddening model from this study and those of \citet{bouwens07} at
$z\ga 3.8$, compared with the star formation history assuming a
constant dust-correction of $4.5$ ({\em red line}), along with lower
redshift determinations compiled in \citet{hopkins04}.  For
consistency with the latter study, we have integrated to zero
luminosity (see \S~\ref{sec:cautions} for a discussion of the
systematic effects associated with the limits of integration).  The
luminosity-dependent reddening-corrected star formation history points
to a factor of $8-9$ increase in the star formation rate density
between $z\sim 6$ and $z\sim 2$ (e.g., see also R08,
\citealt{bouwens07}), significantly steeper than the factor of $4$
that we would have inferred in the case of a luminosity-invariant
(constant) dust correction.  As discussed in \S~\ref{sec:mstarevol},
the evolution in the unobscured star formation rate density is
connected to the increase in number, and hence luminosity, density of
galaxies brighter than $L^{\ast}$.  Figure~\ref{fig:sfrd} demonstrates
that the comparable evolution in the bolometric star formation rate
density is driven both by an increase in the number density of bright
galaxies and an evolving dust correction.

The elevated star formation rate densities predicted in the
luminosity-invariant reddening model result in a stellar mass growth
of $\Omega_{\ast} = 0.0026\pm0.0007$ between $2.3\le z\le 5.9$,
compared to $\Omega_{\ast} = 0.0013\pm0.0003$ for the
luminosity-dependent reddening model over the same redshifts, a factor
of two difference between the two dust correction scenarios.  More
importantly, a constant dust correction model predicts stellar mass
buildup between $2.3\le z\le 5.9$ that exceeds the local measurement.
Notwithstanding the noted disagreement between stellar mass density
measurements and the integral of the star formation history
(Figure~\ref{fig:sfrd} and discussion below), these results suggest
that we may be able to rule out the elevated star formation rates
predicted by constant dust correction models, although we caution that
the differences in total stellar mass accumulated by $z\sim 2.3$ based
on a constant versus declining star formation history are small
compared to the uncertainties in the star formation rate and stellar
mass density measurements.  In the next section, we discuss this
disparity between the integrated star formation history and stellar
mass density measurements and possible resolutions.

\subsection{Reconciling the Star Formation History with Stellar Mass Density Measurements:
Luminosity-Dependent Dust Corrections and Missing Stellar Mass}

Figure~\ref{fig:sfrd} shows the star formation history inferred by
differentiating measurements of the stellar mass density (integrated
to zero) as a function of redshift ({\em purple hashed region}). The
results imply that there is a maximum disparity of $\approx 0.5$~dex
in this inference and actual observations of the star formation rate
density at $z\sim 2-3$.  It is of general interest to determine
whether this discord is due to some lack of understanding of the
fundamental physical processes that govern star formation and/or to
the mundane nature of the uncertainties that seemingly plague SFR and
stellar mass estimates, including sample incompleteness and the limits
to which one integrates to obtain the star formation rate and stellar
mass densities.

In light of the steep faint-end slopes of the UV LF advocated at $z\ga
2$, it is worthwhile to consider the possibility that the stellar mass
density measurements at these redshifts are too low, primarily because
they do not account for low mass galaxies that may escape stellar mass
selected samples but, even with their low stellar masses, are
sufficiently numerous to add appreciably to the total budget of
stellar mass.  The comparison drawn in Figure~\ref{fig:sfrd}
implicitly assumes that all the galaxies contributing to the estimate
of the star formation rate density are in some way also represented in
the estimate of the stellar mass density.  In practice, the problem is
that unlike SFR-limited samples, mass-selected samples at high
redshift do not probe far enough down the stellar mass function due to
the significant amount of time required to assemble the requisite
near-IR data.  Hence, such studies may underestimate the low-mass
slope of the stellar mass function.

\subsubsection{Stellar Mass Density in UV-Bright ($\rs \le 25.5$) Galaxies}

The slope of the stellar mass function at $z\sim 2-3$ is not
well-constrained.  However, if we are able to estimate the average
stellar mass of LBGs, then knowing their number density from the UV LF
will enable us to estimate their contribution to the stellar mass
density.  We compiled stellar mass estimates for BXs and LBGs in the
GOODS-N and Q1700 fields \citep{reddy06b, shapley05}.  Briefly,
stellar masses are computed for spectroscopically-confirmed BXs and
LBGs by fitting \citet{bruzual03} model templates to the observed
$\ugr$+$J\ks$+IRAC photometry, and allowing the star formation history
$\tau$ and $\ebmv$ to vary freely.  The star formation rate and
stellar mass are determined by the normalization of the model SED to
the broadband photometry \citep{shapley05, erb06b, reddy06b}.

\begin{figure*}[t]
\plottwo{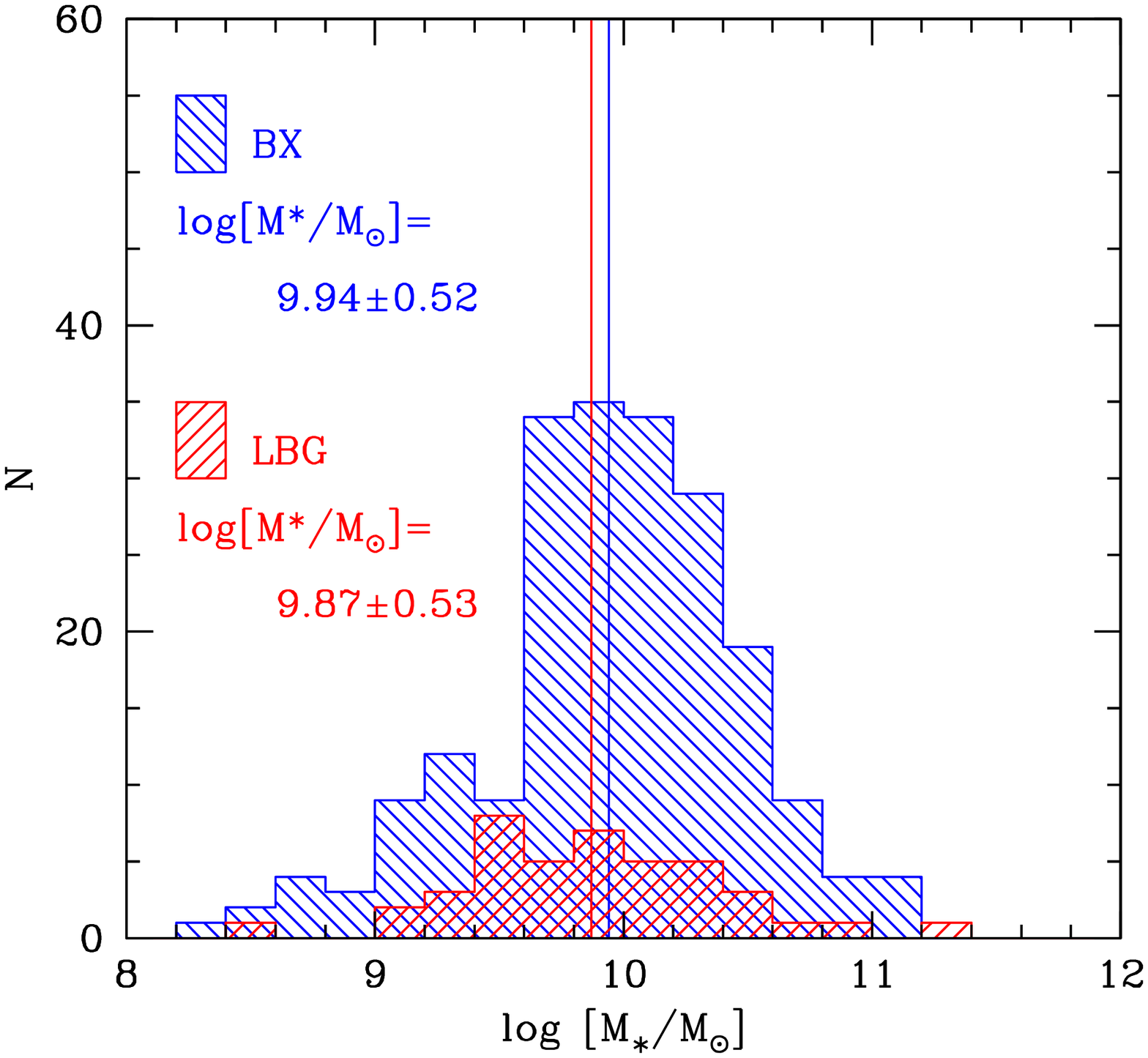}{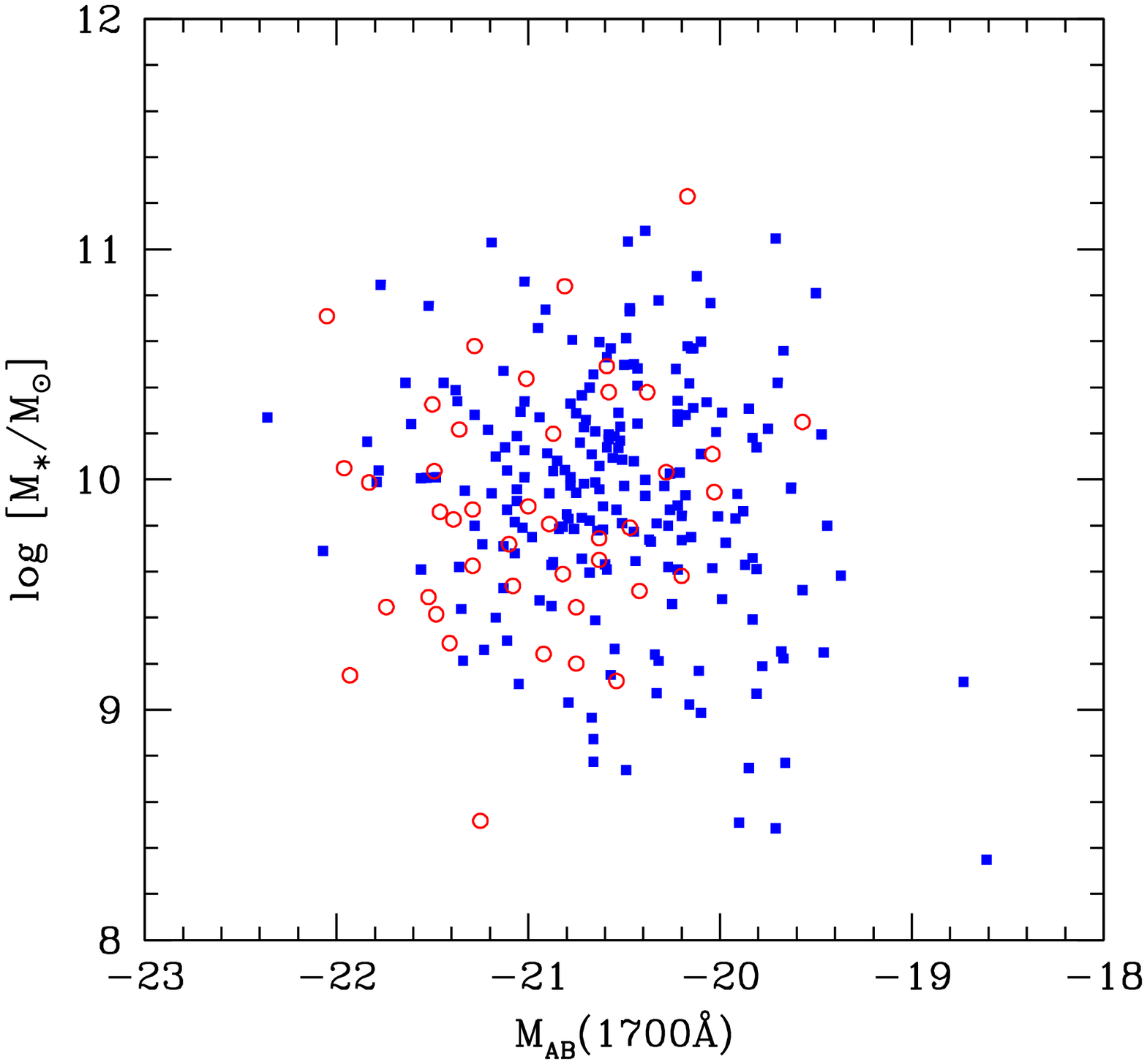}
\caption{({\em Left}): Distribution of stellar mass for spectroscopically-confirmed
BX/LBG galaxies, excluding AGN, with mean values indicated in the panel and by the vertical
lines.  ({\em Right}): Stellar mass as a function of unobscured absolute UV magnitude for
BX galaxies ({\em filled squares}) and LBGs ({\em open circles}).}
\label{fig:smdist}
\end{figure*}

Excluding spectroscopically identified AGN, the distributions of
stellar mass for 208 BXs and 42 LBGs from the 2 aforementioned fields
are shown in Figure~\ref{fig:smdist}.  Since most moderately
star-forming galaxies escape BX/LBG selection due to stochastic
effects (e.g., photometric scatter; \S~\ref{sec:maxlik}), we adopt the
reasonable premise that galaxies to $\rs=25.5$ that do not satisfy the
BX/LBG criteria have a similar distribution in stellar masses to those
that do.  Note that, as discussed previously, this may not be the case
for the most massive galaxies at these redshifts (e.g.,
\citealt{vandokkum06}).  However, in the present context we are
interested primarily in the contribution of {\em typical} star-forming
galaxies --- which outnumber by far the most massive galaxies at these
redshifts (\S~\ref{sec:nature}) --- to the stellar mass density.
Further, we show in the appendix that adopting a young stellar
population does not affect appreciably the incompleteness corrections
to the BX/LBG samples.  Consequently, it is unlikely that large
numbers of young galaxies with low stellar masses are scattered out
the sample relative to the frequency with which more massive galaxies
are scattered out of the sample, particularly among UV-bright
galaxies.  Therefore, we make the reasonable assumption that the
stellar mass distributions for UV-bright BX/LBGs are representative of
UV-bright star-forming galaxies in general.  Figure~\ref{fig:smdist}
also shows the distribution in absolute UV magnitude for galaxies with
stellar mass estimates, spanning the full range of magnitudes
represented in the spectroscopic sample.  As first shown by
\citet{shapley05}, we find no significant correlation between
unobscured UV magnitude and stellar mass, perhaps not surprising since
the two quantities are related only peripherally.  We will revisit
this issue below.

\begin{figure*}[t]
\plottwo{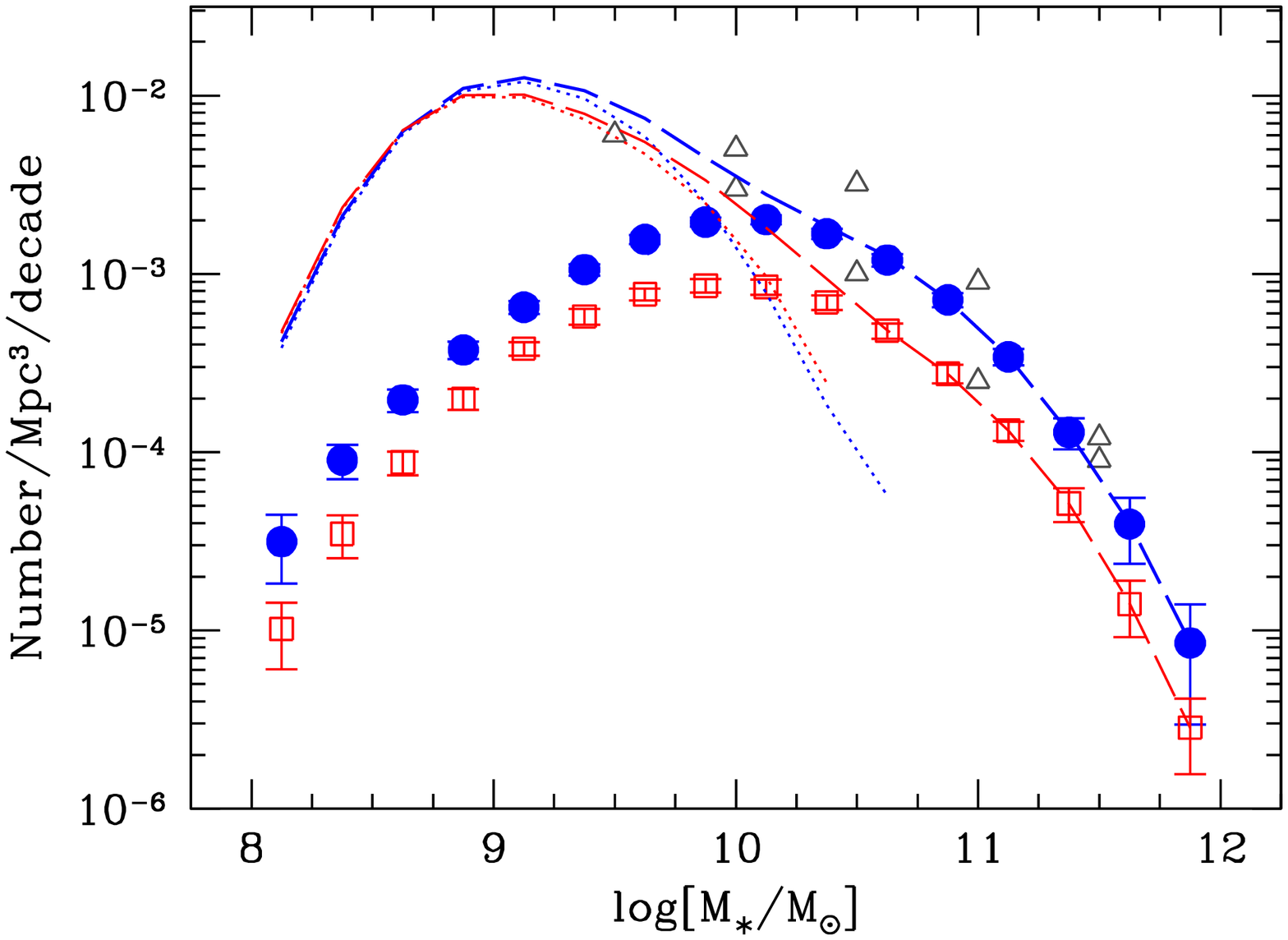}{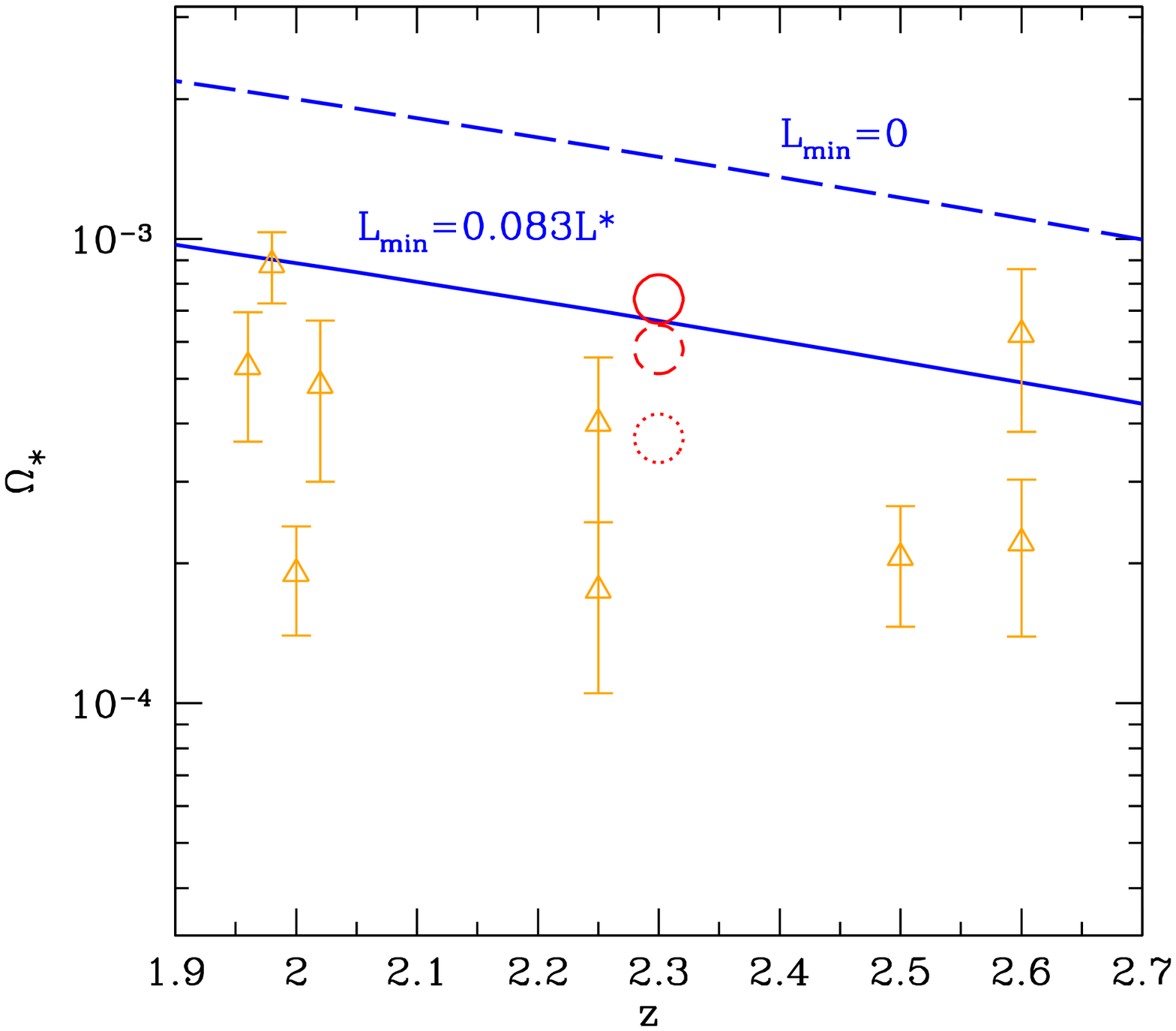}
\caption{({\em Left}): Stellar mass functions for $\rs \le 25.5$
star-forming galaxies at $1.9\le z < 2.7$ ({\em filled circles}) and
$2.7\le z < 3.4$ ({\em open squares}), based on combining the number
density computed from the UV LF and the stellar mass distribution
measured for BXs and LBGs (Figure~\ref{fig:smdist}).  The dotted lines
indicate the inferred contribution from galaxies fainter than
$\rs=25.5$, based on the trend between SFR and stellar mass for
UV-selected galaxies (see text).  The dashed lines indicate the total
contribution from both UV-bright and faint galaxies.  For comparison,
the GOODS and FDF results at $2.25<z<3.00$ from \citet{drory05} are
denoted by the open triangles. ({\em Right}): Stellar mass density
measurements at $1.9<z<2.7$ ({\em open triangles}) from the following
sources: \citet{rudnick03, drory05, pozzetti07, fontana03,
dickinson03, fontana06}.  All of these studies constrain the SMD over
areas that are significantly smaller than the almost $1$~square degree
probed in this study, and most rely on photometric redshifts.  Our
estimates at $z\sim 2.3$ are shown by the large circles: {\em dotted}
shows the estimate for UV-bright ($\rs \le 25.5$) galaxies with
$M_{\ast}<10^{11}$~M$_{\odot}$; {\em dashed} shows the estimate
including UV-faint galaxies with $M_{\ast}<10^{11}$~M$_{\odot}$ to a
faint limit of $M_{\rm AB}(1700\AA) = -18.0$; and {\em solid} denotes
the total contribution including massive
($M_{\ast}>10^{11}$~M$_{\odot}$) galaxies.  The stellar mass density
inferred by integrating the star formation history
(Figure~\ref{fig:sfrd}) to a limit of $0.083L^{\ast}$ (same as that
used to compute the stellar mass density) is denoted by the solid
line.  For comparison, the dashed line shows the result when
integrating the star formation history to zero luminosity.}
\label{fig:massfcn}
\end{figure*}

Based on these distributions, let us proceed to estimate the stellar
mass density contributed by UV-bright galaxies.  To do this, we
generated many random realizations of the UV LF as allowed by the
errors, and drew random absolute magnitudes from each of these
realizations.  We assigned a stellar mass drawn randomly from the
observed distribution (Figure~\ref{fig:smdist}), which is then
perturbed by $0.3$~dex to account for {\em random} uncertainties
\citep{shapley05}.  The masses are then binned to produce a rough
proxy for the stellar mass function.  The resulting Gaussian
distributions for $\rs \le 25.5$ galaxies at $z\sim 2$ and $3$ --
corresponding to galaxies with $M_{\rm AB}(1700\AA) \le -19.53$ and
$-20.05$ and $z=2.3$ and $z=3.05$, respectively --- are shown in
Figure~\ref{fig:massfcn}.  Integrating these distributions for those
galaxies with $M_{\ast}<10^{11}$~M$_{\odot}$ yields:
\begin{eqnarray}
\Omega_{\ast}(M_{\rm AB}(1700\AA) \le -19.53; <10^{11}\, {\rm M}_{\odot}; z\sim 2.3) = \\
(3.72\pm0.28)\times 10^{-4} \nonumber \\
\Omega_{\ast}(M_{\rm AB}(1700\AA) \le -20.05; <10^{11}\, {\rm M}_{\odot}; z\sim 3.05) =\nonumber \\
(1.53\pm0.15)\times 10^{-4}
\end{eqnarray}
in units of the critical density (Table~\ref{tab:smopt}).  These
estimates are meant to reflect the stellar mass densities contributed
by UV-bright star-forming galaxies.  Note that the stellar mass
densities computed here differ from those derived in
\S~\ref{sec:smdcont}; the latter are based on integrating the star
formation rate density, whereas the former are based on masses
determined from broadband fitting of galaxy SEDs, and so are subject
to somewhat different systematics.  The important result of this
section is that even without corrections for (1) the most massive and
dusty galaxies at these redshifts for which the BX/LBG criteria are
incomplete (\S~\ref{sec:nature}) and (2) UV-faint galaxies with $\rs >
25.5$, we already find a stellar mass density at $z\sim 2$ comparable
to estimates from rest-frame optically-selected samples
(Figure~\ref{fig:massfcn}; \citealt{fontana03, dickinson03, rudnick03,
drory05}).  Of course, direct comparisons between our measurements and
those from optically-selected samples are fraught with significant
biases, both random (e.g., field-to-field variations in the optical
samples) and systematic (adopted rest-frame optical limits,
underestimates of stellar mass by assuming a single component SF
model, or more generally systematics in the assumed $M/L$ ratio and
differences in stellar population models).  Some of the random
uncertainties are constrained by taking values from different surveys
conducted in spatially disjoint fields, and at face value, the results
above suggest that typical star-forming galaxies already contain an
amount of stellar mass comparable to that detected in rest-frame
optically-selected surveys.

\begin{deluxetable}{lcc}
\tabletypesize{\footnotesize}
\tablewidth{0pc}
\tablecaption{Stellar Mass Density Budget at $1.9\le z<3.4$}
\tablehead{
\colhead{} &
\colhead{$1.9\le z < 2.7$} &
\colhead{$2.7\le z < 3.4$}}
\startdata
$\Omega_{\ast}(\rs \le 25.5; < 10^{11}\, {\rm M}_{\odot})$\tablenotemark{a} & $3.72\pm0.28$ & $1.53\pm0.15$ \\
$\Omega_{\ast}(\rs > 25.5; < 10^{11}\, {\rm M}_{\odot})$\tablenotemark{b} & $2.06\pm0.26$ & $1.86\pm0.21$ \\
$\Omega_{\ast}(> 10^{11}\, {\rm M}_{\odot})$\tablenotemark{c} & $1.64\pm0.45$ & ... \\
$\Omega_{\ast}({\rm Total})$\tablenotemark{d} & $7.42\pm 0.59$ & $>3.39\pm 0.26$
\enddata
\tablenotetext{a}{Stellar mass density, in units of the critical density $\times 10^{-4}$, in galaxies
with $\rs\le 25.5$ and stellar masses $<10^{11}$~M$_{\odot}$ assuming a Kroupa IMF.}
\tablenotetext{b}{Same as (a), but includes the contribution inferred for $\rs>25.5$
galaxies based on the correlation between SFR and stellar mass for UV-selected galaxies (see text).}
\tablenotetext{c}{Stellar mass density in galaxies with stellar masses $> 10^{11}$~M$_{\odot}$,
based on the data of \citet{vandokkum06}.}
\tablenotetext{d}{Total stellar mass density, computed by adding the numbers from the first
three rows, including the contribution of UV-bright and faint galaxies, as well as those
with stellar masses $> 10^{11}$~M$_{\odot}$.}
\label{tab:smopt}
\end{deluxetable}

\subsubsection{Massive Galaxies}

From the survey results of \citet{vandokkum04}, after converting to a
common IMF, the mass density contributed by galaxies with stellar
masses $> 10^{11}$~M$_{\odot}$ at $2.0<z<3.0$ is
$\Omega_{\ast}(>10^{11}\, {\rm M}_{\odot}) = (1.64\pm 0.45)\times
10^{-4}$ (Table~\ref{tab:smopt}), where the uncertainty does not
include potentially large systematic errors in photometric redshifts
(e.g., R08, \citealt{shapley05}).  For this calculation, we have
assumed that the mass density does not evolve over redshifts
$2.0<z<3.0$, although it most likely does, and have assumed the
aforementioned value is valid at $z\sim 2.3$.  Adding this to the
contribution from UV-bright galaxies yields a mass density of
$\Omega_{\ast}(z\sim 2.3) = (5.36\pm 0.53)\times 10^{-4}$.  Do
UV-faint galaxies contain enough stellar mass to add appreciably to
this number?  We explore this question in the next section.

\subsubsection{Stellar Mass in UV-faint Galaxies}

How might the stellar mass distribution be expected to change for
UV-faint sub-$L^{\ast}$ galaxies?  \citet{shapley05} and
\citet{reddy06a} highlight the biases inherent in photometric redshift
estimates for star-forming galaxies at $z\sim 2$, perhaps even more so
for UV-faint galaxies.  We are at present targeting UV-faint galaxies
with deep spectroscopy to remedy this situation.  Such spectroscopy,
combined with deep multi-wavelength data in several of our survey
fields, should allow us to constrain the stellar populations and
masses of sub-$L^{\ast}$ to as much confidence as one can obtain with
such an analysis.  A full SED analysis of such galaxies is beyond the
scope of this paper, yet we can make some progress in determining the
stellar mass content of sub-$L^{\ast}$ galaxies based on observations
of UV-bright galaxies.

To do this, we exploited the log-linear relation between SFR and
stellar mass found at $z\sim 2.3$ from an analysis of deep HDF data by
\citet{sawicki07}: $\log (M_{\rm
stars}/M_{\odot})=9.0+0.86\log[SFR/(M_{\odot}yr^{-1})]$.  They present
evidence that this correlation remains valid for galaxies with SFRs of
$\approx 1$~M$_{\odot}$~yr$^{-1}$, corresponding to unobscured UV
magnitudes of $M_{\rm AB}(1700\AA)\sim -18.0$.  Several other
correlations between SFR and stellar mass have been published,
including most recently by \citet{reddy06a} and \citet{daddi07a}.
Adopting these latter relations results in a slightly larger
contribution of stellar mass from UV-faint galaxies.  Therefore, as a
conservative estimate, we have adopted the \citet{sawicki07}
correlation in the subsequent discussion.\footnote{The zeropoint of
the SFR-stellar mass relation appears to evolve with redshift
\citep{noeske07}.  For our analysis, we have assumed the correlation
found at $z\sim 2.3$.}  Although we do not observe a significant
correlation between unobscured UV luminosity and stellar mass for
UV-bright ($\rs \le 25.5$) galaxies, the log-linear behavior of SFR
with stellar mass implies that such a correlation must exist when
examined over a large dynamic range in unobscured UV luminosity.  In
particular, since UV-faint galaxies are likely to have lower reddening
than their brighter counterparts (\S~\ref{sec:nature}), the UV
luminosity for these galaxies is expected to track the bolometric
luminosity given the tight relation between SFR and reddening (e.g.,
\citealt{reddy06b}).  This, combined with the trend between SFR and
$M_{\ast}$, implies a correlation between UV luminosity and stellar
mass when examined over a large range in luminosity.  Using the
procedure outlined above, we recomputed the stellar mass density by
integrating the UV LF to $M_{\rm AB}(1700\AA) = -18.00$ and allowing
the stellar mass to adjust according to the empirical relation between
SFR and $M_{\ast}$.  The SFR is determined by combining the absolute
magnitude of galaxies with the luminosity-dependent reddening model.
The resulting stellar mass densities (Table~\ref{tab:smopt}) suggest
that roughly as much stellar mass is contained in UV-faint galaxies as
is contained in UV-bright ones, implying a relatively steep low mass
slope of the stellar mass function, a conclusion that appears to be a
generic result of most cosmological simulations \citep{nagamine04,
finlator07}.  Note that this computation includes only those galaxies
that are brighter than $M_{\rm AB}(1700\AA) = -18.00$, since it is
down to this limit that the correlation between SFR and $M_{\ast}$ has
been verified empirically.  Assuming the correlation is valid at
fainter magnitudes results in a $64\%$ larger stellar mass density
contribution from UV-faint galaxies when integrated to $M_{\rm
AB}(1700\AA) = -16.00$.  For the subsequent discussion, however, we
assume the numbers that result from integrating to the brighter limit.
More importantly, this discussion highlights the considerable leeway
in adjusting the stellar mass density estimates upwards even with
conservative assumptions of the stellar mass distribution for UV-faint
galaxies.

\subsubsection{Comparisons with the Integrated Star Formation History}

Can our revised estimate of the stellar mass density at $z\sim 2.3$
account for the star formation that has occurred until then?
Figure~\ref{fig:massfcn} shows a compilation of stellar mass density
estimates from the literature, along with our determinations and the
stellar mass density inferred by integrating the luminosity-dependent
reddening-corrected star formation history.  Recall that our
calculation of the stellar mass density includes galaxies brighter
than $M_{\rm AB}(1700\AA) = -18.0$ at $z=2.3$, corresponding to a
luminosity of $0.083L^{\ast}$, where $L^{\ast}$ is the unobscured
characteristic luminosity at $z=2.3$.  When integrating the star
formation history, we must keep track of how these galaxies evolved
with time.  We have already shown that $L^{\ast}$ evolves strongly
with redshift, such that $>0.083L^{\ast}_{z=2.3}$ galaxies at $z=2.3$
would have been fainter on average at higher z.  To account for this
fading with increasing redshift, we adopt a lower limit to the
integral of the UV LF that evolves in the same way as $L^{\ast}$.
Specifically, to find the star formation rate density, we integrate
the UV LF to a limit of $0.083L^{\ast}$ where, in the integral,
$L^{\ast}$ varies with redshift.  By doing this, we are in effect
keeping track of these galaxies' location on the UV LF as a function
of time.  Note that there may be individual galaxies that contribute
to the stellar mass density prior to $z=2.3$, but then fall below a
luminosity of $0.083L^{\ast}$ at some later epoch.  However, galaxies
will of course accumulate most of their stellar mass when they are
forming stars at higher rather than lower rate.  Further, virtually
all of the evolution of the UV LF at $z>2.3$ can be accommodated by a
brightening of $L^{\ast}$ with increasing cosmic time
(\S~\ref{sec:mstarevol}) and, therefore, it is unlikely that there are
large numbers of galaxies fading beyond a fixed fraction of
$L^{\ast}$.  We noted in \S~\ref{sec:lfevol} that there is little
evolution in the UV LF between $z\sim 4$ and $z\sim 2.3$, but
presumably most of the galaxies that are fading are destined to become
massive galaxies by $z\sim 2.3$, and recall that our estimate of the
stellar mass density at $z\sim 2.3$ {\em includes} that from massive
galaxies with $M_{\ast}>10^{11}$~M$_{\odot}$.  Given these reasons, it
is reasonable to assume that integrating to $0.083L^{\ast}(z)$ should
approximate the stellar mass accumulated by galaxies brighter than
this limit at all previous epochs.

With this premise, we show the integrated star formation history in
Figure~\ref{fig:massfcn}.  Accounting for the stellar mass content of
both UV-bright and faint galaxies results in a stellar mass density at
$z=2.3$ that agrees well with our inference from integrating the star
formation history.  In fact, it is perhaps remarkable that with a
simple calculation where we account for the fading of galaxies with
increasing cosmic time, we are able to resolve the integrated star
formation history with the global stellar mass density at the very
epoch where their supposed disparity reaches its greatest amplitude
(Figure~\ref{fig:sfrd}).  A careful analysis of the stellar mass
density contributed by galaxies over the bulk of the LF, combined with
an integration of the luminosity-dependent reddening-corrected star
formation history to an appropriate limit, may obviate the need to
invoke some other mechanism, such as an evolving IMF, to explain the
discrepancy.  Of course, with the present analysis we cannot rule out
that there may be {\em some} redshift evolution of the IMF, and there
are theoretical arguments as to why this may be the case
\citep{larson98, larson05}.  Indeed, such an evolution may plausibly
explain the shift in zeropoint of the trend between SFR and stellar
mass as a function of redshift \citep{dave08}.  All we have shown here
is that there is a simpler explanation for the discrepancy between the
integrated star formation history and stellar mass measurements at
$z\sim 2$, namely that the former must take into account an evolving
dust correction and the latter are likely to be incomplete for
galaxies with low stellar masses.  In point of fact, incompleteness of
stellar mass density measurements and an evolving dust correction are
physically well-motivated by observations of high redshift galaxies,
as we have shown here and elsewhere (\citealt{reddy06a}, R08), whereas
IMF evolution has yet to be verified observationally.  The results of
the last few sections highlight the subtleties that without proper
accounting may lead to the types of discrepancies reported in the
past.  Our findings favor a more nuanced view of the purported
discrepancy between the integral of the star formation history and
stellar mass density measurements.

Note that we have not measured directly the stellar mass function at
$z\sim 2$, but have inferred it by combining our knowledge of the UV
LF (which gives the number density of galaxies) with stellar masses
determined from broadband SED fitting.  Our analysis suggests that an
appreciable fraction of stellar mass is hosted by sub-$L^{\ast}$
galaxies and that the steepness of the slope of stellar mass function
may have been underestimated in previous studies based on near-IR
data.  The robustness of our conclusions should be verified by
significantly deeper rest-frame near-IR observations that constrain
the low mass end of the stellar mass function.  A more direct sampling
of the stellar masses of UV-faint galaxies is required.

\subsection{Concluding Remarks}
\label{sec:cautions}

We conclude this section with a few cautionary remarks.  As stated
previously, we have adopted the $z\ga 4$ measurements of the UV LF
that are based on a maximum-likelihood analysis that is most analogous
to the method we have used, and that are based on data that extend to
comparable depths as achieved here, albeit over an area an order of
magnitude smaller, in order to make the most consistent comparison
between star formation rate density estimates.  Obvious effects that
can contribute to both random and systematic error in the star
formation history include cosmic variance and the limit to which the
UV LF is measured.  Large-scale multi-field surveys at $z\ga 4$
analogous to the present survey at $z\sim 2-3$ will provide better
constraints on the random errors associated with cosmic variance.
Further, it may be of interest to determine if the similarity in the
UV LF from the HDF studies versus the universal one measured at $z\sim
2-3$ (\S~\ref{sec:hdf}) extends to higher redshifts.

Another point of consideration is the expected turnover in the
faint-end slope at very faint luminosities.  This turnover is likely
dictated by the threshold of cold gas surface density in halos
required to trigger star formation \citep{schmidt59, kennicutt98}.  In
the context of the present analysis, the magnitude of the systematic
effect that results from integrating the LF to zero luminosity will
depend on the steepness of the faint-end slope.  For a fixed
$\phi^{\ast}$, the difference between integrating the LF to
$0.04L^{\ast}$ versus zero luminosity is a factor of 1.85 for
$\alpha=-1.73$, 1.44 for $\alpha=-1.6$, and 1.19 for $\alpha=-1.4$.
However, at present there are no empirical constraints on the turnover
of the faint-end slope of the LF.  Even locally, where surveys can
probe to luminosities significantly fainter than $L^{\ast}$, there is
no evidence for a fall-off in number density of dwarf galaxies.  The
local $u$-band LF from the SDSS, for instance, appears to abide by a
log-linear relationship between number density and magnitude down to
$\approx 0.02L^{\ast}$ \citep{baldry05} and, in fact, surveys of the
local group of star-forming dwarf galaxies suggest an increasing slope
to $0.0001L^{\ast}$ (\citealt{mateo98}).  Of course, there is no
reason why the local rest-frame optical LFs should have the same slope
as the rest-frame UV LF, particularly if the LF represents a sequence
in mass-to-light ratio.  Recall that we have adopted a zero luminosity
limit for consistency with the SFRD compilation of \citet{wilkins08}
and \citet{hopkins04}.  Given the steep faint-end slopes found at
$z\ga 2$ and the accommodation of a significant fraction of the
luminosity density by faint galaxies, we should bear in mind the
possible systematic effects of integrating to zero luminosity, both in
terms of the unobscured UV luminosity density and the average dust
corrections in the case of luminosity-dependent reddening.  A further
systematic in the star formation history may be caused by a
redshift-dependency in the turnover of the faint-end slope of the UV
LF.  We note, however, that these systematics will not affect our
comparison of the integrated star formation history and stellar mass
density measurements at $z\sim 2$ given that we restricted our
analysis to luminosities ($>0.083L^{\ast}$) where we do have empirical
constraints on the LF.

It is also worthwhile to mention that the \citet{kennicutt98} relation
for converting UV luminosity to star formation rate is valid only for
a stellar population age that is $\ga 100$~Myr, since it is after this
time that the mix of O and B stars stabilizes assuming a constant star
formation history.  For galaxies much younger than this, the UV
luminosity will underpredict the SFR based on this relation.  Hence,
discerning trends in the stellar population age as a function of
unobscured UV luminosity is a necessary step in computing accurately
the star formation rate density, particularly since the UV luminosity
density appears to be dominated by UV-faint galaxies, if such UV-faint
galaxies are systematically younger than their brighter counterparts.

Finally, we note that because the stellar mass density is an
integrated quantity, we cannot add arbitrarily large amounts of
stellar mass at high redshift without violating the local constraints,
assuming the latter are complete in stellar mass.  Given that the
local measurements can be systematically uncertain by up to $30\%$
\citep{bell03}, and the additional contribution from faint galaxies to
$M_{\rm AB}(1700\AA)=-18.0$ at $z\ga 2$ is only $\approx 9\%$ of the
local value, then our finding of significant stellar mass in UV-faint
galaxies at $z>2$ does not pose a problem in terms of the budget of
stellar mass in the local universe.  In practice, our preliminary
estimates of the number density of galaxies with low stellar masses
($M_{\ast}\la 10^{10}$~M$_{\odot}$) may be used to constrain
cosmological models that currently predict low-mass slopes of the
stellar mass function at $z\ga 3$ that are comparably steep as the
slope of the halo mass function \citep{nagamine08}.  Ultimately, this
issue may be resolved through detailed clustering analysis of
sub-$L^{\ast}$ galaxies and inferences as to their local descendants.
Alternatively, if LAEs represent a short phase in the lifetimes of
UV-faint galaxies but are otherwise unremarkable sub-$L^{\ast}$
galaxies, then the clustering of LAEs may provide clues to the
descendants of UV-faint galaxies (e.g., \citealt{kovac07, gawiser07}).

\section{Discussion: Evolution of the Faint-End Slope}
\label{sec:alpha}

So far, the discussion has focused on what the UV LFs can tell us
about the star formation rate density and buildup of stellar mass.
The modulation of the LF with respect to the underlying halo mass
distribution also yields important information regarding the processes
that regulate star formation, such as supernovae-driven or radiative
winds, and energy injection from AGN, mechanisms generically referred
to as ``feedback.''  For example, the sharp cutoff at the bright-end
of the UV LF may be partly attributable to AGN feedback suppressing
star formation in high mass halos \citep{croton06, scannapieco05,
granato04}, even after taking into account the saturation of UV light
with respect to the total star formation rates of galaxies with large
SFRs \citep{adel00, reddy06a}.  Similarly, the shallowness of the
faint-end slope of the LF relative to that of the halo mass function
suggests some regulating mechanism associated with star formation
itself, such as through reionization \citep{kravtsov04, gnedin00},
supernovae winds, or radiatively-driven winds \citep{martin99,
springel03}.

One important conclusion from our analysis is that the faint-end slope
of the UV LF is relatively constant and steep between $z\sim 2$ and
the highest redshifts where $\alpha$ can be measured well, around
$z\sim 6$ (Figure~\ref{fig:alpha}).  At the same time, $L^{\ast}$
evolves strongly between these redshifts (\S~\ref{sec:mstarevol}).
Because the average galaxy is brightening, the invariance of $\alpha$
over these redshifts is not likely reflective of an equilibrium
between fading and brightening galaxies.  Rather, whatever
sub-$L^{\ast}$ galaxies at $z\sim 6$ brighten to become $L^{\ast}$
galaxies by $z\sim 2$ are made up in number by halos in which gas has
newly condensed to form stars by $z\sim 2$.

The steep $\alpha \sim -1.7$ at $z\ga 2$ stands in contrast with the
shallower values of $\alpha \sim -1.1$ measured locally
\citep{wyder05, budavari05}.  The redshift evolution of $\alpha(z)$ is
summarized in Figure~\ref{fig:alpha} and suggests that most of the
change in $\alpha$ occurs mainly below $z\sim 2$.  What may be the
cause of this change?  It has been suggested recently that the
evolution in $\alpha(z)$, such that $\alpha$ is shallower at lower
redshifts, may reflect the delayed onset of feedback from Type Ia SN
\citep{khochfar07}.  However, even if such feedback is energetically
important, it is unclear whether it would have any perceivable effect
on $\alpha(z)$ given that the faint-end population evolves strongly
between the redshifts in question.  A more likely explanation is that
the evolution in $\alpha(z)$ is dictated simply by the availability of
low mass halos with cold gas at redshifts $z\la 2$.  Perhaps it is not
surprising that the apparent shift from steep to shallow faint-end
slopes occurs at an epoch ($z\sim 2$) that is marked by a confluence
of other important transitions, including the reversal in the
evolution of the cosmic star formation density.

\section{Conclusions}
\label{sec:conclusions}

We have used the spectroscopic redshifts and photometric data in all
of the fields of the Lyman Break Galaxy (LBG) survey to make the most
robust determination of the UV luminosity functions (LFs) at $1.9\le
z<2.7$ and $2.7\le z<3.4$.  Our sample includes over $2000$
spectroscopic redshifts, and $\approx 31000$ LBGs spread across $31$
spatially-independent fields over a total area of $3261$~arcmin$^{2}$.
The depth of these data allow us to select LBGs to $0.07L^{\ast}$ and
$0.1L^{\ast}$ at redshift $z\sim 2$ and $3$, respectively.  The LFs
are constrained using a maximum-likelihood procedure that includes the
effects of photometric errors, contaminants, and
perturbation of galaxy colors due to Ly$\alpha$.  The principle
conclusions of this work are as follows:

1. We have quantified the effects of a luminosity dependent reddening
and Ly$\alpha$ equivalent width ($W_{\rm Ly\alpha}$) distribution on
the incompleteness corrections to our sample.  Allowing for a larger
fraction of galaxies with large $W_{\rm Ly\alpha}$ among UV-faint
galaxies results in a $3-4\%$ increase in the faint-end number
densities relative to those obtained by assuming a
luminosity-invariant $W_{\rm Ly\alpha}$ distribution as constrained
from our spectroscopic sample.  Similarly, adopting a
luminosity-dependent reddening distribution where the mean reddening
of galaxies decreases to fainter UV magnitudes results in an up to
$10\%$ increase in the inferred number density of UV-faint galaxies.
While these differences in the number density are not negligible,
accounting for these luminosity-dependent systematics does little to
alter the Schechter parameters, in particular the faint-end slope
($\alpha$), and it suggests that the UV-color criteria are robust to
such systematics and that our derived LF must be reasonably complete
for UV-faint galaxies.  Adopting reasonable assumptions for the
luminosity dependence of $W_{\rm Ly\alpha}$ and reddening, we derive
faint-end slopes of $\alpha(z=2)=-1.73\pm0.07$ and
$\alpha(z=3)=-1.73\pm0.13$.

2. A comparison indicates that our determination yields sub-$L^{\ast}$
number densities that are significantly larger, and faint-end slopes
that are somewhat steeper, than those published previously.  We
believe our results are robust given (a) the large number of
spectroscopic redshifts used to constrain the bright-end of the UV LF,
(b) photometry over a large area spread across many
spatially-independent fields to mitigate cosmic variance, and (c) a
careful analysis of the systematics (Ly$\alpha$ line perturbations and
luminosity-dependent reddening) that are important for computing the
faint-end slope.  Our analysis suggests that LFs based on HDF-N data
alone are not biased significantly from the universal value of the LF
determined here, at least at redshifts $z\sim 2-3$.

3. There is very little evolution in the UV LF in the redshift range
$1.9\le z<3.4$.  However, examined over a larger baseline in redshift
and using the published results at $z\sim 6$, we find a brightening of
the characteristic unobscured UV magnitude of $\sim 1.2$~mag between
$z\sim 6$ and $z\sim 2$.  The faint-end slope remains relatively
constant and steep between these redshifts, with a value of
$\alpha\sim -1.7$ to $-1.8$.

4. To examine the frequency of atypical galaxies on the faint-end of
the UV LF, we compared the number density of sub-$L^{\ast}$ galaxies
to those of dusty ultraluminous infrared galaxies and galaxies with
large stellar masses $>10^{11}$~M$_{\odot}$.  With conservative
assumptions regarding their UV-magnitude distributions, we find that
galaxies with large stellar masses and bolometrically-luminous
galaxies comprise $\la 2\%$ of the total space density of galaxies
fainter than $\rs = 25.5$.  This small fraction underscores not only
the rarity of these objects, but also the large number of UV-faint
galaxies implied by the steep faint-end slope.

5. Integrating the UV LFs at $z\sim 2-3$ to zero implies that $93\%$
of the unobscured luminosity density resides in galaxies fainter than
$L^{\ast}$.  Adopting our prescription for the luminosity-dependence
of reddening, we construct bolometric luminosity functions to estimate
that $>70\%$ of the bolometric luminosity density arises from galaxies
fainter than the characteristic bolometric luminosity at these
redshifts.  The luminosity-dependent reddening model combined with a
steep $\alpha$ imply that the average dust corrections needed to
recover the bolometric luminosity density from the unobscured UV
luminosity density will depend sensitively on the limit of integration
used to compute the luminosity density.  Of course, these corrections
will depend also on whether they include only the reddening
corrections for galaxies routinely selected by their rest-UV colors or
if they also include corrections for galaxies that escape UV selection
altogether.

6. Assuming a constant reddening correction of $4.5$ to the
UV-determined star formation history results in a factor of two
overestimate of star formation rates and stellar mass densities
accumulated at $z\sim 2-3$ relative to the values obtained by assuming
a luminosity-dependent reddening correction to the star formation
history.  Integrating the latter indicates that at least $25\%$ of the
present-day stellar mass density was formed in sub-ultraluminous
galaxies between redshifts $z=3.4$ and $z=1.9$.

7. The luminosity-dependent reddening-corrected star formation history
points to a factor of $8-9$ increase in the star formation rate
density (integrated to zero luminosity) between $z\sim 6$ and $z\sim
2$, significantly steeper than the factor of $4$ that we would have
inferred in the case of constant dust reddening.  The evolution in the
bolometric star formation rate density is driven equally by an
evolution in the unobscured characteristic luminosity and an evolving
(luminosity-dependent) dust correction.

8. We have examined the offset between the integral of the star
formation history and previously published determinations of the
stellar mass densities at $z\sim 2$, the epoch where this discrepancy
appears to peak in amplitude and where our data are most sensitive.
Given the steep faint-end slopes observed at $z\sim 2$, we have
explored whether UV-faint galaxies could plausibly account for the
observed differences.  By summing the stellar mass from all galaxies
brighter than $0.083L^{\ast}_{z=2}$, we find a stellar mass density
that is in remarkable agreement with the luminosity-dependent
reddening-corrected star formation history when the latter is
integrated to the same $0.083L^{\ast}$ limit that accounts for the
fading of galaxies with increasing cosmic time.  This exercise
highlights the importance of UV-faint galaxies in the total budget of
stellar mass, and suggests that computing the integral of the star
formation history in a way that reflects how galaxies evolve may
obviate the need to invoke other mechanisms (e.g., an evolution of the
IMF) to reconcile the integrated star formation history and the global
stellar mass density at $z\sim 2$.

9.  Finally, while the faint-end slope at any given redshift is likely
to be regulated by feedback, discerning the signatures of delayed
feedback (e.g., from Type Ia SN) in the redshift evolution of $\alpha$
is not trivial, particularly given the strong evolution of the UV LF
at $z\ga 2$.  Our results suggest that $\alpha$ is roughly constant at
$z\ga 2$, contrasting with the shallower values found locally.  This
evolution may be dictated simply by the availability of low mass halos
capable of supporting star formation at $z\la 2$.

\acknowledgements

This work benefited from discussions with Arjun Dey and Max Pettini.
We thank Max Pettini for useful suggestions regarding the organization
of the paper.  N. A. R. thanks Romeel Dav\'{e} for sending the data
included in Figure~\ref{fig:sfrd}, and Pieter van Dokkum for an
electronic version of the stellar mass data from \citet{vandokkum06}.
We thank the staff of the Keck Observatory for their help in obtaining
the data presented here.  Support for N. A. R. was provided by NASA
through Hubble Fellowship grant HST-HF-01223.01 awarded by the Space
Telescope Science Institute, which is operated by the Association of
Universities for Research in Astronomy, Inc., for NASA, under contract
NAS 5-26555.  Additional support has been provided by research funding
for the {\em Spitzer} Space Telescope Legacy Science Program, provided
by NASA through contracts 1224666 and 1287778, issued by the Jet
Propulsion Laboratory, California Institute of Technology.
C. C. S. has been supported by grants AST 03-07263 and AST 06-06912
from the National Science Foundation and by the David and Lucile
Packard Foundation.



\appendix

\section{Testing for Systematic Effects}

As discussed in \S~\ref{sec:maxlik}, the transitional probability
function $\xi$ (Eq.~\ref{eq:xi}) is sensitive to changes in the
intrinsic properties of galaxies.  Our goal is to determine if the
resulting modulation of $\xi$ is significant enough to induce a
noticeable difference in the maximum-likelihood LF.  Since we are
interested primarily in computing the UV luminosity distribution of
galaxies at these redshifts, we will concentrate on how the reddening,
$N(E[B-V])$, and Ly$\alpha$ equivalent width, $N(W_{\rm Ly\alpha})$,
distributions might change with luminosity.  More generally, the
distributions may also be a function of redshift, but for this
analysis we ignore such redshift evolution since there is little
published evidence for it over the redshifts considered here (see
below).  We conclude by discussing the fraction of stellar objects and
galaxies outside the redshift ranges of interest.

The most direct approach for testing systematic changes in $N(E[B-V])$
and $N(W_{\rm Ly\alpha})$ is to examine the LBG color distribution.
However, it is difficult from an analysis of the candidates' colors
alone to separate the selection effects imposed by the color criteria
from those induced by other systematics, such as real changes in the
reddening and/or $W_{\rm Ly\alpha}$ distributions.  A different
approach takes advantage of the fact that the transitional probability
function $\xi$ encapsulates all of the information regarding the
selection biases imposed by the color criteria.  Therefore, rather
than examine directly the color distribution of BXs and LBGs, we chose
to make various assumptions of how the reddening and $W_{\rm
Ly\alpha}$ distributions vary as a function of magnitude.  Then, based
on these assumptions, we recalculated $\xi$ using Monte Carlo
simulations and repeated the maximum-likelihood procedure to find the
best-fit LF.

\subsection{A. $W_{\rm Ly\alpha}$ Distribution}

\begin{figure}[t]
\epsscale{0.5}\plotone{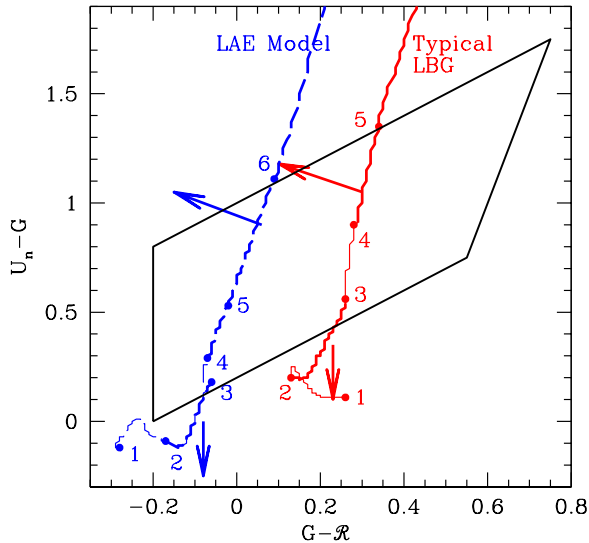}
\caption{Color tracks for a typical LBG with constant star formation
for $100$~Myr and $\ebmv = 0.15$ ({\em solid line}) and our model for
Ly$\alpha$ emitters (LAEs) with constant star formation for $50$~Myr
and $\ebmv = 0$ (no reddening; {\em dashed line}).  The attenuation of
colors due to the IGM has been accounted for following
\citet{madau95}.  The labels along each track indicate particular
redshifts as follows: (1) $z=1.00$, (2) $z=1.68$, (3) $z=2.17$, (4)
$z=2.48$, (5) $z=2.65$, and (6) $z=2.91$.  The Ly$\alpha$ line falls
in the $U_{\rm n}$ band at $1.68\le z<2.17$ (between points 2 and 3)
and in the $G$-band at $z>2.48$ (above point 4), as indicated by the
thicker lines.  The effect of {\em adding} Ly$\alpha$ emission to the
spectrum is shown by the arrows, tending to scatter galaxies out of
the BX selection window ({\em trapezoid}).}
\label{fig:lyaprob}
\end{figure}

There is an increasing body of work that indicates that Ly$\alpha$
emitters (LAEs), those objects selected by narrowband techniques,
exhibit significantly larger rest-frame $W_{\rm Ly\alpha}$, but are
much fainter in the continuum, on average, than traditional
color-selected galaxies that are restricted to $\rs\la 25.5$.  Most of
these LAEs will lie on the faint-end of the UV LF.  Adding Ly$\alpha$
emission to a star-forming galaxy's spectrum will tend to scatter such
a galaxy out of the BX selection window
(Figure~\ref{fig:lyaprob}).\footnote{For galaxies at $z>2.48$, where
  Ly$\alpha$ lies in the $G$-band, the presence of emission will
  scatter them out of the BX window and into the LBG window (see also
  Figure~4 of R08).}  Therefore, for a fixed observed number of faint
BX candidates, the incompleteness corrections will be {\em larger} if
they have a distribution skewed towards high $W_{\rm Ly\alpha}$, thus
increasing the inferred number density of faint galaxies.  The degree
to which the faint-end slope $\alpha$ changes will depend on the
$W_{\rm Ly\alpha}$ distribution, the stellar population distribution,
and the number density of high $W_{\rm Ly\alpha}$ systems.  For
example, there may be little difference in the faint-end slope derived
assuming a high $W_{\rm Ly\alpha}$ (LAE) population relative to that
derived assuming a constant $W_{\rm Ly\alpha}$ distribution
(\S~\ref{sec:results}) if LAEs constitute a small and constant
percentage of galaxies as a function of magnitude on the faint-end of
the UV LF.  We will now consider in detail how a luminosity-dependent
$W_{\rm Ly\alpha}$ distribution affects our analysis.

\subsubsection{A.1. Assumptions}

Here we quantify the effects of a luminosity dependent $W_{\rm
  Ly\alpha}$ distribution by making the following assumptions.  First,
for ease of discussion, we assume that an ``LAE'' is any star-forming
galaxy with rest-frame $W_{\rm Ly\alpha}\ge 50$~\AA, corresponding
roughly to the observational lower limits of typical LAE surveys
(e.g., \citealt{ouchi08}) and upper $\la 10\%$ of continuum-selected
galaxies to $\rs=25.5$ (e.g., \citealt{steidel03, shapley03}, R08).
In a strict sense, an LAE is any galaxy with $W_{\rm Ly\alpha}>0$, but
here we limit ourselves to those that are easily identified using
narrowband techniques, to distinguish them from emission line galaxies
that are routinely identified from spectroscopy of continuum-selected
galaxies.  The adoption of the $W_{\rm Ly\alpha}=50$~\AA\, cutoff is
for reference purposes only, and does not affect the subsequent
analysis since a separate assumption is made regarding the median
value of $W_{\rm Ly\alpha}$ for LAEs.  In particular, the simulations
are performed using different values of $W_{\rm Ly\alpha}$ ranging
from $50$~\AA\, to $250$~\AA\, (rest-frame), the latter being a
canonical upper limit for standard assumptions of the IMF
\citep{salpeter55} and solar metallicity \citep{charlot93}.

Second, we adopt an average stellar population consistent with the
most recent analyses of LAEs at high redshift.  The range of ages
found for LAEs is $\sim 10$~Myr at the low end to $\sim 1$~Gyr at the
high end, with typical ages of $\sim 100-200$~Myr, and low metallicity
and reddening (e.g., \citealt{gronwall07, ouchi08}).  We
conservatively assume an average LAE age of $50$~Myr, metallicity of
$1/20$~Z$_{\odot}$, and zero dust reddening for the purposes of our
simulation.  For illustrative purposes, Figure~\ref{fig:lyaprob} shows
the colors as a function of redshift for such a stellar population
compared to a model that represents the typical LBG with age
$>100$~Myr and $\ebmv \sim 0.15$ (assuming the BC06 model).  As
discussed above, irrespective of whether Ly$\alpha$ lies in the
$U_{\rm n}$ or $G$ band, the effect of adding emission is to scatter
galaxies at $1.9\le z<2.7$ out of the BX selection window.

Third, we must determine the frequency of LAEs among the general
star-forming population as a function of continuum magnitude.
\citet{ouchi08} determine the UV LF of LAEs with $W_{\rm Ly\alpha}\ga
50$~\AA\, at $z\approx 3.1$ and find it to be fit adequately with a
Schechter form of the LF with $\phi^{\ast} = (5.6^{+6.7}_{-3.1})\times
10^{-4}$~Mpc$^{-3}$ and $M^{\ast}_{\rm 1500\AA} = -19.8\pm 0.4$ with a
fixed faint-end slope of $\alpha_{\rm LAE} = -1.6$.  Integrating the
continuum-UV LF of LAEs and comparing with the UV LF determined above
for all star-forming galaxies implies an LAE fraction ($\epsilon$)
that is a strong function of magnitude, ranging from less than
$0.02\%$ in the brightest magnitude bin ($-22.83\le
M(1700\AA)<-22.33$) to $\approx 9.3\%$ in the faintest bin ($-18.33\le
M(1700\AA)<-17.83$).  Note that the fraction of LAEs on the faint-end
is based on an extrapolation of the UV LF of LAEs assuming
$\alpha=-1.6$ \citep{ouchi08}.  Further, the fractions will go up (or
down) depending on whether we decrease (or increase) the limiting
$W_{\rm Ly\alpha}$ that segregates LAEs from other galaxies (e.g.,
using a limit of $W_{\rm Ly\alpha}=20$~\AA\, instead of $50$~\AA).  In
our simulations, we assume (1) no evolution in the UV LF of LAEs
between $z\sim 2-3$ and (2) a fraction of LAEs that varies with
absolute magnitude in accordance with our findings above, with a
fraction of $10\%$ for $W_{\rm Ly\alpha}>50$~\AA\, in the faintest bin
considered here.

To recap, the main assumptions going forward are that LAEs are
described by a $50$~Myr stellar population with no reddening and
comprise anywhere from $<0.02\%$ to $10\%$, respectively, of galaxies
within the bright and faint magnitude bins of our analysis.  In the
next section, we consider how different values of $W_{\rm Ly\alpha}$
among LAEs affects the faint-end of the UV LF of all star-forming
galaxies.

\subsubsection{A.2. Effect of a Changing $W_{\rm Ly\alpha}$ Distribution}

\begin{figure*}[t]
\plotone{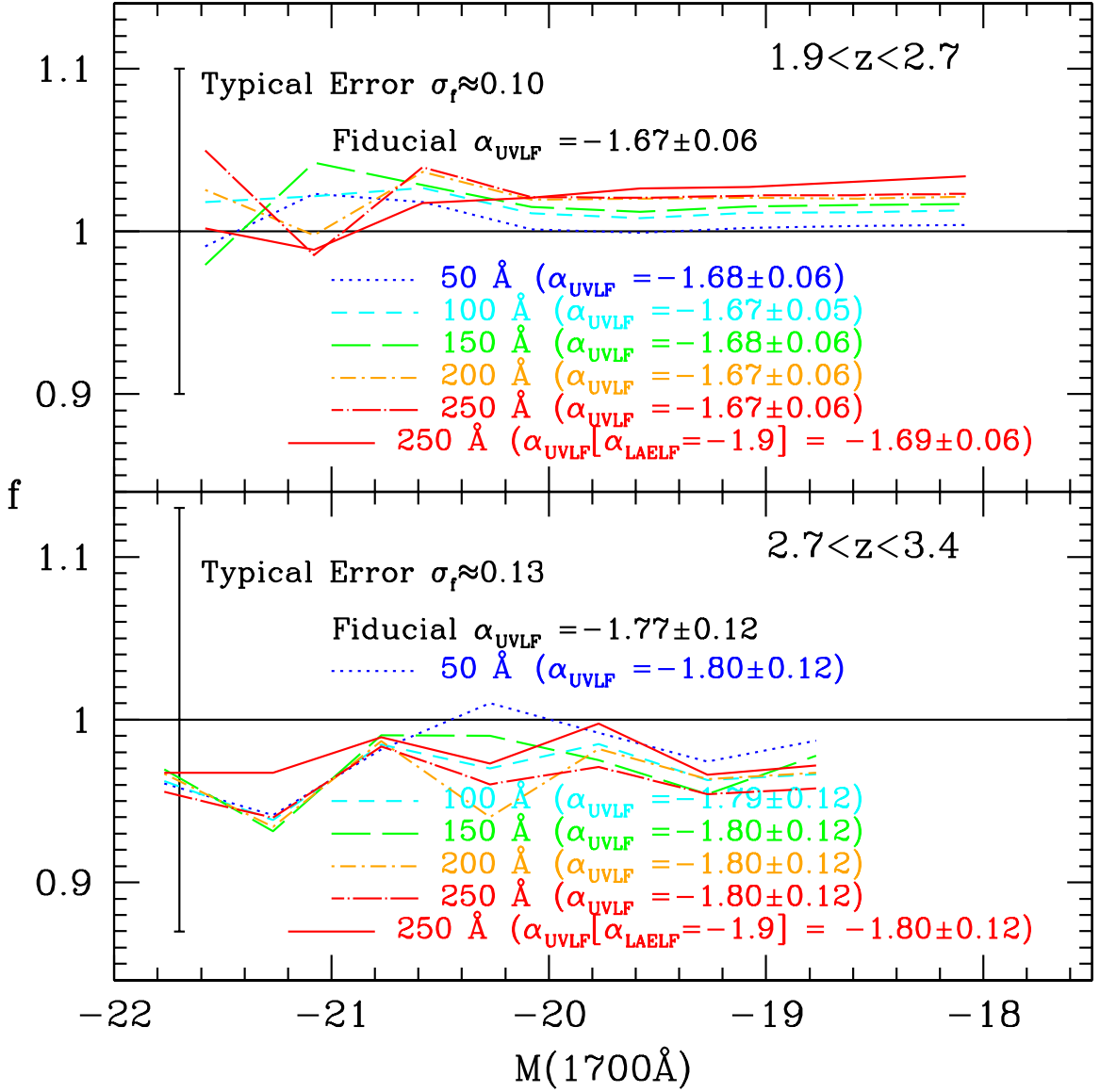}
\caption{Comparison of maximum-likelihood number density of galaxies
as a function of magnitude for (1) our fiducial UV LFs assuming that
{\em all} galaxies abide by the $W_{\rm Ly\alpha}$ distribution seen
for bright ($\rs\le 25.5$) continuum-selected galaxies
(Figure~\ref{fig:lyaew}) and (2) the UV LFs derived assuming a
population of Ly$\alpha$ emitters (LAEs) with high $W_{\rm Ly\alpha}$
at faint magnitudes.  The different lines show the ratio of the LFs
determined from (2) to that determined from (1), and correspond to
differing amounts of Ly$\alpha$ emission attributed to the LAE
population.  The typical error in this ratio ($\sigma_{f}$) is shown
by the vertical errorbar.  In all cases, we find that such a
population of high $W_{\rm Ly\alpha}$ systems does little to alter the
faint-end slope of the UV LF.}
\label{fig:lyalf}
\end{figure*}

With the aforementioned premises, the UV LF is computed for varying
amounts of emission among the LAEs, with equivalent widths from
$50$~\AA\, to $250$~\AA.  In our simple model, the luminosity
dependent $W_{\rm Ly\alpha}$ distribution can be expressed as:
\begin{eqnarray}
N(W_{Ly\alpha},\rs) = [1-\epsilon(\rs)]N_{\rm o}(W_{\rm Ly\alpha}) + \epsilon(\rs)\delta(\omega),
\end{eqnarray}
where $N_{\rm o}(W_{\rm Ly\alpha})$ is the distribution for $\rs \le
25.5$ galaxies (Figure~\ref{fig:lyaew}), $\epsilon(\rs)$ is the
$W_{\rm Ly\alpha}>50$~\AA\, fraction as a function of magnitude as
determined above, and $\delta(\omega)$ is a delta function with center
at $\omega =$ 50, 100, 150, 200, and 250 \AA.\footnote{The intrinsic
distribution for UV-bright galaxies (Figure~\ref{fig:lyaew}) also
includes a small fraction of continuum-selected galaxies with $W_{\rm
Ly\alpha}\ge 50$~\AA.  We ignore this small overlap between the
continuum and LAE $W_{\rm Ly\alpha}$ distributions, with the obvious
consequence of slightly increasing the total fraction of galaxies with
$W_{\rm Ly\alpha}\ge 50$~\AA.}  The results at $z\sim 2$ and $z\sim 3$
(Figure~\ref{fig:lyalf}) are presented in terms of the ratio of the
number density ($\eta^{\ast}$) for different values of $W_{\rm
Ly\alpha}$ for LAEs to the number density ($\eta$) derived using the
fiducial $W_{\rm Ly\alpha}$ distribution for $\rs\le 25.5$
continuum-selected galaxies (e.g., Figure~\ref{fig:lyaew}),
\begin{eqnarray}
f={\eta^\ast \over \eta},
\label{eq:lyarat}
\end{eqnarray}
with error given by
\begin{eqnarray}
\sigma_{f} = {[\eta^{2}\sigma_{\eta^{\ast}}^{2} + (\eta^{\ast})^{2}\sigma_{\eta}^2]^{1/2}
\over \eta^{2}}.
\label{eq:lyaerr}
\end{eqnarray}
The error in number density $\sigma_{\eta^{\ast}}$ is determined in
exactly the same manner as the error in the LF ($\sigma_{\eta}$;
\S~\ref{sec:results}).  Even with the most conservative assumption for
the $W_{\rm Ly\alpha}$ attributed to LAEs (an assumed value of $W_{\rm
Ly\alpha}=250$~\AA), we find that inclusion of such a population
alters the faint-end number densities at a $2-4\%$ level depending on
the redshift.  Not surprisingly, at lower redshifts, $1.9\le z<2.7$,
the inclusion of LAEs increases the inferred number densities at the
faint-end by $\la 3\%$, since such galaxies would be preferentially
scattered out of the BX selection window.  The opposite is found at
higher redshifts ($2.7\le z<3.4$), where the number density is
systematically lower by up to $4\%$, primarily because there are more
LAEs scattered from $z<2.7$ into the LBG selection window than are (at
any redshift $2.7\le z<3.4$) scattered out of the window, assuming no
evolution of the UV LF of LAEs at the redshifts of concern.  Hence,
for a fixed number of $z\sim 3$ LBG candidates, the tendency would be
to over-estimate the number density had we not accounted for the LAE
population.

The expectation of a steepening $\alpha$ for the universal UV LF
(i.e., for all star-forming galaxies) at $z\sim 2$ when including LAEs
is not borne out with our simulations for several reasons.  First, the
overall fraction of LAEs, even at the faint-end of the UV LF, is small
($\la 10\%$).  Their effect on the LF is further diminished because
they will only affect the broadband colors if they are at particular
redshifts (Figure~\ref{fig:lyaprob}).  Second, the current
determination of the UV LF of LAEs assumed a fixed $\alpha_{\rm
  LAE}=-1.6$ \citep{ouchi08}, which is not too different from the
fiducial $\alpha$ (\S~\ref{sec:results}).  Thus, while the fraction of
LAEs among the general population is a strong function of magnitude
between our brightest and faintest magnitude bin, it is in fact a
relatively constant $\approx 8-9\%$ for bins fainter than $M(1700\AA)
\approx -20$.  Adopting a steeper faint-end slope of the UV LF of LAEs
of $\alpha_{\rm LAELF} = -1.9$ still results in a universal faint-end
slope ($\alpha_{\rm UVLF}$) that is consistent with the fiducial value
(Figure~\ref{fig:lyalf}).

Future studies that constrain more robustly $\alpha_{\rm LAELF}$ over
the {\em entire} redshift range probed here will be useful for
assessing the overall impact of high $W_{\rm Ly\alpha}$ systems on the
faint-end slope inferred for all star-forming galaxies.  It is also
not unreasonable to suspect disparate absorption properties between UV
bright versus faint galaxies, so {\em spectroscopic} studies, while
difficult to carry out at present, are crucial for assessing the
variation of $W_{\rm Ly\alpha}$ with luminosity (e.g.,
\citealt{shapley03}).

Nonetheless, while the inclusion of high $W_{\rm Ly\alpha}$ systems
among the UV continuum-faint population may be a small systematic
effect ($3-4\%$), it is not negligible compared to the error in number
density at the faint-end of the UV LF ($10-15\%$), and so should be
included in any proper assessment of the UV LF.  The critical point,
and one that is demonstrated unambiguously with our simulations
(Figure~\ref{fig:lyalf}), is that such a systematic effect does little
to alter the faint-end slope of the universal LF.\footnote{While the
small change in the $W_{\rm Ly\alpha}$ distribution brought on by the
inclusion of LAEs does little to alter the faint-end slope,
significant discrepancies in the faint-end of the UV LF arise when not
accounting at all for the $W_{\rm Ly\alpha}$ distribution of galaxies
at these redshifts (R08).}  In our final determination of the LF, we
have made the conservative assumption of a median value of $W_{\rm
Ly\alpha} = 150$~\AA\, for the LAE population with $W_{\rm Ly\alpha} >
50$~\AA\, (although, as noted above, the exact value does little to
alter the LF).  The resulting best-fit Schechter parameters are listed
in Table~\ref{tab:schechter}.

\begin{deluxetable*}{lcccc}
\tabletypesize{\footnotesize}
\tablewidth{0pc}
\tablecaption{Best-fit Schechter Parameters for UV LFs of $1.9\la z\la 3.4$ Galaxies}
\tablehead{
\colhead{Redshift Range} &
\colhead{Systematic Effect} &
\colhead{$\alpha$} &
\colhead{M$^{\ast}_{\rm AB}(1700{\rm \AA})$} &
\colhead{$\phi^{\ast}$ ($\times 10^{-3}$~Mpc$^{-3}$)}}
\startdata
$1.9\le z<2.7$ & Fiducial\tablenotemark{a} & $-1.67\pm0.06$ & $-20.65\pm0.08$ & $2.84\pm0.42$ \\
& $W_{\rm Ly\alpha}=150$~\AA\tablenotemark{b} & $-1.68\pm0.06$ & $-20.68\pm0.08$ & $2.83\pm0.42$ \\
& $[\ebmv]_{\rm grad}$\tablenotemark{c} & $-1.72\pm0.06$ & $-20.71\pm0.09$ & $2.64\pm0.46)$ \\
& {\bf Combined}\tablenotemark{d} & $-1.73\pm0.07$ & $-20.70\pm0.11$ & $2.75\pm0.54$ \\
\\
$2.7\le z<3.4$ & Fiducial\tablenotemark{a} & $-1.77\pm0.12$ & $-20.98\pm0.13$ & $1.54\pm0.43$ \\
& $W_{\rm Ly\alpha}=150$~\AA\tablenotemark{b} & $-1.80\pm0.12$ & $-21.06\pm0.13$ & $1.34\pm0.42$ \\
& $[\ebmv]_{\rm grad}$\tablenotemark{c} & $-1.78\pm0.12$ & $-20.97\pm 0.13$ & $1.55\pm0.43$ \\
& {\bf Combined}\tablenotemark{d} & $-1.73\pm0.13$ & $-20.97\pm0.14$ & $1.71\pm0.53$ \\
\enddata
\label{tab:schechter}
\tablenotetext{a}{Fiducial LF assumes no change in the $W_{\rm Ly\alpha}$ and $\ebmv$ 
distributions as a function of UV apparent magnitude.}
\tablenotetext{b}{LF derived assuming an LAE population at the faint-end with 
$W_{\rm Ly\alpha}=150$~\AA\, (see text).}
\tablenotetext{c}{LF derived assuming a linearly declining mean $\ebmv$ for galaxies
with $\rs > 25.5$ (see text).}
\tablenotetext{d}{LF derived combining the effects of a changing $W_{\rm Ly\alpha}$ and $\ebmv$ 
distribution as a function of UV apparent magnitude.}
\end{deluxetable*}

\section{B. Reddening Distribution}

\subsubsection{B.1. Test Cases}

In the prior section, we assumed a young stellar population and no
reddening when modeling LAEs.  In this section, we test for modulation
in the LF if {\em all} UV-faint galaxies are characterized with a
young stellar population and lower reddening than UV-bright galaxies.
In this case, the LAEs would simply represent a phase of UV-faint
galaxies with a short duty cycle of $\la 10\%$, based on the number
density of LAEs compared to the general continuum population (e.g.,
\citealt{kovac07, nagamine08, verhamme08, gawiser07}).  Note that here
we are concerned with the changing distribution of reddening of
galaxies as a function UV magnitude {\em at a given epoch}.  A
somewhat related issue is how the reddening distribution in general
shifts to lower values at higher redshift for galaxies of a given star
formation rate (R08; see also \S~\ref{sec:sfrd}).

The first case under consideration is if the average $\ebmv$ of
galaxies with $\rs>25.5$ is zero.  This scenario would be the most
conservative one we can make for two reasons.  First, R08 demonstrate
using UV continuum-slopes and {\em Spitzer} MIPS data that the
reddening distribution of UV-selected galaxies does not vary
significantly to the spectroscopic limit of $\rs\sim 25.5$.  Because
this limit is arbitrary with respect to galaxy properties, we would
not expect the reddening to change so suddenly for galaxies fainter
than this limit and indeed it would be unphysical.  A more meaningful
model is one in which the average reddening asymptotes to zero
proceeding to fainter luminosities (R08).  Second, there is a
non-negligible fraction of galaxies at $z\sim 2-3$ on the faint-end of
the UV LF that are bolometrically-luminous and dusty (e.g.,
\citealt{chapman05, vandokkum04}).  Yet, a large fraction of these
galaxies have colors and UV opacity that do not differ significantly
from those of UV continuum-bright objects (\citealt{chapman05,
  reddy05a, reddy06a}, R08).  By assuming an average reddening that
falls to zero for galaxies fainter than $\rs=25.5$, we are effectively
seeking the {\em maximal} change in the LF under such a scenario.  We
also consider a more physical reddening distribution whose mean
$\langle\ebmv\rangle$ decreases monotonically from a value of $\langle
\ebmv\rangle \sim 0.13$ at $\rs=25.5$ to zero at the faintest apparent
magnitude bin of our analysis (case 2).  With this model, galaxies at
the faint-end of the UV LF ($M(1700\AA)\approx -18.00$) will have
close to zero reddening, similar to the mean reddening (as inferred
from the rest-UV slopes) of dropout galaxies at $z\sim 6$ with
comparable unobscured UV luminosities \citep{bouwens06}.  If we define
the $N(\ebmv)$ distribution for galaxies brighter than $\rs=25.5$ as
\begin{eqnarray}
N[\ebmv,\rs\le 25.5] \equiv N_{\rm o},
\end{eqnarray}
then our model for
the luminosity dependence of $N(\ebmv,\rs)$ can be written
as 
\begin{eqnarray}
N(\ebmv,\rs) & = & N_{\rm o}, \,\,\,\,\,\rs\le 25.5 \nonumber \\
& = & f(\langle\ebmv\rangle,\sigma(N_{\rm o})), \nonumber \\
& & \rs > 25.5,
\end{eqnarray}
where the function $f(\langle\ebmv\rangle,\sigma(N_{\rm o}))$ is a
Gaussian with mean $\langle \ebmv\rangle = 0$ (Case 1) and
$\langle\ebmv\rangle = -0.09\rs+2.43$ (Case 2) and dispersion
equivalent to that observed for $N_{\rm o}$ (i.e., $\sigma(N_{\rm
  o})$).  Note that while $N_{\rm o}$ is not in fact distributed
normally, the differences that arise by assuming a Gaussian are
negligible.  Further, for simplicity we have assumed that the
dispersion of the $\ebmv$ distribution is independent of magnitude.
R08 argue for an increased dispersion at faint magnitudes due to the
mixing of galaxies with intrinsically low star formation rates and
those that are UV-faint because of high extinction.  The effect of
such an increasing dispersion is to reduce the effective volume of the
survey and thus the incompleteness corrections will be larger at the
faint-end.  In general, however, because the number density of
UV-faint galaxies with little reddening is inferred to be much larger
than that of heavily reddened UV-faint galaxies (\S~\ref{sec:nature}),
the increase in dispersion attributable to the latter is likely to be
negligible.  In addition, at the faintest magnitudes where $\langle
\ebmv\rangle$ approaches zero, the dispersion will be dominated not by
reddening but by the intrinsic variation in SEDs of galaxies.
Consequentially, the dispersion will likely be smaller than the
$\sigma(N_{\rm 0})$ measured at brighter magnitudes.  Note that there
is a non-negligible number of galaxies in the spectroscopic sample
that have measured $\ebmv<0$.  Since we use $\ebmv$ as an indicator of
dust, we set a lower limit of $\ebmv = 0$ for any galaxies that happen
to be assigned a negative value.  Finally, R08 demonstrate that the
mean extinction among galaxies above a particular unobscured
luminosity is roughly constant with redshift between $1.9\le z<3.4$.
Motivated by this, we adopt non-evolution of reddening in the
simulations; i.e., $N(\ebmv,\rs,z)\approx N(\ebmv,\rs)$.  For brevity
in the subsequent discussion, the abbreviation ZR refers to the case
of a discontinuous reddening distribution such that galaxies with
$\rs>25.5$ have $\ebmv = 0$.  Similarly, LDR refers to our analytical
model for the luminosity-dependent reddening distribution that has
reddening decreasing monotonically with UV magnitude.

\subsubsection{B.2. Results}

\begin{figure*}[t]
\plottwo{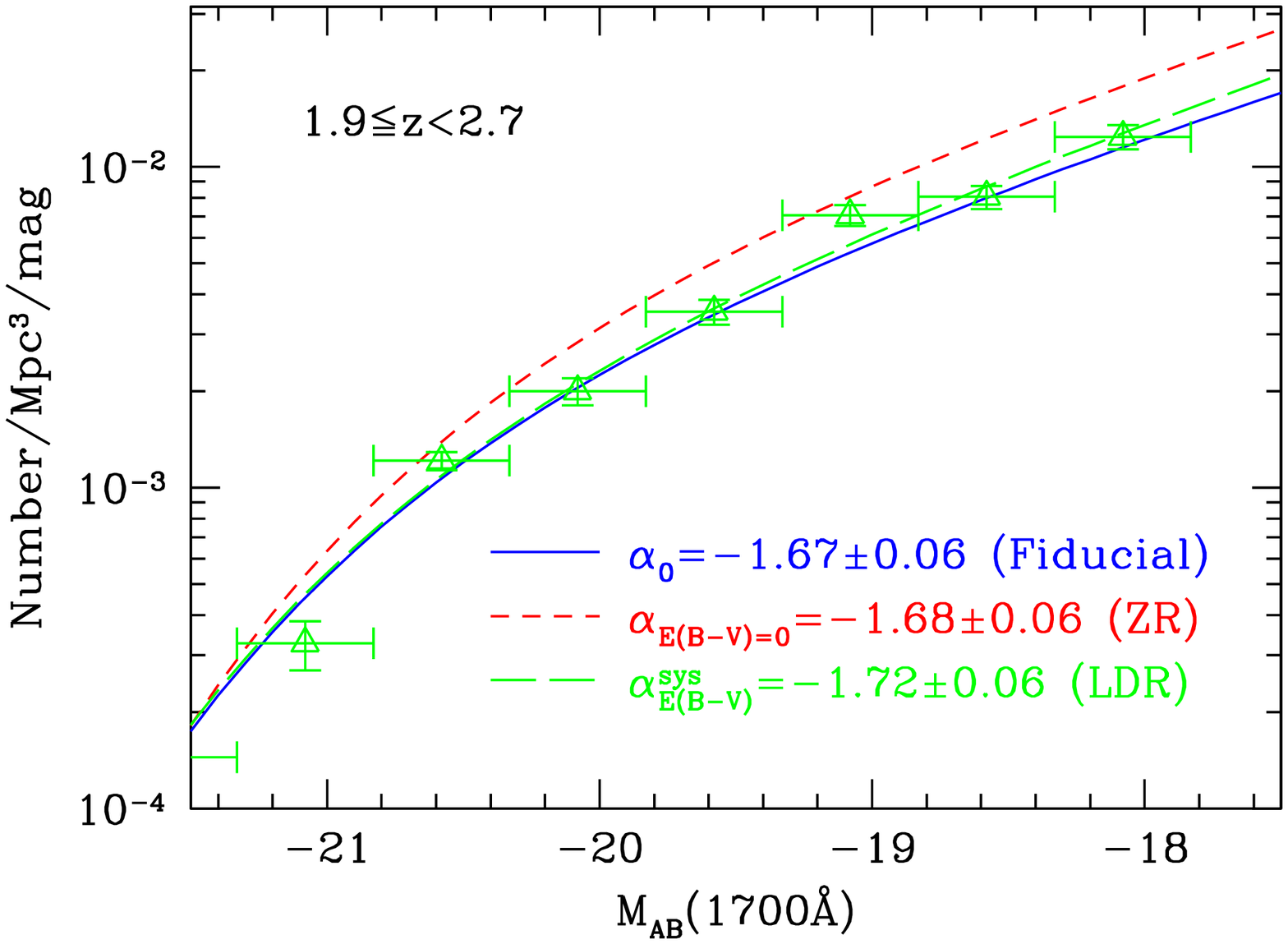}{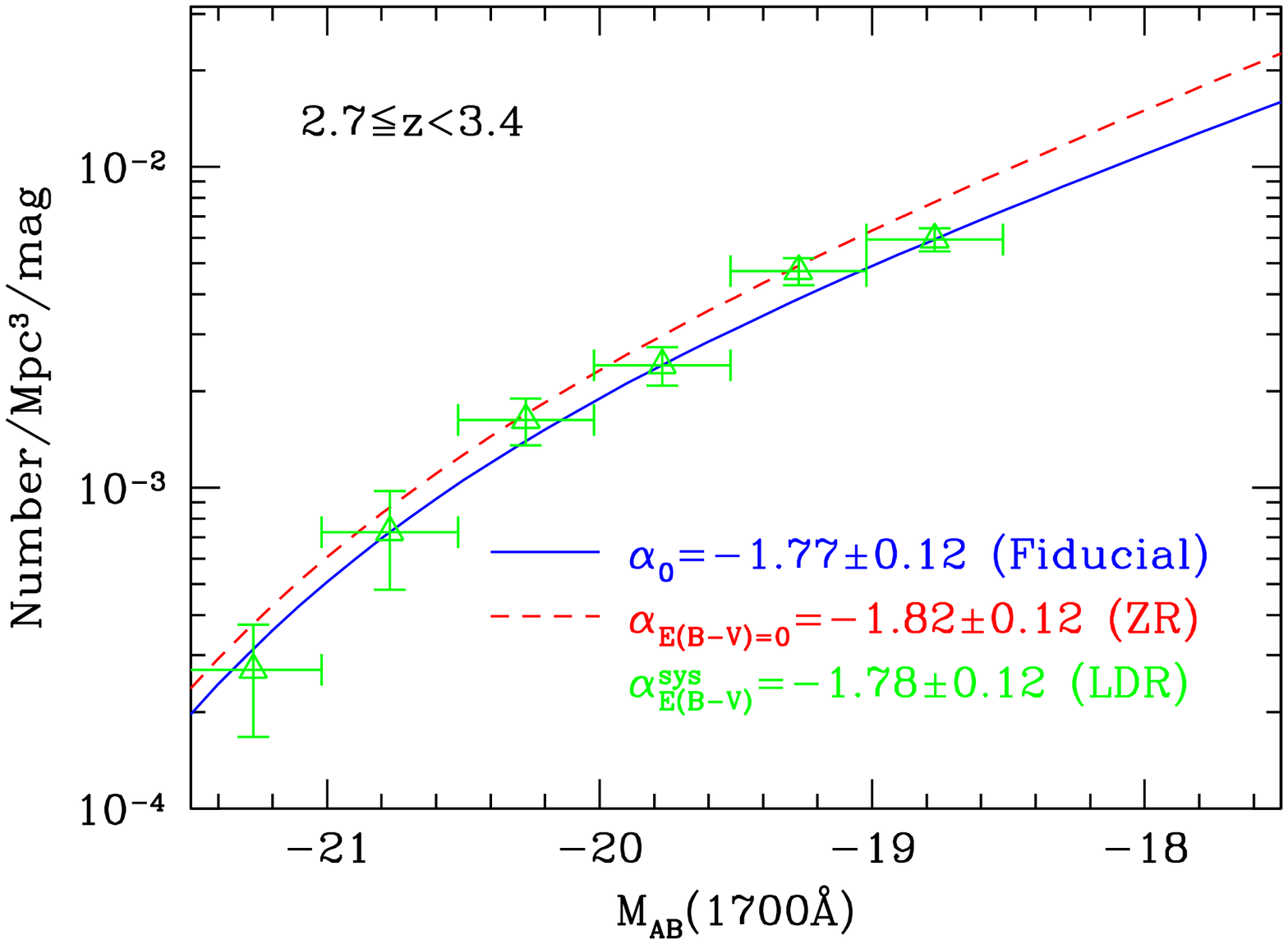}
\caption{Change in faint-end slope of the UV luminosity function
assuming that galaxies with $\rs>25.5$ have (1)
$\langle\ebmv\rangle=0$ ($\alpha_{\rm \ebmv=0}$) and (2)
$\langle\ebmv\rangle$ that falls off linearly with magnitude
($\alpha^{sys}_{\rm \ebmv}$), compared to the fiducial value that
assumes galaxies have the same $\langle\ebmv\rangle\approx 0.13$
irrespective of apparent magnitude ($\alpha_{\rm 0}$).  For clarity,
data points are shown only for case (2) in both panels and the
Schechter fit for case (2) is excluded from the right
panel.}
\label{fig:ebmvlf}
\end{figure*}

The results are summarized in Figure~\ref{fig:ebmvlf} for both case 1
(ZR), where all galaxies with $\rs>25.5$ have $\ebmv = 0$, and case 2
(LDR), where the mean reddening decreases with magnitude.  There are
several conclusions of import.  Focusing on the lower redshift
galaxies, in the ZR case we find a significant increase of up to
$70\%$ in the inferred number density of galaxies with
$M(1700\AA)>-20$ relative to that inferred from the fiducial model.
This can be understood by examining Figure~\ref{fig:lyaprob}.
Galaxies with no reddening (e.g., as in the case of the LAE model)
have bluer colors that approach the boundary of the BX color selection
window.  Thus, photometric errors preferentially scatter galaxies out
of the selection window compared to typical BXs.  The net effect is
that for a fixed number of observed BX candidates, the incompleteness
corrections will be larger, leading to larger number densities.  We
find a statistically insignificant difference between the fiducial and
ZR faint-end slopes, attributable to the fact that the reddening
distribution is fixed to have $\langle\ebmv\rangle=0$.  A gradually
declining distribution (case 2; LDR), results in number densities that
are up to $\approx 10\%$ larger and a slightly steeper faint-end
slope, although the parameters of the Schechter function are still
consistent with those of the fiducial case within their respective
marginalized errors.

The luminosity-dependent variation of the $\ebmv$ distribution has
less of an effect on the faint-end number densities in the higher
redshift range $2.7\le z<3.4$, primarily because the LBG criteria
include colors that are much bluer than those expected for even a
young and unreddened stellar population.  However, because we account
also for the intrinsic scatter in $N(\ebmv)$, the differences in
faint-end number densities are still somewhat larger than we would
have obtained had we modeled the $N(\ebmv)$ distribution as a $\delta$
function at a given $\rs$-band magnitude (i.e., assuming all galaxies
at a given $\rs$ have the same reddening).  In any case, the LBG
criteria are somewhat more robust to extreme assumptions regarding the
$\ebmv$ distribution of UV-faint galaxies compared to the BX criteria.
Simply altering the BX criteria to include candidates with bluer
$\gmr$ colors could alleviate some of this difference, but we note
that our deep photometry indicates that galaxies with such blue colors
($\gmr \la -0.2$) are rare.  For the purposes of our present analysis,
the exact placement of the color criteria is not important as long as
the effect of the criteria is modeled appropriately and incompleteness
is accounted for using a likelihood analysis.

Deep spectroscopy combined with multi-wavelength indicators of the
extinction of sub-$L^{\ast}$ galaxies is required to more robustly
constrain the $\ebmv$ distribution as a function of unobscured UV
luminosity.  For the time being, however, we have shown that adopting
simple scenarios for how the distribution changes with UV luminosity
results in a faint-end slope that can be potentially steeper at $z\sim
2$ than we would have obtained by assuming a constant $\ebmv$
distribution.  For our final determination of the UV LF, we have
adopted our prescription for the luminosity dependence of reddening,
namely one in which the reddening declines linearly with apparent
magnitude, as discussed above.  The resulting best-fit Schechter
parameters are listed in Table~\ref{tab:schechter}.

\subsection{C. Objects Outside the Redshift Ranges of Interest}

Intrinsic variations in the SEDs of star-forming galaxies and
photometric errors lead to significant wings of the redshift selection
functions, $N(z,L)$, for color-selected samples.  Here, the selection
function is parameterized in terms of both redshift and luminosity.
In this analysis, we have computed the LFs specifically over the
redshift ranges $1.9\le z<2.7$ and $2.7\le z<3.4$, although the exact
redshift interval used is unimportant since the LF does not evolve
over these redshifts (see R08 and \S~\ref{sec:lfevol}).  More
importantly, we estimate the fraction of photometric candidates that
fall in the redshift ranges over which the LFs are computed,
information that is provided directly from the spectroscopic sample.

In the faintest apparent magnitude bin of the spectroscopic sample,
$25.0\le \rs <25.5$, the observed fractions of galaxies (excluding
AGN/QSOs) that fall in the ranges $1.9\le z<2.7$ and $2.7\le z<3.4$
are $77\%$ and $72\%$, respectively (virtually all of the objects that
are outside these redshift ranges still have $z>1$ since the
contamination due to $z\le 1$ objects is very small at these faint
magnitudes --- see R08).  In the previous analysis, the fractions of
$77\%$ and $72\%$ are assumed to remain constant for galaxies fainter
than $\rs=25.5$.  There are at least a couple of reasons why this is
likely to be a reasonable assumption.  First, the contamination from
$z<1$ objects to the photometric sample is a strong function of
magnitude.  If this trend continues to fainter magnitudes, then we
would expect the $z<1$ contamination rate to be less than $1\%$ for
objects fainter than $\rs=25.5$.  Adopting zero contamination from
$z<1$ objects at $\rs>25.5$ results in number densities at the
faint-end that are $\approx 1\%$ larger.  This is an insignificant
change given the magnitude of the other systematics discussed above.
In theory, the larger photometric errors for UV-faint objects may
result in an increase of the contamination rate at the faint-end.  A
small sample of 15 spectroscopically-confirmed galaxies with
$25.0<\rs<26.0$ in the Q1422 field (\citealt{steidel03}; R08) implies
a zero contamination fraction.  Larger spectroscopic samples of
UV-faint objects will be required to obtain a statistically-robust
estimate of contamination at the faint-end.  However, all of the
previous studies that have attempted to constrain the faint-end of the
UV LF \citep{steidel99, sawicki06a} also assume negligible
contamination at the faint-end.  Hence, the difference between our
determination of a steep-faint end slope and the shallower values
found elsewhere (see \S~\ref{sec:context}) cannot be attributable to
differences in the assumed contamination rate as a function of UV
luminosity.

Second, the redshift distribution for UV-selected galaxies will be
modulated primarily by systematics that affect the overall colors of
galaxies at these redshifts, namely Ly$\alpha$ perturbations and the
$\ebmv$ distribution, such that $N(z,L)=f(N[W_{\rm
Ly\alpha}],N[E(B-V)])$.  From this discussion, we conclude that the
redshift distribution likely remains similar between UV-bright and
faint galaxies (i.e., $N(z,L)\approx N(z)$).  It is easy to see,
however, the potential danger of assuming that $N(z)$ is insensitive
to luminosity: even gradual changes in the stellar population and
reddening of galaxies as a function of magnitude may result in an
artificial suppression of the faint-end of the UV LF with respect to
the bright-end.  This motivates the need for selection criteria that
are efficient at targeting galaxies with a wide range of intrinsic
properties at the desired redshifts (see next section).

For the selection criteria adopted here, the presence of high $W_{\rm
Ly\alpha}$ systems and/or a bluer population of UV-faint galaxies does
little to alter the parameterization of the maximum-likelihood LF
under reasonable assumptions for LAE number densities and the
reddening distribution.  In other words, the modulation of $N(z,L)$,
and more generally $\xi$, the transitional probability function, due
to these systematic effects do not affect appreciably our LF
determination.  Substantial changes in the redshift distribution {\em
can} arise from a rapid evolution of the LAE number density and
$\ebmv$ distributions over the redshifts of concern.  However, lacking
evidence that such evolution is occurring --- and, as discussed above,
there is little evolution in the $\ebmv$ distribution over these
redshifts; R08) --- it is likely that the redshift distributions of
BXs and LBGs with $\rs>25.5$ is similar to those of $\rs<25.5$
galaxies.

\subsection{D. Implications for Color Selection at High Redshift}
\label{sec:selcrit}

It is instructive to take a broader view and examine the significance
of the tests described here in the context of the color selection
criteria.  The primary result of this section is that the systematics
brought about by reasonable assumptions for the unobscured UV
luminosity dependence of the $W_{\rm Ly\alpha}$ and reddening
distributions do little to alter our inference of the UV LF {\em for
the selection criteria used here}.  We emphasize the latter part since
obviously some sets of criterion will be more susceptible to
modulations of $N(W_{\rm Ly\alpha})$ and $N(E[B-V])$ than others.  As
has been discussed elsewhere \citep{steidel03, steidel04,
adelberger04}, the goal of observing efficiency dictates that a
balance be struck between the inclusion of as many galaxies at the
redshifts of interest as possible and the exclusion of as many
foreground or background contaminants.  In the context of the present
analysis, luminosity-dependent properties of galaxies should also be
taken into account when designing color criteria.  UV-dropout criteria
are in general the most efficient method for selecting high-redshift
galaxies.  Because they target preferentially bluer galaxies, the
luminosity-dependent systematics expected for UV-faint galaxies works
in favor of their selection via rest-frame UV emission.  In contrast,
near-IR selections that target redder galaxies (either because they
are dusty, have large stellar masses, or both), may potentially miss
an appreciable fraction of galaxies that populate the faint-end of the
UV LF (\S~\ref{sec:nature}).  Hence, the aggregate of these selection
methods provides a complementary view and are necessary for obtaining
an unbiased census of star formation.

Obviously, incompleteness corrections for criteria that target blue
galaxies are not completely immune to extreme luminosity-dependent
gradients in $N(E[B-V],\rs)$, for example (Figure~\ref{fig:ebmvlf}).
The power of our simulations and maximum-likelihood method is that
they can be used to quantify and correct for even severe biases (e.g.,
in the faint-end slope) imposed by high redshift galaxy selection in a
way that is not possible with the traditional methods of computing
luminosity functions (see discussion in R08).  The keystone of our
entire method is the spectroscopy: while not extending fainter than
the typical ground-based magnitude limit, spectroscopy for UV-bright
galaxies does provide a critical foundation, or ``zero-point,'' for
assessing how luminosity dependent systematics may affect our
inferences of the faint-end.  Using these techniques, we showed in R08
that, after correcting for low redshift objects and AGN/QSOs based on
extensive spectroscopy, the UV LF inferred by magnitude limited
surveys is similar to that derived from color-selected surveys.
Hence, we argued that we must be complete for UV-bright galaxies.  The
slight modifications of the Schechter parameterization of the LF after
taking into account $N(W_{\rm Ly\alpha},\rs)$ and $N(E[B-V],\rs)$
implies that our determination must also be reasonably complete for
UV-faint galaxies.

\end{document}